\documentclass[fleqn,usenatbib]{mnras}
\usepackage{newtxtext,newtxmath}

\usepackage[T1]{fontenc}

\DeclareRobustCommand{\VAN}[3]{#2}
\let\VANthebibliography\thebibliography
\def\thebibliography{\DeclareRobustCommand{\VAN}[3]{##3}\VANthebibliography}

\usepackage{graphicx}	
\usepackage{amsmath}	
\usepackage{amssymb}	
\usepackage{mathtools}
\usepackage{booktabs}
\usepackage{multirow}
\usepackage{adjustbox}
\usepackage{longtable,tabularx}

\usepackage{nomencl}
\makenomenclature

\title[King-Hele theory for periodic variations]{King-Hele orbit theory for periodic orbit and attitude variations}

\author[Ray and Scheeres]{
Vishal Ray,$^{1}$\thanks{vishal.ray@colorado.edu}
Daniel J. Scheeres $^{2}$
\\
$^{1,2}$Ann and H.J Smead Aerospace Engineering Sciences, University of Colorado Boulder, 3775 Discover Dr, 80303, USA
}

\date{Accepted XXX. Received YYY; in original form ZZZ}

\pubyear{2020}

\begin{document}
\label{firstpage}
\pagerange{\pageref{firstpage}--\pageref{lastpage}}
\maketitle

\begin{abstract}
The analytical theory of satellite orbits in an atmosphere developed by King-Hele remains widely in use for satellite mission design because of its accurate approximation to numerical integration under simplifying assumptions. Over the course of six decades, modifications to the theory have addressed many of its weaknesses. However, in all subsequent modifications of the original theory, the assumption of a constant drag-coefficient has been retained. The drag-coefficient is a dynamic parameter that governs the physical interaction between the atmosphere and the satellite and depends on ambient as well as satellite specific factors. In this work, Fourier series expansion models of the drag-coefficient are incorporated in the original King-Hele theory to capture time-variations of the drag-coefficient in averaging integrals. The modified theory is validated through simulations that demonstrate the attained improvements in approximating numerical results over the original King-Hele formulation. 
\end{abstract}



\mbox{}
\nomenclature[01]{$e$}{eccentricity}
\nomenclature[02]{$a$}{semi-major axis}
\nomenclature[03]{$f_T$}{drag-force magnitude (absolute velocity direction)}
\nomenclature[04]{$\rho$}{atmospheric density}
\nomenclature[05]{$\delta$}{drag-parameter}
\nomenclature[06]{$\text{v}$}{magnitude of inertial velocity}
\nomenclature[07]{$F$}{wind-factor}
\nomenclature[08]{$m_s$}{mass of satellite}
\nomenclature[09]{$C_D$}{drag-coefficient}
\nomenclature[10]{$S$}{cross-sectional area}
\nomenclature[11]{$r$}{orbital radius}
\nomenclature[12]{$w$}{angular velocity of the atmosphere}
\nomenclature[13]{$i$}{inclination}
\nomenclature[14]{$\Omega$}{right-ascension of ascending node}
\nomenclature[15]{$\omega$}{argument of perigee}
\nomenclature[16]{$\mu$}{gravitational parameter}
\nomenclature[17]{$f_N$}{cross atmospheric force magnitude (perpendicular to orbital plane)}
\nomenclature[18]{$\theta$}{true anomaly}
\nomenclature[19]{$x$}{focal-length (product of semi-major axis and eccentricity)}
\nomenclature[20]{$E$}{eccentric anomaly}
\nomenclature[21]{$\delta ' $}{$\frac{\delta}{C_D}$}
\nomenclature[22]{$H$}{scale-height}
\nomenclature[23]{$\beta$}{inverse scale-height}
\nomenclature[24]{$n$, $m$}{orders of Fourier model}
\nomenclature[25]{$\overline{\mathbb{A}}_n$, $\overline{\mathbb{B}}_n$}{cosine and sine orbit-fixed Fourier coefficients respectively}
\nomenclature[26]{$\overline{\mathcal{A}}_n$, $\overline{\mathcal{B}}_n$}{cosine and sine body-fixed Fourier coefficients respectively}
\nomenclature[27]{$\overline{\mathcal{A}}_{mn}$, $\overline{\mathcal{B}}_{mn}$}{cosine and sine body-orbit double Fourier coefficients}
\nomenclature[28]{$D_c$}{constant factor in King-Hele integrals}
\nomenclature[29]{$I_n$}{modified Bessel function of the first kind with imaginary argument}
\nomenclature[30]{$\Delta$}{ change over an orbit}
\nomenclature[31]{$\lambda$}{auxiliary variable in high eccentricity regime}
\nomenclature[32]{$z$}{ratio of focal-length and scale-height}
\nomenclature[33]{$\Gamma$}{Gamma function}
\nomenclature[34]{$\phi$}{angle of rotation of velocity vector in body frame}
\nomenclature[35]{$\text{v}_P$}{velocity magnitude along P-axis in perifocal frame}
\nomenclature[36]{$\text{v}_Q$}{velocity magnitude along Q-axis in perifocal frame}
\nomenclature[37]{Subscript $0$}{initial value}
\nomenclature[38]{Subscript $p0$}{initial value at perigee}
\printnomenclature

\section{Introduction}

The motion of satellites in an atmosphere is governed by parameters with complex time-dependent profiles such as the atmospheric density and the drag-coefficient that represents the gas-surface interactions between the satellite surface and atmosphere. A prediction of the satellite states, the position and velocity or the orbital elements, can be most accurately obtained by numerical integration of the equations of motion. But numerical integration is a computationally expensive process and does not prove feasible for analysis of long-term evolution of the orbital elements. A time profile of the semi-major axis and eccentricity evolution is required to obtain an estimate of the satellite lifetime which is indispensable for mission design and maneuver planning. Therefore, a closed-form analytical solution of the change in semi-major axis and eccentricity, i.e., orbit contraction, is essential. Fortunately, a closed-form solution is possible under some simplifying assumptions regarding the atmosphere and was outlined by \citet{kh} in his comprehensive treatise on the subject. Assuming an exponentially decaying atmosphere with constant density surfaces at any altitude, expressions for changes in semi-major axis and eccentricity, averaged over an orbital period, were derived for both spherically symmetrical and oblate atmospheres. The integration of the Lagrange planetary equations led to separate series formulations for three eccentricity regimes - circular, low eccentricity ($e<0.2$) and high eccentricity ($e > 0.2$). 

Several improvements to the original King-Hele formulation have been developed in the literature. Whereas King-Hele developed separate formulations for low and high eccentricity regimes with an empirical boundary condition of $e=0.2$ using heuristic methods for integration, \citet{vinh} provided a rigorous analytic solution using Poincaré's method for small parameters \citep{poincare}. The singularities arising in classical orbital elements for nearly circular orbits were removed by \citet{sharma1} by formulating the theory in non-singular elements. One of the weakest assumptions in the original King-Hele formulation is the stationary and exponentially decaying nature of the atmospheric density. The effect of diurnal and latitudinal variation of atmospheric density on the orbital elements have been addressed by various authors \citep{kh}. The assumption of a fixed scale height for the exponentially decaying density can introduce large errors as the distance from the perigee increases. \citet{kh} tried to address this approximation by assuming a linear variation of scale height. More recently, significant advances were made in incorporating a generic atmospheric model in the theory by fitting multiple exponentially decaying partial atmospheres to the model \citep{Frey}. \citet{Frey} were also able to arrive at a variable boundary condition for the eccentricity regime that was fixed by King-Hele at 0.2. 

\subsection{Proposed improvement to the original King-Hele theory}
In the original King-Hele theory and all the modifications thereafter, the drag-coefficient has been considered constant. Advances in the study of the drag-coefficient for low altitude satellites in the past few decades have revealed the dynamic nature of the parameter with variations correlated with the atmospheric density as well as independent of it. Several numerical \citep{meh,resp,sun} and analytical methods \citep{ walker,sesam,Moe2005} exist to capture the time-variation of the drag-coefficient tied to ambient parameters such as atmospheric composition and satellite-specific factors such as attitude. All these methods assume a functional dependence of the drag-coefficient on various input parameters through a gas-surface interaction model. It is not possible to incorporate these complex drag-coefficients models in their original form in the King-Hele theory and obtain a closed-form solution. But a parameterization of these models in terms of a high-frequency orbital element such as true anomaly or eccentric anomaly can allow the drag coefficient to vary in the King-Hele theory. The variation of the drag coefficient is periodic with the velocity vector in the body frame of the satellite and nearly-periodic in the orbital frame. This allows the drag-coefficient to be expanded as a Fourier series in the body frame and orbit frame of the satellite. The authors proposed body-dependent and orbit-dependent Fourier drag-coefficient models and demonstrated their improved performance in orbit determination and prediction over the standard `cannonball' model that estimates the drag-coefficient as a constant \citep{jastr,jgcd_drag}. In this work, the original King-Hele theory for a spherically symmetric atmosphere is expanded upon by allowing the drag-coefficient to vary in time using the Fourier drag-coefficient models. This extended King-Hele theory will not only be useful in improving lifetime estimates of the satellite but also improve derivation of atmospheric densities from satellite decay data using general perturbation methods since the drag-coefficient is assumed to be constant in such studies \citep{tle}. The extended theory also reveals the value of the constant drag-coefficient that should be used in the original King-Hele formulation to obtain an accurate approximation to numerical results. Additionally, this extension of the King-Hele formulation can be combined with the other improvements outlined previously to obtain a generalized analytical theory of satellite orbits in the presence of atmospheric drag.
\subsection{Outline}
The analytical change in the semi-major axis, eccentricity and argument of perigee over an orbital period with the proposed extension is derived in detail while noting that the other orbital elements remain constant under the given assumptions. The orbit and body-fixed Fourier models are discussed in section \ref{Fourier}. Section \ref{lpe} outlines the Lagrange planetary equations for the orbital elements following King-Hele's formulation. In section \ref{off}, the orbit-fixed Fourier (OFF) model is applied to the King-Hele theory and the orbital element changes are derived for both low eccentricity and high eccentricity regimes. The drag-coefficient is assumed to vary solely due to ambient parameters and the orientation w.r.t the atmosphere is assumed to be constant. Section \ref{bff} carries out the procedure for the body-fixed Fourier (BFF) model for a nadir-pointing and an inertially stabilized satellite where the drag-coefficient is assumed to vary solely due to changes in orientation of velocity vector in the body-frame. But the drag-coefficient in an actual scenario varies due to both the factors. An approximate method to capture the dependence of the drag-coefficient on both the factors is provided. In section \ref{circ}, it is demonstrated that the higher-order Fourier coefficients do not contribute to the change in orbital elements for a circular orbit under the assumptions of the King-Hele theory. The theory is validated using simulated satellite trajectories in section \ref{valid}. Finally, section \ref{conc} discusses and summarizes the developed theory. 

\section{Fourier drag-coefficient models}
\label{Fourier}
This section outlines the Fourier drag-coefficient models developed in \citet{jastr,jgcd_drag} that is used in this work to model the drag-coefficient in the King-Hele theory. The previous papers captured the dependence of the drag-coefficient on ambient parameters by carrying out a Fourier series expansion around the argument of latitude in the orbit frame - the OFF model. The dependence on satellite orientation was modeled using a Fourier series expansion around the orientation of the velocity vector in the body frame - the BFF model. In order to derive closed-form analytical solutions of the change in orbital elements, the Fourier drag-coefficient models need to be expressed in terms of the eccentric anomaly, such that the drag-coefficient becomes a function of the eccentric anomaly, $C_D(E)$. The transformation of the time variation of the drag-coefficient to eccentric anomaly simplifies the integration of the Lagrange planetary equations as will be seen in section \ref{lpe}. It should be noted that the time variations of the orbit are still being accurately tracked using Kepler's equation. The inclusion of drag-coefficient variation in the perturbation equations is an added layer of complexity that improves the approximation to the true variation of the orbits.
\subsection{Orbit-fixed Fourier (OFF) model}
\label{off_intro}
The drag-coefficient varies with ambient parameters such as the partial pressure of atomic oxygen, atmospheric composition and ambient temperature. Under the assumptions of a symmetric exponentially decaying atmosphere, the atmospheric composition and partial pressure of oxygen are both periodic in the orbit since they are functions of only altitude in this case while the ambient temperature is constant. The drag-coefficient is also a function of the velocity of the satellite that is periodic as well. Therefore, the drag-coefficient can be expressed as a Fourier series expansion around the eccentric anomaly in the orbit frame of the satellite as follows
\begin{equation}
\label{orb}
    C_D =  \sum_{n=0}^{\infty} (\overline{\mathbb{A}}_{n}\cos{nE} + \overline{\mathbb{B}}_{n}\sin{nE}). 
\end{equation}
where $\overline{\mathbb{A}}_{n}$ and $\overline{\mathbb{B}}_{n}$ are Fourier coefficients that are calculated by integrating the drag coefficient over one period as follows
\begin{equation}
\label{An}
    \overline{\mathbb{A}}_{n} = \frac{1}{\pi}\int_{0}^{2\pi}C_D(E)\cos{nE}dE,
\end{equation}
\begin{equation}
\label{Bn}
      \overline{\mathbb{B}}_{n} = \frac{1}{\pi}\int_{0}^{2\pi}C_D(E)\sin{nE}dE,
\end{equation}
for $n>0$ and, 
\begin{equation}
\label{A0}
      \overline{\mathbb{A}}_{0} = \frac{1}{2\pi}\int_{0}^{2\pi}C_D(E) dE,
\end{equation}

for $n=0$. Any drag-coefficient model can be parameterized in this manner by numerically evaluating the integrals given by Eqs. \ref{An}-\ref{A0}. 

\subsection{Body-fixed Fourier (BFF) model}
\label{bff_intro}
In order to capture the variation of the drag-coefficient with attitude, the drag-coefficient can be expressed as a Fourier-series in the body frame. The drag-coefficient is expanded as a Fourier series around the orientation of the inertial velocity vector in the body frame. It is assumed that the variation is around a single axis, i.e., the axis of rotation is fixed in the body frame. Therefore, the following theory is valid only for specific attitude profiles such as for a nadir pointing profile or inertially stabilized profile where the change in angle due to the rotating atmosphere is neglected. The drag-coefficient can be written as, 
\begin{equation}
\label{bodcd}
    C_D =  \sum_{n=0}^{\infty} (\overline{\mathcal{A}}_n\cos{n\phi} + \overline{\mathcal{B}}_n\sin{n\phi}). 
\end{equation}
where $\phi$ is the angle that the velocity vector makes in the body frame with a reference axis perpendicular to the axis of rotation. The Fourier coefficients $\overline{\mathcal{A}}_n$ and $\overline{\mathcal{B}}_n$ are given by,
\begin{equation}
\label{An1}
    \overline{\mathcal{A}}_{n} = \frac{1}{\pi}\int_{0}^{2\pi}C_D(\phi)\cos{n\phi}d\phi,
\end{equation}
\begin{equation}
\label{Bn1}
      \overline{\mathcal{B}}_{n} = \frac{1}{\pi}\int_{0}^{2\pi}C_D(\phi)\sin{n\phi}d\phi,
\end{equation}
for $n>0$ and, 
\begin{equation}
\label{A01}
      \overline{\mathcal{A}}_{0} = \frac{1}{2\pi}\int_{0}^{2\pi}C_D(\phi) d\phi,
\end{equation}
The drag-coefficient can be expressed as a function of the eccentric anomaly for specific cases where a transformation exists between angle of velocity vector in the body frame and the eccentric anomaly. Note the fundamental difference between Eqs. \ref{An} - \ref{A0} and Eqs. \ref{An1}-\ref{A01}. The first set integrates the drag-coefficient over an orbital period, taking into account its dependence on parameters that are periodic in orbit and are therefore functions of the orbital elements. The second set of equations integrates the drag coefficients over a rotation in the body frame, with orbit-dependent parameters considered constant. Whereas the OFF coefficients are functions of attitude that is considered constant, the BFF coefficients are functions of orbital parameters that are considered constant. 
\section{Lagrange planetary equations for air drag}
\label{lpe}
In this section, the Lagrange planetary equations for the classical orbital elements in terms of the eccentric anomaly are repeated using King-Hele's notation. The magnitude of the drag force acting tangential to the orbit is given by
\begin{equation}
    f_T = -\frac{1}{2}\rho \text{v}^2 \delta.
\end{equation}
The variation in velocity direction due to the rotating atmosphere is neglected since the angle between the absolute velocity vector and relative velocity vector never exceeds $\approx 4^\circ$ \citep{jastr} but the magnitude is accounted for in the drag-parameter that is assumed to be constant,  
\[\delta = \frac{FSC_D}{m_s}.\]
$F$ is the wind-factor that accounts for the relative speed w.r.t the atmosphere, given by
\[F = (1-\frac{r_{p0}w}{\text{v}_{p0}}\cos{i_0})^2.\]
The atmosphere in this theory is assumed to be symmetric and exponentially decaying with a constant scale-height,
\begin{equation}
    \rho = \rho_{p0}\exp{\{(r_{p0}-r)/H\}}.
\end{equation}
A closed-form analytical theory is possible for a higher-fidelity density model accounting for oblateness of the atmosphere, day-night and solar activity variations, meridional winds and varying scale-height \citep{kh2,Frey}. But the modified theory is developed for the simplest case in this work and can be extended to incorporate other refinements.  

Following \citet{kh}, the theory is derived for the semi-major axis ($a$), the focal-length ($x = ae$) and the argument of periapsis ($\omega$) of the satellite orbit. The orientation of the orbital plane is affected by atmospheric rotation leading to time-variations in the inclination, right-ascension of ascending node and argument of perigee. Whereas the inclination and right-ascension vary solely due to atmospheric forces perpendicular to the orbit plane, the argument of periapsis depends on the forces in the orbital plane. In this work, the forces perpendicular to the velocity direction are neglected. Therefore, the variation of inclination and right-ascension is considered to be zero. The Lagrange planetary equations for semi-major axis and eccentricity expressed in the tangential ($T$) and orbit inward normal in the orbit plane ($N$) directions are as follows-
\begin{equation}
\label{dota}
    \dot{a} = \frac{2a^2\text{v}}{\mu}f_T,
\end{equation}
\begin{equation}
\label{dote}
    \dot{e} = \frac{1}{\text{v}}\left\{2f_T(e+\cos{\theta})-f_N\frac{r}{a}\sin{\theta}\right\}.
\end{equation}
For the argument of perigee, the derivation deviates a little from King-Hele, since the normal forces due to atmospheric rotation are neglected here. The Lagrange Planetary equation for argument of perigee expressed in the radial ($r$)-transverse ($t$) direction is given by \citet{kh}
\begin{equation}
\label{omg1}
    \dot{\omega} + \dot{\Omega}\cos{i} = \frac{1}{\text{e}}\sqrt{\dfrac{p}{\mu}}\left\{-f_r\cos{\theta} + f_t\left(1+\dfrac{r}{p}\right)\sin{\theta}\right\}.
\end{equation}
The forces in the radial-transverse direction can be expressed in the tangential-inward normal directions as \cite{kh}
\[
    f_r = \dfrac{1}{\text{v}}\sqrt{\dfrac{\mu}{p}}\{f_Te\sin{\theta}-f_N(1+e\cos{\theta})\}
\]
\begin{equation}
\label{tn}
    f_t = \dfrac{1}{\text{v}}\sqrt{\dfrac{\mu}{p}}\{f_T(1+e\cos{\theta})+f_Ne\sin{\theta}\}    
\end{equation}
Substituting Eq. \ref{tn} in \ref{omg1} and simplifying,
\begin{equation}
   \begin{split}
\label{omg2}
   \dot{\omega} = & \dfrac{1}{\text{v}e}\left[f_T\left\{\sin{\theta} + \dfrac{r}{p}(1+e\cos{\theta})\sin{\theta} \right\} \right.\\
   & \left. + f_N\left\{e+\cos{\theta} +\dfrac{r}{p}e\sin^2{\theta}\right\}\right]
   \end{split}
\end{equation}
The rate of right-ascension is neglected here, as discussed before. Considering only the drag force, Eqs. \ref{dota}, \ref{dote} and \ref{omg2} can be re-written as,
\begin{equation}
    \dot{a} = -\frac{a^2\rho\delta \text{v}^3}{\mu},
\end{equation}
\begin{equation}
    \dot{e} = -\rho\delta \text{v}(e+\cos{\theta}).
\end{equation}
\begin{equation}
\label{omg3}
   \dot{\omega} = \dfrac{-\rho \text{v}\delta}{2e}\left[\sin{\theta} + \dfrac{r}{p}(1+e\cos{\theta})\sin{\theta} \right]
\end{equation}
It is desirable to transform the time variable into eccentric anomaly as that simplifies the integration of these equations. After transforming the time variable to eccentric anomaly, the final form of the Lagrange planetary equations is \citep{kh}
\begin{equation}
\label{a_der}
    \frac{da}{dE} = -a^2\rho\delta\frac{(1+e\cos{E})^{3/2}}{(1-e\cos{E})^{1/2}},
\end{equation}
\begin{equation}
\label{e_der}
    \frac{dx}{dE} = -a^2\rho\delta\left(\frac{1+e\cos{E}}{1-e\cos{E}}\right)^{1/2}(\cos{E}+e),
\end{equation}
\begin{equation}
\label{omg_der}
    \frac{d\omega}{dE} = -\dfrac{a\rho\delta}{e}\sqrt{1-e^2}\left(\frac{1+e\cos{E}}{1-e\cos{E}}\right)^{1/2}\sin{E}.
\end{equation}
In the original formulation \citep{kh}, the equations are integrated over an orbital period by assuming $\delta$ to be constant. This is modified by allowing the drag-coefficient to vary in orbit as a function of the eccentric anomaly, i.e., $\delta = \delta'C_D(E)$. The modified integrated equations are given by,
\begin{equation}
\label{a_lpe}
    \Delta a = -a^2\delta'\int_0^{2\pi}\frac{(1+e\cos{E})^{3/2}}{(1-e\cos{E})^{1/2}}C_D(E)\rho dE,
\end{equation}
\begin{equation}
\label{e_lpe}
    \Delta x = -a^2\delta'\int_0^{2\pi}\left(\frac{1+e\cos{E}}{1-e\cos{E}}\right)^{1/2}(\cos{E}+e)C_D(E)\rho dE.
\end{equation}
\begin{equation}
\label{omg_lpe}
   \Delta \omega = -\dfrac{a\delta'}{e}\sqrt{1-e^2}\int_0^{2\pi}\left(\frac{1+e\cos{E}}{1-e\cos{E}}\right)^{1/2}\sin{E} C_D(E)\rho dE.
\end{equation}
The density transformed to eccentric anomaly is given by,
\begin{equation}
\label{den}
    \rho = \rho_{p0}\exp{\{\beta(a_0-a-x_0) + \beta x\cos{E}\}}.
\end{equation}
Substituting Eq. \ref{den} in Eqs. \ref{a_lpe}-\ref{omg_lpe}, 
\begin{equation}
\label{af}
   \begin{split}
    \Delta a  = & -\delta'a^2\rho_{p0}\exp{\{\beta(a_0-a-x_0)\}}  \int_0^{2\pi}\left[\frac{(1+e\cos{E})^{3/2}}{(1-e\cos{E})^{1/2}}\right.\\ & \times \left.C_D(E)\exp{(\beta x\cos{E})}\right] dE
    \end{split}
\end{equation}
\begin{equation}
\label{ef}
\begin{split}
    \Delta x = & -\delta'a^2\rho_{p0}\exp{\beta(a_0-a-x_0)\}}  \int_0^{2\pi}\left[\left(\frac{1+e\cos{E}}{1-e\cos{E}}\right)^{1/2} \right. \\ & \left. \times(\cos{E}+e)C_D(E)\exp{(\beta x\cos{E})}\right] dE
\end{split}
\end{equation}

\begin{equation}
\label{omgf}
\begin{split}
    \Delta \omega = & -\dfrac{\delta'a}{e}\sqrt{1-e^2}\rho_{p0}\exp{\beta(a_0-a-x_0)\}}\\ & \times \int_0^{2\pi}\left[\left(\frac{1+e\cos{E}}{1-e\cos{E}}\right)^{1/2}\sin{E}C_D(E)\exp{(\beta x\cos{E})}\right] dE
\end{split}
\end{equation}
In order to integrate these equations, the integrand (without the time-varying drag-coefficient) is expressed as a power series expansion in $e$ and truncated at the third order by King-Hele. A similar approach is followed here after expressing the drag-coefficient as an analytical function in eccentric anomaly using the Fourier models of section \ref{Fourier}. An important point to note here is that the R.H.S of Eq. \ref{omg_der} is an odd function unlike Eqs. \ref{a_der} and \ref{e_der}. Therefore, when it is integrated from $0$ to $2\pi$, Eq. \ref{omg_der} should integrate to zero for a spherically symmetric, non-rotating atmosphere as noted by \citet{kh}. But since the drag-coefficient is considered time-varying here, the argument of perigee variation can integrate to a non-zero value as will be demonstrated in subsequent sections. 

\section{Re-deriving the King-Hele theory using OFF model}
\label{off}
The dependence of the drag-coefficient on eccentric anomaly in the OFF model is introduced through input ambient parameters to the chosen drag-coefficient model such as the partial pressure of oxygen and mean molecular mass that vary in the orbit. As noted in section \ref{off_intro}, for an exponentially decaying spherically symmetric atmosphere, they are dependent only on the altitude and therefore symmetric about $E =0, \pi$. Therefore $C_D(E)$ is an even function of eccentric anomaly and Eq. \ref{Bn} reduces to 
\[\overline{\mathbb{B}}_{n} = 0.\]
Note that this is, in general, not true for an arbitrary atmosphere. Since there is no odd component to the drag-coefficient, the integrand in Eq. \ref{omgf} is odd and therefore, the argument of perigee change is zero over an orbit. The following sections provide details of the derivation for low and high eccentricities with the OFF drag-coefficient model. Note that the derivations are independent of the gas-surface interaction model considered.

\subsection{Low eccentricity regime, e<0.2}
For low eccentricities, Eqs. \ref{af} and \ref{ef} can be integrated by expanding the integrands as power series in $e$. The power series for the integrand in Eq. \ref{af} truncated at order 3 is given by,
\begin{equation}
\label{apow}
    \dfrac{(1+e\cos{E})^{3/2}}{(1-e\cos{E})^{1/2}} = 1+2e\cos{E}+\frac{3}{2}e^2\cos^2{E} + e^3\cos^3{E} + \mathbb{O}(e^4)
\end{equation}
Substituting Eqs. \ref{orb} and \ref{apow} in Eq. \ref{af} and rearranging,
\begin{equation}
\label{a1}
   \begin{split}
     \Delta a  = & -\delta'a^2\rho_{p0}\exp{\{\beta(a_0-a-x_0)\}}\int_0^{2\pi}\left[\sum_{n=0}^{\infty} (\overline{\mathbb{A}}_{n}\cos{nE} \right.\\ & \left. \times(1+2e\cos{E}+\frac{3}{2}e^2\cos^2{E} + e^3\cos^3{E})
     \exp{(\beta x\cos{E})}\right] dE
    \end{split}
\end{equation}
In order to integrate the equation, the following multiple angle formulae are used.
\begin{equation}
\label{cosmul1}
    \cos^2{E} = \frac{1+\cos{2E}}{2}
\end{equation}
\begin{equation}
\label{cosmul}
    \cos^3{E} = \frac{3\cos{E}+\cos{3E}}{4}
\end{equation}
Substituting Eqs. \ref{cosmul1} and \ref{cosmul} in Eq. \ref{a1},  
\begin{equation}
\label{delaA}
\begin{split}
    \Delta a = & D_c\int_0^{2\pi}\left[\sum_{n=0}^{\infty} \overline{\mathbb{A}}_{n}\cos{nE}(1+2e\cos{E}+\frac{3}{4}e^2(1+\cos{2E}) \right.\\ & \left. + e^3\frac{3\cos{E}+\cos{3E}}{4})
     \exp{(\beta x\cos{E})}\right] dE
\end{split}
\end{equation}
where $D_c = -\delta'a^2\rho_{p0}\exp{\{\beta(a_0-a-x_0)\}}$. Using the following cosine product formula in Eq. \ref{delaA},
\[\cos{A}\cos{B} = \frac{\cos{(A+B)}+\cos{(A-B)}}{2}\] 
\begin{equation}
\label{delaA2}
\begin{split}
    \Delta a = & D_c\int_0^{2\pi}\left[\sum_{n=0}^{\infty} \overline{\mathbb{A}}_{n}\{\cos{nE} + e(\cos{(n+1)E}+\cos{(n-1)E})\right. \\ & \left.+\frac{3}{4}e^2\left(\cos{nE}+\frac{\cos{(n+2)E} + \cos{(n-2)E}}{2}\right) \right. \\ & \left. +  \frac{e^3}{4}\left(\frac{3}{2}(\cos{(n+1)E}+\cos{(n-1)E}) \right.\right. \\ & \left.\left. +\frac{\cos{(n+3)E}+\cos{(n-3)E}}{2}\right)\}
     \exp{(\beta x\cos{E})}\right] dE
\end{split}
\end{equation}
Now, the integral can be expressed as a sum of modified Bessel functions of the first kind with imaginary argument,
\begin{equation}
\label{bessel}
I_n(z) = \frac{1}{2\pi}\int_0^{2\pi}\cos{nx}\exp{(z\cos{x})}    
\end{equation}

Therefore, Eq. \ref{delaA2} is written as, 
\begin{equation}
\label{delaf}
\begin{split}
    \Delta a = & 2\pi D_c\left[\sum_{n=0}^{\infty} \overline{\mathbb{A}}_{n}\{I_{n} + e(I_{n+1}+I_{n-1})+\frac{3}{4}e^2(I_n+\frac{I_{n+2} + I_{n-2}}{2})\right. \\&  \left.+  \frac{e^3}{4}(\frac{3}{2}(I_{n+1}+I_{n-1}) +\frac{I_{n+3}+I_{n-3}}{2})\}\right]
\end{split}
\end{equation}
where $I_n = I_n(\beta x)$ is implicit. The derivation of $\Delta x$ follows similar steps. The integrand in Eq. \ref{ef} can be expanded as a power series in $e$. Truncating the power series at the third order and substituting Eq. \ref{orb},  Eq. \ref{ef} can be written as 
\begin{equation}
\label{delx1}
    \begin{split}
        \Delta x = &  D_c\int_{0}^{2\pi}\left [ \sum_{n=0}^{\infty}(\overline{\mathbb{A}}_{n}\cos{nE})  \{ \cos{E}  +\frac{1}{2}e(3+\cos{2E})\right. \\ & \left. +\frac{1}{8}e^2(11\cos{E}+\cos{3E})+\frac{1}{16}e^3(7+8\cos{2E}+\cos{4E})\right.\\ & \left.+\mathbb{O}(e^4) \}\exp{(\beta x\cos{E})}dE  \right]
    \end{split}
\end{equation}
Carrying out the trigonometric simplifications outlined in Eqs. \ref{a1}-\ref{delaf}, the final form of Eq. \ref{delx1} is
\begin{equation}
\label{delxf}
    \begin{split}
        \Delta x = &  2\pi D_c\left [ \sum_{n=0}^{\infty}\overline{\mathbb{A}}_{n} \left\{ \frac{1}{2}(I_{n+1}+I_{n-1}) +\frac{1}{4}e\{6I_n+(I_{n+2}+I_{n-2})\} \right.\right. \\& 
        \left.\left.+\frac{1}{16}e^2\{11(I_{n+1}+I_{n-1})+(I_{n+3}+I_{n-3})\}\right.\right.\\& \left.\left.+\frac{1}{32}e^3\{14I_n+8(I_{n+2}+I_{n-2})+ (I_{n+4}+I_{n-4})\}\right \} \right]
    \end{split}
\end{equation}
Eqs. \ref{delaf} and \ref{delxf} calculate the change in semi-major axis and focal-length over an orbital period for the modified King-Hele theory. Note that for $n=0$, the equations reduce to the original forms derived by King-Hele as follows since $I_n = I_{-n}$, 
\begin{equation}
\label{delakh}
\begin{split}
     \Delta a & = 2\pi D_c\overline{\mathbb{A}}_0[I_{0} + 2eI_1+\frac{3}{4}e^2(I_0+I_2)+\frac{1}{4}e^3(3I_1 +I_3)]
\end{split}     
\end{equation}
\begin{equation}
\label{delxkh}
\begin{split}
     \Delta x & = 2\pi D_c\overline{\mathbb{A}}_0[I_{1} + \frac{1}{2}e(3I_0+I_2) +\frac{1}{8}e^2(11I_1+I_3) \\ &+\frac{1}{16}e^3(7I_0+8I_2 +I_4)]
\end{split}     
\end{equation}
For the original formulation, the drag-coefficient is assumed to be constant, denoted by $\overline{\mathbb{A}}_{0}$ in Eqs. \ref{delakh} and \ref{delxkh}. The constant drag-coefficient can be assumed to be the zeroth-order order Fourier coefficient ($\overline{\mathbb{A}}_{0}$), the drag-coefficient evaluated at perigee or a weighted average of the orbital drag-coefficient variation. The value that will approximate the results of the Fourier theory given by Eqs. \ref{a_lpe} and \ref{e_lpe} can be calculated by equating the original King-Hele $\Delta a$ and $\Delta e$ to the Fourier theory given by Eqs. \ref{af} and \ref{ef} as follows,
\begin{equation}
\label{acd}
\begin{split}
   C_{D0}\int_0^{2\pi}\frac{(1+e\cos{E})^{3/2}}{(1-e\cos{E})^{1/2}}\rho dE = & \int_0^{2\pi}\frac{(1+e\cos{E})^{3/2}}{(1-e\cos{E})^{1/2}} \\ & \times C_D(E)\rho dE
\end{split}
\end{equation}

\begin{equation}
\label{ecd}
\begin{split}
    & C_{D0}\int_0^{2\pi}\left(\frac{1+e\cos{E}}{1-e\cos{E}}\right)^{1/2}(\cos{E}+e)\rho dE = \\&
     \int_0^{2\pi}\left(\frac{1+e\cos{E}}{1-e\cos{E}}\right)^{1/2}(\cos{E}+e)C_D(E)\rho dE
\end{split}    
\end{equation}
A drag-coefficient that approximates both Eqs. \ref{acd} and \ref{ecd} can be calculated by considering only the density and the Fourier drag-coefficient inside the integral. This results in a weighted average of the Fourier drag-coefficient as follows-
\begin{equation}
\label{cdavg}
    C_{D0} = \frac{\int_0^{2\pi}\rho C_D(E)dE}{\int_0^{2\pi}\rho dE}
\end{equation}
Using Eqs. \ref{den} and \ref{bessel}, an analytical form of the constant drag-coefficient can be found as 
\begin{equation}
\label{cdrho}
    C_{D0} = \frac{\sum_{n=0}^{\infty}\overline{\mathbb{A}}_{n} I_n}{I_0}
\end{equation}

\subsection{High eccentricity regime, $0.2 \leq e<1$}
For large values of eccentricity, expanding the integrands in Eqs. \ref{af} and \ref{ef} as a power series in $e$ is not appropriate. \citet{kh} introduced the auxiliary variable $\lambda$ in order to integrate Eqs. \ref{af} and \ref{ef} and carried out the following transformation of variables,
\begin{equation}
\label{subs}
    \cos{E} = 1-\lambda^2/z
\end{equation}
such that,
\begin{equation}
\label{subsd}
    dE = \sqrt{\frac{2}{z(1-\lambda^2/2z)}}d\lambda
\end{equation}
where $z = \beta x = \frac{ae}{H}$, where $z$ goes to infinity as $e$ approaches 1. Replacing the integrals in Eqs. \ref{af} and \ref{ef} from 0 to $2\pi$ by twice the integrals from $0$ to $\pi$ and substituting Eqs. \ref{subs} and \ref{subsd}, 
\begin{equation}
\label{aeh1}
\begin{split}
\Delta a = & 2\exp{(z)}\sqrt{2/z}D_c\int_{0}^{\sqrt{2z}}\frac{(1+e-e\lambda^2/z)^{3/2}}{(1-e+e\lambda^2/z)^{1/2}}\exp{(-\lambda^2)} \\ & \times \sqrt{\frac{1}{(1-\lambda^2/(2z))}}C_d(\lambda)d\lambda
\end{split}
\end{equation}

\begin{equation}
    \label{xeh1}
    \begin{split}
    \Delta x =& 2\exp{(z)}\sqrt{2/z}D_c\int_{0}^{\sqrt{2z}}(e+1-\lambda^2/z)\left (\frac{1+e-e\lambda^2/z}{1-e+e\lambda^2/z} \right )^{1/2} \\ & \times \exp{(-\lambda^2)}\sqrt{\frac{1}{(1-\lambda^2/(2z))}}C_d(\lambda)d\lambda
     \end{split}
\end{equation}
The drag-coefficient in Eq. \ref{orb} needs to be first expressed in the new variable before substituting in Eqs. \ref{aeh1} and \ref{xeh1}. The following formulae for multiple angles are used for that purpose,
\begin{equation}
\label{cosn}
    \cos{nE} = \sum_{k=0}^{\lfloor{n/2}\rfloor}(-1)^k\binom{n}{2k}\sin^{2k}{E}\cos^{n-2k}{E}
\end{equation}
\begin{equation}
\label{sinn}
    \sin{nE} = \sum_{k=0}^{\lfloor{(n-1)/2}\rfloor}(-1)^k\binom{n}{2k+1}\sin^{2k+1}{E}\cos^{n-2k-1}{E}
\end{equation}
where $\lfloor{n/2}\rfloor$ denotes the floor function and $\binom{n}{2k}$ denotes the binomial coefficient. Therefore, Eq. \ref{orb} can be expressed as
\begin{equation}
\label{cdlam}
    C_D(\lambda) = \sum_{n=0}^{\infty}\overline{\mathbb{A}}_n\left[\sum_{k=0}^{\lfloor{n/2}\rfloor}(-1)^k\binom{n}{2k}\left\{\frac{\lambda^2}{z}(2 - \frac{\lambda^2}{z})\right\}^{k}\left(1-\frac{\lambda^2}{z}\right)^{n-2k}\right]
\end{equation}
since $\overline{\mathbb{B}}_n = 0$ as noted before. Substituting Eq. \ref{cdlam} in Eq. \ref{aeh1} and carrying out a power series expansion in $\lambda^2/z$,
\begin{equation}
\label{aeh2}
    \begin{split}
        \Delta a =& 2\exp{(z)}\sqrt{2/z}D_c\frac{(1+e)^{3/2}}{(1-e)^{1/2}}\int_{0}^{\sqrt{2z}} \sum_{n=0}^{\infty}\sum_{k=0}^{\lfloor n/2\rfloor}\overline{\mathbb{A}}_n(-1)^k\binom{n}{2k}\\ & \times 2^k \left [(\lambda^2/z)^k + K_1(\lambda^2/z)^{k+1}+ K_2(\lambda^2/z)^{k+2} \right. \\& \left. + \mathbb{O}((\lambda^2/z)^{k+3}) \right ] \exp{(-\lambda^2)}d\lambda
    \end{split}
\end{equation}
where $K_1$ and $K_2$ are functions of the summation indices $n$ and $k$, and the eccentricity $e$ and are given by
\begin{equation}
    K_1 = \frac{1}{4(1-e^2)}[(-4n+6k+1)-8e+(4n-6k+3)e^2]
\end{equation}
\begin{equation}
\begin{split}
    K_2 = & \frac{1}{32(1-e^2)^2}[(4n-6k)(4n-6k-6)+(4k+3) +16(4n-6k\\ & -1)e 
    + \{(4n-6k)(-8n+12k+4) -(8k-50)\}e^2  -16(4n\\ &-6k-1)e^3 + \{(4n-6k)(4n-6k+2)+(4k-5)\}e^4] 
\end{split}
\end{equation}
Approximating the upper limit of the integral as $\infty$ since the integrand becomes very small as $\lambda$ becomes large and $\sqrt{2z}>6$ \citep{kh}, the integrals can be expressed as a sum of Gamma functions that are given by
\begin{equation}
\label{gammfunc}
    \int_0^{\infty}\lambda^{k}\exp{(-\lambda^2)}d\lambda = \frac{1}{2}\Gamma\left(\frac{k+1}{2}\right)
\end{equation}
Therefore, the final form of Eq. \ref{aeh2} is given by
\begin{equation}
\begin{split}
    \Delta a = & D_c'\sum_{n=0}^{\infty}\sum_{k=0}^{\lfloor n/2\rfloor} \overline{\mathbb{A}}_n(-1)^k\binom{n}{2k}(\frac{2}{z})^k\left [\Gamma\left(\frac{2k+1}{2}\right) \right. \\ & \left. + \frac{K_1}{z}\Gamma\left(\frac{2k+3}{2}\right)\right.  \left. + \frac{K_2}{z^2}\Gamma\left(\frac{2k+5}{2}\right)\right ]
\end{split}
\end{equation}
where $D_c' = \exp{(z)}\sqrt{2/z}\dfrac{(1+e)^{3/2}}{(1-e)^{1/2}}D_c$.
Similarly, the equation for $\Delta x$ can be derived by substituting Eq. \ref{cdlam} in Eq. \ref{xeh1} and carrying out a power series expansion,
\begin{equation}
\label{xeh2}
    \begin{split}
        \Delta x =& 2\exp{(z)}\sqrt{2/z}D_c\dfrac{(1+e)^{3/2}}{(1-e)^{1/2}}\int_{0}^{\sqrt{2z}} \sum_{n=0}^{\infty}\sum_{k=0}^{\lfloor n/2\rfloor} \overline{\mathbb{A}}_n(-1)^k\binom{n}{2k} \\ & \times 2^k\left [(\lambda^2/z)^k + M_1(\lambda^2/z)^{k+1}+ M_2(\lambda^2/z)^{k+2} \right. \\ & \left.+ \mathcal{O}((\lambda^2/z)^{k+3}) \right ] \exp{(-\lambda^2)}d\lambda
    \end{split}
\end{equation}
where $M_1$ and $M_2$ are functions of the summation indices $n$ and $k$, and the eccentricity $e$ and are given by
\begin{equation}
    M_1 = \frac{1}{4(1-e^2)}[(-4n+6k-3)+(4n-6k-1)e^2]
\end{equation}
\begin{equation}
    \begin{split}
        M_2 = & \frac{1}{32(1-e^2)^2}[(4n-6k)(4n-6k+2)+(4k-5)+32e \\ & -2\{(4n-6k)(4n-6k-2) +(4k+7)\}e^2 +32e^3 \\& + \{(4n-6k)(4n-6k-6)+(4k+3)\}e^4] 
    \end{split}
\end{equation}
The final form of Eq. \ref{xeh2} in terms of Gamma functions is as follows
\begin{equation}
\begin{split}
    \Delta x = & D_c'\sum_{n=0}^{\infty}\sum_{k=0}^{\lfloor n/2\rfloor} \overline{\mathbb{A}}_n(-1)^k\binom{n}{2k}(\frac{2}{z})^k\left [\Gamma\left(\frac{2k+1}{2}\right) \right. \\ & \left. + \frac{M_1}{z}\Gamma\left(\frac{2k+3}{2}\right) + \frac{M_2}{z^2}\Gamma\left(\frac{2k+5}{2}\right)\right ].
\end{split}
\end{equation}

For $n = 0$, the equations reduce to the original KH formulation,
\begin{equation}
   \begin{split}
    \Delta a = & D_c' A_0\sqrt{\pi}\left [1+ \frac{K_1}{2z} + \frac{3K_2}{4z^2}\right ],
\end{split} 
\end{equation}
where \[K_1 = \frac{1}{4(1-e^2)}[1-8e+3e^2],\]
\[    K_2 =  \frac{1}{32(1-e^2)^2}[3-16e + 50e^2 +16e^3 -5e^4], \]
and,
\begin{equation}
\begin{split}
    \Delta x = & D_c'\sqrt{\pi} A_0\left [1+ \frac{M_1}{2z} + \frac{3M_2}{4z^2}\right ],
\end{split}
\end{equation}
where \[ M_1 = -\frac{1}{4(1-e^2)}[3+e^2],\]
\[  M_2 =  \frac{1}{32(1-e^2)^2}[-5+32e  -14e^2 +32e^3 + 3e^4] .\]

The density-averaged constant drag-coefficient derived for the low eccentricity regime can be used for high-eccentricity regime as well.

\section{Re-deriving the King-Hele theory using BFF model}
\label{bff}
In this section, the theory is developed for two attitude profiles for which $\phi$ can be expressed as a function of the eccentric anomaly. Unlike the OFF model, $\overline{\mathcal{B}}_n$ is not generally zero for BFF since the drag-coefficient may not be symmetric about $\phi = 0, \pi$. If the satellite shape is symmetric about $\phi =0, \pi$, then $\overline{\mathcal{B}}_n = 0$.

\subsection{Nadir-pointing profile}
For a nadir-pointing profile, the angle between the velocity vector and the body axis is equal to the flight path angle that can be expressed in terms of the eccentric anomaly as 
\begin{equation}
\label{cosp}
    \cos{\phi} = \sqrt{\frac{1-e^2}{1-e^2\cos^2{E}}}
\end{equation}
\begin{equation}
\label{sinp}
    \sin{\phi} = \frac{e\sin{E}}{\sqrt{1-e^2\cos^2{E}}}
\end{equation}
Using Eqs. \ref{cosn}, \ref{sinn}, \ref{cosp} and \ref{sinp} in Eq. \ref{bodcd}, 
\begin{equation}
\label{cdnad}
    \begin{split}
        & C_D(E) =  \sum_{n=0}^{\infty}\left [ \overline{\mathcal{A}}_n\left \{ \sum_{k=0}^{\lfloor \frac{n}{2}\rfloor}(-1)^k\binom{n}{2k}\left ( \frac{e\sin{E}}{\sqrt{1-e^2\cos^2{E}}} \right)^{2k}\right.\right. \\& \left.\left. \times \left( \sqrt{\frac{1-e^2}{1-e^2\cos^2{E}}} \right )^{n-2k}  \right\} + \overline{\mathcal{B}}_n\left \{ \sum_{k=0}^{\lfloor \frac{n-1}{2}\rfloor}(-1)^k\binom{n}{2k+1}\right.\right. \\ & \left.\left. \times \left ( \frac{e\sin{E}}{\sqrt{1-e^2\cos^2{E}}} \right )^{2k+1} \left ( \sqrt{\frac{1-e^2}{1-e^2\cos^2{E}}} \right )^{n-2k-1}  \right \} \right ]
    \end{split}
\end{equation}
\par
\par
\textbf{\emph{Low eccentricity regime}}
\par
Substituting Eq. \ref{cdnad} in Eq. \ref{af} and noting that the integrand corresponding to $\overline{\mathcal{B}}_n$ is an odd function, the equation reduces to 
\begin{equation}
    \begin{split}
        \Delta a  & =  D_c\int_0^{2\pi}\sum_{n=0}^{\infty}\sum_{k=0}^{\lfloor \frac{n}{2}\rfloor} \overline{\mathcal{A}}_n (-1)^k\binom{n}{2k}\left ( \frac{e\sin{E}}{\sqrt{1-e^2\cos^2{E}}} \right )^{2k} \\ & \times \left ( \sqrt{\frac{1-e^2}{1-e^2\cos^2{E}}} \right )^{n-2k} \frac{(1+e\cos{E})^{3/2}}{(1-e\cos{E})^{1/2}}\exp{(\beta x\cos{E})}dE 
    \end{split}
\end{equation}
Expanding the integrand as a power series in $e$, 
\begin{equation}
\label{a1nad}
    \begin{split}
        &\Delta a  =  D_c\int_0^{2\pi}\sum_{n=0}^{\infty}\sum_{k=0}^{\lfloor \frac{n}{2}\rfloor} \overline{\mathcal{A}}_n (-1)^k\binom{n}{2k}[1+2\cos{E}e+\{k-\frac{n}{2} \\ &  +\frac{1}{2}(n+3)\cos^2{E}\} e^2 + \cos{E}\{2k-n+(n+1)\cos^2{E}\}e^3]\\ & \times(e\sin{E})^{2k} \exp{(\beta x\cos{E})}dE
    \end{split}
\end{equation}
Truncating the series at $\mathcal{O}(e^3)$, $k$ can only be 0 and 1. Therefore, Eq. \ref{a1nad} can be written as, 
\begin{equation}
\label{a1nad2}
    \begin{split}
        \Delta a  & =  D_c\int_0^{2\pi}\left[\sum_{n=0}^{\infty} \overline{\mathcal{A}}_n\{1+2\cos{E}e+\{-\frac{n}{2}  +\frac{1}{2}(n+3)\cos^2{E}\} e^2 \right. \\ & \left.+ \cos{E}\{-n+(n+1)\cos^2{E}\}e^3\} + \sum_{n=2}^{\infty} -\overline{\mathcal{A}}_n\binom{n}{2}\{\sin^2{E}e^2 \right.\\& \left. +2\sin^2{E}\cos{E}e^3\}\right]\exp{(\beta x\cos{E})}dE 
    \end{split}
\end{equation}
The trigonometric powers can be written as 
\begin{equation}
\label{sin2e}
    \sin^2{E} = \frac{1-\cos{2E}}{2}
\end{equation}
\begin{equation}
\label{sin2cose}
    \sin^2{E}\cos{E} = \cos{E}-\frac{1}{4}(3\cos{E}+\cos{3E})
\end{equation}
Using Eqs. \ref{cosmul1}, \ref{cosmul}, \ref{bessel}, \ref{sin2e} and \ref{sin2cose}, Eq. \ref{a1nad2} is given by,
\begin{equation}
\label{a1nad3}
    \begin{split}
       & \Delta a  =  2\pi D_c\sum_{n=0}^{\infty} \overline{\mathcal{A}}_n\left[I_0+2I_1e+\left\{\left(\frac{n+3}{4}\right)(I_0+I_2)-\frac{n}{2}I_0\right\} e^2 \right.\\& \left.+ \left\{\left(\frac{n+1}{4}\right)(3I_1+I_3)-nI_1\right\}e^3\right]  - \sum_{n=2}^{\infty} \overline{\mathcal{A}}_n\binom{n}{2}\left[\frac{(I_0-I_2)}{2}e^2 \right.\\ &\left. +\frac{(I_1-I_3)}{2}e^3\right]
    \end{split}
\end{equation}

In order to derive $\Delta x$, a similar procedure can be followed. Substituting Eq. \ref{cdnad} in Eq. \ref{ef},
\begin{equation}
\label{xnad}
   \begin{split}
    & \Delta x  =  D_c\int_0^{2\pi}\left(\frac{(1+e\cos{E})}{1-e\cos{E}}\right)^{1/2}(\cos{E}+e)\exp{\{\beta x\cos{E}\}}\\ &\times \sum_{n=0}^{\infty} \overline{\mathcal{A}}_n  \left [ \sum_{k=0}^{\lfloor \frac{n}{2}\rfloor}(-1)^k\binom{n}{2k}\left ( \frac{e\sin{E}}{\sqrt{1-e^2\cos^2{E}}} \right )^{2k} \right. \\& \left. \times \left ( \sqrt{\frac{1-e^2}{1-e^2\cos^2{E}}} \right )^{n-2k}  \right ] dE
    \end{split}
\end{equation}
Expanding the integrand as a power series in $e$,
\begin{equation}
\label{x1nad}
    \begin{split}
        &\Delta x  =  D_c\int_0^{2\pi}\sum_{n=0}^{\infty}\sum_{k=0}^{\lfloor \frac{n}{2}\rfloor} \overline{\mathcal{A}}_n (-1)^k\binom{n}{2k}[\cos{E}+\left(\frac{3+\cos{2E}}{2}\right)e \\ & +\frac{1}{4}\cos{E}\{5+4k-n  +(n+1)\cos{2E}\} e^2 + \frac{1}{8}(3+\cos{2E})\{1+4k \\&-n+(n+1)\cos{2E}\}e^3](e\sin{E})^{2k}\exp{(\beta x\cos{E})}dE
    \end{split}
\end{equation}
Truncating the series at $\mathcal{O}(e^3)$, $k$ can only be 0 and 1. Therefore, Eq. \ref{a1nad} can be written as, 
\begin{equation}
\label{x1nad2}
    \begin{split}
        \Delta x  & =  D_c\int_0^{2\pi}\left[\sum_{n=0}^{\infty} \overline{\mathcal{A}}_n[\cos{E}+\frac{3+\cos{2E}}{2}e+\frac{1}{4}\{(5-n)\cos{E} \right.\\& \left. +\frac{(n+1)}{2}(\cos{3E}+\cos{E})\} e^2 + \frac{1}{8}[3\{(1-n)+(n+1)\cos{2E}\} \right.\\& \left. + \{(1-n)\cos{2E}  + (n+1)\frac{(1+\cos{4E})}{2}\}e^3]]\exp{(\beta x\cos{E})} \right.\\& \left.+ \sum_{n=2}^{\infty} -\overline{\mathcal{A}}_n\binom{n}{2}\right. \left. [\frac{(\cos{E}+\cos{3E})}{4}e^2 - (\cos{2E}+\frac{1}{4}\cos{4E} \right.\\& \left. -\frac{5}{4})\frac{e^3}{2}]\exp{(\beta x\cos{E})}\right]dE 
    \end{split}
\end{equation}
Integrating the equation,
\begin{equation}
\label{x1nad3}
    \begin{split}
        \Delta x = & 2\pi D_c\left[\sum_{n=0}^{\infty} \overline{\mathcal{A}}_n[I_1+\frac{3I_0+I_2}{2}e+\frac{1}{4}\{(5-n)I_1  +\frac{(n+1)}{2}\right. \\ & \left. \times (I_3  +I_1)\} e^2   + \frac{1}{8}[3\{(1-n)I_0+(n+1)I_2\}  + \{(1-n)I_2 \right.\\& \left. + (n+1)\frac{(I_0+I_4)}{2}\}  e^3]]  + \sum_{n=2}^{\infty} -\overline{\mathcal{A}}_n\binom{n}{2} [\frac{(I_1+I_3)}{4}e^2 \right.\\& \left. - (I_2+\frac{1}{4}I_4-\frac{5}{4}I_0)\frac{e^3}{2}]\right]
    \end{split}
\end{equation}

For a satellite with an arbitrary shape, $\Delta \omega \neq 0$ since $\overline{\mathcal{B}}_n \neq  0$. To derive $\Delta \omega$, it should be noted that the even part of the drag-coefficient will integrate out to zero unlike $\Delta a$ and $\Delta x$. Therefore, Eq. \ref{omgf} can be written as,
\begin{equation}
\label{omgnad1}
   \begin{split}
    & \Delta \omega  =  D_w\int_0^{2\pi}\left[\left(\frac{(1+e\cos{E})}{1-e\cos{E}}\right)^{1/2}\sin{E}\exp{(\beta x\cos{E})}
     \sum_{n=0}^{\infty}\overline{\mathcal{B}}_n \right.\\& \times\left. \left \{ \sum_{k=0}^{\lfloor \frac{n-1}{2}\rfloor}(-1)^k\binom{n}{2k+1}\left ( \frac{e\sin{E}}{\sqrt{1-e^2\cos^2{E}}} \right )^{2k+1} \right. \right. \\& \times \left. \left. \left ( \sqrt{\frac{1-e^2}{1-e^2\cos^2{E}}} \right )^{n-2k-1}  \right \} \right ] dE
    \end{split}
\end{equation}
where $D_w =  -\dfrac{\delta'a}{e}\sqrt{1-e^2}\rho_{p0}\exp{\{\beta(a_0-a-x_0)\}}$. On expanding the integrand as a power series in $e$ and truncating at order 3, 
\begin{equation}
\label{omgnad}
    \begin{split}
        \Delta \omega = & 2\pi D_w\sum_{n=0}^{\infty} n\overline{\mathcal{B}}_n\left[\frac{I_0-I_2}{2}e+\frac{I_1-I_3}{4} e^2 \right. \left. + \frac{1}{4}\{(n+1)  \right. \\ & \left.  \times \frac{2I_1-I_4-1}{4}  -(n-3)\frac{I_0-I_2}{2}\}e^3 \right]
    \end{split}
\end{equation}

For $n = 0$, Eqs. \ref{a1nad3} and \ref{x1nad3} reduce to the original KH formulation given by Eqs. \ref{delakh} and \ref{delxkh} while Eq. \ref{omgnad} reduces to zero. The average drag-coefficient that best approximates the higher order Fourier theory given by Eqs. \ref{a1nad3} and \ref{x1nad3} can be calculated using Eq. \ref{cdavg} as follows,
\begin{equation}
\label{cdrho1}
\begin{split}
    C_{D0} & = \frac{\int_0^{2\pi}\rho C_D(E)dE}{\int_0^{2\pi}\rho dE}\\
    & = \dfrac{1}{{\int_0^{2\pi}\text{exp}(\beta x \cos{E})}dE} \left[\int_0^{2\pi}\text{exp}(\beta x \cos{E}) \sum_{n=0}^{\infty}\sum_{k=0}^{\lfloor \frac{n}{2}\rfloor}\overline{\mathcal{A}}_n \right. \\ & \left. \times (-1)^k\binom{n}{2k}\left( \dfrac{e\sin{E}}{\sqrt{1-e^2\cos^2{E}}} \right)^{2k}\left( \sqrt{\dfrac{1-e^2}{1-e^2\cos^2{E}}} \right )^{n-2k} dE\right] \\
    & = \dfrac{1}{I_0}\left[\sum_{n=0}^{\infty}\overline{\mathcal{A}}_n\{I_0 + \dfrac{n}{4}(I_2-I_0)e^2\}+ \sum_{n=2}^{\infty}\dfrac{\overline{\mathcal{A}}_n}{2}\binom{n}{2}(I_2-I_0)e^2 \right]
\end{split}
\end{equation}

\par
\textbf{\emph{High eccentricity regime}, $0.2\leq e<1$}
\par
Similar to the OFF model in high eccentricity regime, the eccentric anomaly is transformed to the auxiliary variable $\lambda$. The flight path angle in the new variable is given by
\begin{equation}
    \sin{\phi} = e\sqrt{\frac{1-(1-\lambda^2/z)^2}{1-e^2(1-\lambda^2/z)^2}}
\end{equation}
\begin{equation}
    \cos{\phi} = \sqrt{\frac{1-e^2}{1-e^2(1-\lambda^2/z)^2}}
\end{equation}
The drag-coefficient in the transformed variable is given by
\begin{equation}
\label{cdnadlam}
    \begin{split}
        & C_D(\lambda) =  \sum_{n=0}^{\infty}\left[ \overline{\mathcal{A}}_n\left \{ \sum_{k=0}^{\lfloor \frac{n}{2}\rfloor}(-1)^k\binom{n}{2k}\left ( e\sqrt{\frac{1-(1-\lambda^2/z)^2}{1-e^2(1-\lambda^2/z)^2}} \right)^{2k}\right.\right. \\&  \left. \left. \times \left( \sqrt{\frac{1-e^2}{1-e^2(1-\lambda^2/z)^2}} \right )^{n-2k}  \right\}  + \overline{\mathcal{B}}_n\left \{ \sum_{k=0}^{\lfloor \frac{n-1}{2}\rfloor}(-1)^k\binom{n}{2k+1} \right. \right. \\ & \left. \left. \times \left ( e\sqrt{\frac{1-(1-\lambda^2/z)^2}{1-e^2(1-\lambda^2/z)^2}} \right)^{2k+1} \left( \sqrt{\frac{1-e^2}{1-e^2(1-\lambda^2/z)^2}} \right )^{n-2k-1}  \right\} \right]
    \end{split}
\end{equation}
Substitute Eq. \ref{cdnadlam} in Eq. \ref{aeh1} and carrying out a power series expansion in $\lambda^2/z$,
\begin{equation}
\label{aehnad2}
    \begin{split}
        \Delta a =& 2\exp{(z)}\sqrt{2/z}D_c\frac{(1+e)^{3/2}}{(1-e)^{1/2}}\int_{0}^{\sqrt{2z}} \sum_{n=0}^{\infty}\sum_{k=0}^{\lfloor n/2\rfloor} \overline{\mathcal{A}}_n(-1)^k\binom{n}{2k}\\ & \times \left(\frac{2e^2}{1-e^2}\right)^k \left [(\lambda^2/z)^k + P_1(\lambda^2/z)^{k+1}+ P_2(\lambda^2/z)^{k+2} \right.\\ & \left.+ \mathcal{O}((\lambda^2/z)^{k+3}) \right ]\exp{(-\lambda^2)}d\lambda
    \end{split}
\end{equation}

where $P_1$ and $P_2$ are functions of the summation indices $n$ and $k$, and the eccentricity $e$ and are given by
\begin{equation}
    P_1 = -\frac{1}{4(1-e^2)}[(2k-1)+8e+(4n-2k-3)e^2]
\end{equation}
\begin{equation}
\begin{split}
    P_2 = & \frac{1}{32(1-e^2)^2}[(4k^2-8k+3)  +16(2k-1)e + (8n-8k^2  \\ & +16kn+50)e^2 + 16(4n-2k+1)e^3
    +  \{(4(n-k)(n-k-2) \\ & +4nk-5\}e^4 ] 
\end{split}
\end{equation}
Using Eq. \ref{gammfunc}, 
\begin{equation}
\begin{split}
    \Delta a = & D_c'\sum_{n=0}^{\infty}\sum_{k=0}^{\lfloor n/2\rfloor} \overline{\mathcal{A}}_n(-1)^k\binom{n}{2k}\left(\frac{2e^2}{z(1-e^2)}\right)^k\left [\Gamma\left(\frac{2k+1}{2}\right) \right. \\ & \left. + \frac{P_1}{z}\Gamma\left(\frac{2k+3}{2}\right) + \frac{P_2}{z^2}\Gamma\left(\frac{2k+5}{2}\right)\right ]
\end{split}
\end{equation}

To derive $\Delta x$, substitute Eq. \ref{cdnadlam} in Eq. \ref{xeh1} and expand as a power series in $\lambda^2/z$ to obtain,
\begin{equation}
\label{xehnad2}
    \begin{split}
        \Delta x =& 2\exp{(z)}\sqrt{2/z}D_c\frac{(1+e)^{3/2}}{(1-e)^{1/2}}\int_{0}^{\sqrt{2z}} \sum_{n=0}^{\infty}\sum_{k=0}^{\lfloor n/2\rfloor} \overline{\mathcal{A}}_n(-1)^k\binom{n}{2k}\\ & \times \left(\frac{2e^2}{1-e^2}\right)^k \left [(\lambda^2/z)^k + Q_1(\lambda^2/z)^{k+1}+ Q_2(\lambda^2/z)^{k+2} \right. \\ & \left. + \mathcal{O}((\lambda^2/z)^{k+3}) \right ]\exp{(-\lambda^2)}d\lambda
    \end{split}
\end{equation}

where $Q_1$ and $Q_2$ given by
\begin{equation}
    Q_1 = -\frac{1}{4(1-e^2)}[(2k+3)+(4n-2k+1)e^2]
\end{equation}
\begin{equation}
\begin{split}
    Q_2 = & \frac{1}{32(1-e^2)^2}[(4k^2+8k-5)+32e-2(4k^2-20n-8kn \\ & +7)e^2 
    + 32e^3+  \{4(2n-k)(2n-k+2)+8n+3\}e^4 ] 
\end{split}
\end{equation}
Using Eq. \ref{gammfunc}, 
\begin{equation}
\begin{split}
    \Delta x = & D_c'\sum_{n=0}^{\infty}\sum_{k=0}^{\lfloor n/2\rfloor} \overline{\mathcal{A}}_n(-1)^k\binom{n}{2k}\left(\frac{2e^2}{z(1-e^2)}\right)^k\left [\Gamma\left(\frac{2k+1}{2}\right) \right.  \\ & \left. + \frac{Q_1}{z}\Gamma\left(\frac{2k+3}{2}\right) + \frac{Q_2}{z^2}\Gamma\left(\frac{2k+5}{2}\right)\right ]
\end{split}
\end{equation}

The change in argument of perigee can be similarly derived. Eq. \ref{omgf} in the transformed variable can be written as
\begin{equation}
    \label{omg_nadh1}
    \begin{split}
    \Delta \omega =& \dfrac{4}{z}\exp{(z)}D_w\int_{0}^{\sqrt{2z}}\left (\frac{1+e-e\lambda^2/z}{1-e+e\lambda^2/z} \right )^{1/2}\lambda\exp{(-\lambda^2)}C_d(\lambda)d\lambda
     \end{split}
\end{equation}

On substituting Eq. \ref{cdnadlam} and carrying out a power series expansion, the equation simplifies to the following form
\begin{equation}
\begin{split}
    \Delta \omega = & 4\exp{(z)}D_w\sqrt{\dfrac{1+e}{1-e}}\int_{0}^{\sqrt{2z}}\sum_{n=0}^{\infty}\sum_{k=0}^{\lfloor \frac{n-1}{2}\rfloor}\overline{\mathcal{B}}_n(-1)^k \binom{n}{2k+1}  \\ &  \times \left(\dfrac{2e^2}{1-e^2}\right)^{\frac{2k+1}{2}}\left[(\lambda^2/z)^\frac{2k+3}{2} + W_n(\lambda^2/z)^\frac{2k+5}{2}\right]d\lambda
\end{split}
\end{equation}
where 
\begin{equation}
    W_n = -\frac{1}{4(1-e^2)}[(2k+1)+4e+(4n-2k-1)e^2]
\end{equation}

The integrated change is given by
\begin{equation}
\begin{split}
    \Delta \omega = & 2\exp{(z)}D_w\sqrt{\dfrac{1+e}{1-e}}\sum_{n=0}^{\infty}\sum_{k=0}^{\lfloor \frac{n-1}{2}\rfloor}\overline{\mathcal{B}}_n(-1)^k \binom{n}{2k+1} \\ & \times \left(\dfrac{2e^2}{z(1-e^2)}\right)^{\frac{2k+1}{2}}\left[\frac{1}{z}\Gamma\left(k+2\right) + \frac{W_n}{z^2}\Gamma\left(k+3\right)\right]
    \end{split}
\end{equation}

The constant drag-coefficient to be used with the original King-Hele formulation has to be re-derived for the high eccentricity regime in this case since Eq. \ref{cdrho1} consists of a series truncation in $e$. The drag-coefficient in the auxiliary variable is given by
\begin{equation}
\begin{split}
    C_{D0} & = \dfrac{\exp{z}}{\pi\sqrt{2z}I_0}\left[ \sum_{n=0}^{\infty}\sum_{k=0}^{\lfloor \frac{n}{2}\rfloor}\overline{\mathcal{A}}_n(-1)^k\binom{n}{2k}\left(\frac{2e^2}{z(1-e^2)}\right)^{k} \right. \\ & \left. \times\left\{ \Gamma\left(\frac{2k+1}{2}\right) \right.\left.+ \frac{C_{N1}}{z}\Gamma\left(\frac{2k+3}{2}\right)+ \frac{C_{N2}}{z^2}\Gamma\left(\frac{2k+5}{2}\right)\right\}\right]
\end{split}    
\end{equation}

where 
\begin{equation}
    C_{N1} = -\frac{1}{4(1-e^2)}[(2k-1)+(4n-2k+1)e^2],
\end{equation}
\begin{equation}
\begin{split}
    C_{N2} = & \frac{1}{32(1-e^2)^2}[(4k^2-8k+3)(1-e^2)^2 +8ne^2(3e^2-2ke^2 \\ & +2k+1+2ne^2)] .
\end{split}
\end{equation}

\subsection{Inertially stabilized attitude}
For an inertially stabilized satellite, the angle between the velocity vector and the body axis can be computed from the velocity components in the perifocal frame. The sine and cosine of the angle is given by
\begin{equation}
\label{sinb}
\begin{split}
    \sin{\phi} & = \frac{\text{v}_P}{\sqrt{\text{v}_P^2+\text{v}_Q^2}} \\&
                   = \frac{\sin{\theta}}{\sqrt{1+e^2+2e\cos{\theta}}}
                   = \frac{\sin{E}}{\sqrt{1-e^2\cos^2{E}}}
\end{split}
\end{equation}
\begin{equation}
\label{cosb}
\begin{split}
    \cos{\phi} & = \frac{\text{v}_Q}{\sqrt{\text{v}_P^2+\text{v}_Q^2}} \\&
                   = \frac{e+\cos{\theta}}{\sqrt{1+e^2+2e\cos{\theta}}}
                   = \frac{\sqrt{1-e^2}\cos{E}}{\sqrt{1-e^2\cos^2{E}}}
\end{split}
\end{equation}
Substituting Eqs. \ref{sinb} and \ref{cosb} in Eq. \ref{bodcd} and using Eqs. \ref{sinn} and \ref{cosn}, 
\begin{equation}
\label{cdinere}
    \begin{split}
        & C_D(E) =  \sum_{n=0}^{\infty}\left [ \overline{\mathcal{A}}_n\left \{ \sum_{k=0}^{\lfloor \frac{n}{2}\rfloor}(-1)^k\binom{n}{2k}\left ( \frac{\sin{E}}{\sqrt{1-e^2\cos^2{E}}} \right)^{2k} \right.\right.\\ & \left.\left. \times \left( \frac{\sqrt{1-e^2}\cos{E}}{\sqrt{1-e^2\cos^2{E}}} \right )^{n-2k}  \right\} + \overline{\mathcal{B}}_n\left \{ \sum_{k=0}^{\lfloor \frac{n-1}{2}\rfloor}(-1)^k\binom{n}{2k+1} \right.\right. \\ & \left.\left. \times \left ( \frac{\sin{E}}{\sqrt{1-e^2\cos^2{E}}} \right )^{2k+1}\left ( \frac{\sqrt{1-e^2}\cos{E}}{\sqrt{1-e^2\cos^2{E}}} \right )^{n-2k-1}  \right \} \right ]
    \end{split}
\end{equation}
\par
\emph{\textbf{Low eccentricity regime}}, $e<0.2$
\par
Following the same procedure as the nadir pointing profile, only the $\overline{\mathcal{A}}_n$ terms are considered substituting Eq. \ref{cdinere} in Eq. \ref{af} since the $\overline{\mathcal{B}}_n$ terms integrate to zero,
\begin{equation}
    \begin{split}
        \Delta a  & =  D_c\int_0^{2\pi}\sum_{n=0}^{\infty}\sum_{k=0}^{\lfloor \frac{n}{2}\rfloor} \overline{\mathcal{A}}_n (-1)^k\binom{n}{2k}\left ( \frac{\sin{E}}{\sqrt{1-e^2\cos^2{E}}} \right )^{2k} \\ & \times \left ( \frac{\sqrt{1-e^2}\cos{E}}{\sqrt{1-e^2\cos^2{E}}} \right )^{n-2k} \frac{(1+e\cos{E})^{3/2}}{(1-e\cos{E})^{1/2}}\exp{(\beta x\cos{E})}dE 
    \end{split}
\end{equation}
Expanding as a power series in $e$ and truncating at order 3, 
\begin{equation}
\label{a1iner}
    \begin{split}
        &\Delta a  =  D_c\int_0^{2\pi}\sum_{n=0}^{\infty}\sum_{k=0}^{\lfloor \frac{n}{2}\rfloor} \overline{\mathcal{A}}_n (-1)^k\binom{n}{2k}[1+2\cos{E}e+\{k-\frac{n}{2}  \\&  +\frac{1}{2}(n+3)\cos^2{E}\} e^2 + \cos{E}\{2k-n+(n+1)\cos^2{E}\}e^3]\\&  \times \sin^{2k}{E}\cos^{n-2k}{E}\exp{\{\beta x\cos{E}\}}dE
    \end{split}
\end{equation}
Unlike the nadir pointing profile case, the series cannot be truncated in $k$ as there's no $e^k$ in the expression. The trigonometric powers have to be expressed in multiple angles in order to integrate the equation. The following identities are used to express an arbitrary trigonometric power in multiple angles, 
\begin{equation}
\label{sink}
    \sin^{2k}{E} = \frac{1}{2^{2k}}\binom{2k}{k} + \frac{(-1)^k}{2^{2k-1}}\sum_{j=0}^{k-1}(-1)^j\binom{2k}{j}\cos{[2(k-j)E]}
\end{equation}
\begin{equation}
\label{cosk}
\begin{split}
& \cos^{k}{E}  = \\&
    \begin{dcases}
    \frac{1}{2^k}\binom{k}{k/2} + \frac{1}{2^{k-1}}\sum_{i=0}^{k/2-1}\binom{k}{i}\cos{[k-2i]E}, & \ k \in 2q, q \in \mathbb{Z}^\geq \\
    \frac{1}{2^{k-1}}\sum_{i=0}^{(k-1)/2}\binom{k}{i}\cos{[k-2i]E}, & \ k \in 2q+1, q \in \mathbb{Z}^\geq. 
    \end{dcases}
\end{split}
\end{equation}
In order to simplify the algebra, introduce the following notation.
\begin{equation}
{}^{p}a = 
    \begin{dcases}
     \frac{1}{2^p}\binom{p}{p/2}, & \ p \in 2q, q \in \mathbb{Z}^{\geq}, \\
     0, & \ p \in 2q+1,  q \in \mathbb{Z}^{\geq},
    \end{dcases}
\end{equation}
\begin{equation}
    {}^{p}_{j}\mathcal{S}_1 =
    \begin{dcases}
    \frac{(-1)^{p/2}}{2^{p-1}}\sum_{j=0}^{p/2-1}(-1)^j\binom{p}{j}, & \ p \in 2q, q \in \mathbb{Z}^+, \\
    0, & \ p \in \{0,2q+1\} , q \in \mathbb{Z}^{\geq},
    \end{dcases}
\end{equation}
\begin{equation}
{}^{p}_{i}\mathcal{S}_2 =
    \begin{dcases}
     \frac{1}{2^{p-1}}\sum_{i=0}^{p/2-1}\binom{p}{i}, & \ p \in 2q, q \in \mathbb{Z}^+, \\
     0, & \ p \in \{0,2q+1\}, q \in \mathbb{Z}^{\geq}, \\
    \end{dcases}
\end{equation}

\begin{equation}
\label{nota}
    {}^{p}_{i}\mathcal{S}_3 =
\begin{dcases}    
    \frac{1}{2^{p-1}}\sum_{i=0}^{(p-1)/2}\binom{p}{i}, & \ p \in 2q+1, q \in \mathbb{Z}^\geq, \\
    0, & \ p \in 2q, q \in \mathbb{Z}^{\geq}, \\
\end{dcases}
\end{equation}
Eqs. \ref{sink} and \ref{cosk} can be represented using the notations in \ref{nota} as
\begin{equation}
\label{sink2}
    \sin^{2k}{E} = {}^{2k}a + {}^{2k}_{j}\mathcal{S}_1\cos{[2(k-j)E]}
\end{equation}
\begin{equation}
\label{cosk2}
\cos^{k}{E} = 
    \begin{dcases}
    {}^{k}a + {}^{k}_{i}\mathcal{S}_2\cos{[k-2i]E}, & \  k \in 2q, q\in \mathbb{Z}^{\geq}, \\
     {}^{k}_{i}\mathcal{S}_3\cos{[k-2i]E}, & \ k \in 2q+1, q\in \mathbb{Z}^{\geq}. 
    \end{dcases}
\end{equation}
In order to integrate Eq. \ref{a1iner}, the following integrals are computed.

\begin{equation}
\label{hm}
    \begin{split}
     h_{n,k}(l) & =  \int_{0}^{2\pi}\sin^{2k}{E}\cos^{n-2k+l}{E}\exp{\{\beta x\cos{E}\}}dE \\& = \int_{0}^{2\pi}({}^{2k}a + {}^{2k}_{j}\mathcal{S}_1\cos{[2(k-j)E]})({}^{n+l-2k}a \\ & + {}^{n+l-2k}_{i}\mathcal{S}_2\cos{[(n+l-2k-2i)E]})\exp{(\beta x\cos{E})}dE\\
& = 2\pi[({}^{2k}a)({}^{n+l-2k}a)I_0 + ({}^{2k}a)({}^{n+l-2k}_{i}\mathcal{S}_2)I_{n+l-2k-2i} \\ & + ({}^{n+l-2k}a ) ({}^{2k}_{j}\mathcal{S}_1)I_{2(k-j)} + \frac{({}^{2k}_{j}\mathcal{S}_1)({}^{n+l-2k}_{i}\mathcal{S}_2)}{2} \\&  \times (I_{n+l-2i-2j}+I_{n+l-4k-2i+2j})],
    \end{split}
\end{equation}
\[\text{if} \hspace{0.2cm} (n+l) \in 2q, q \in \mathbb{Z}^\geq,\]
\begin{equation}
\label{gm}
    \begin{split}
         g_{n,k}(l) & = \int_{0}^{2\pi}\sin^{2k}{E}\cos^{n-2k+l}{E}\exp{\{\beta x\cos{E}\}}dE \\& = \int_{0}^{2\pi}({}^{2k}a + {}^{2k}_{j}\mathcal{S}_1\cos{[2(k-j)E]})( {}^{n+l-2k}_{i}\mathcal{S}_3 \\ & \times \cos{[(n+l-2k-2i)E]}) \times \exp{(\beta x\cos{E})}dE
\\& = 2\pi[({}^{2k}a)({}^{n+l-2k}_{i}\mathcal{S}_3)I_{n+l-2k-2i}  + \frac{({}^{2k}_{j}\mathcal{S}_1)({}^{n+l-2k}_{i}\mathcal{S}_3)}{2} \\ & \times(I_{n+l-2i-2j}+I_{n+l-4k-2i+2j})],
    \end{split}
\end{equation}
\[\text{if} \hspace{0.2cm} (n+l) \in 2q+1, q \in \mathbb{Z}^\geq.\]
Using Eqs. \ref{hm} and \ref{gm} to integrate Eq. \ref{a1iner}, 
\begin{equation}
\label{a1iner2}
    \begin{split}
        &\Delta a  =  D_c[\sum_{{}^{l=0}_{n\in 2l}}^{\infty}\sum_{k=0}^{\lfloor \frac{n}{2}\rfloor} A_n (-1)^k\binom{n}{2k}[h_{n,k}(0)+2g_{n,k}(1)e  +\{(k-\frac{n}{2})\\ & \times h_{n,k}(0)   +\frac{1}{2}(n+3)h_{n,k}(2)\} e^2  + \{(2k-n)g_{n,k}(1)+(n+1)\\ & \times g_{n,k}(3)\}e^3] + \sum_{{}^{\hspace{2mm}l=0}_{n\in 2l+1}}^{\infty}\sum_{k=0}^{\lfloor \frac{n}{2}\rfloor} A_n (-1)^k\binom{n}{2k}[g_{n,k}(0) +2h_{n,k}(1)e \\ & +\{(k-\frac{n}{2})g_{n,k}(0)  +\frac{1}{2}(n+3) g_{n,k}(2)\} e^2 + \{(2k-n)h_{n,k}(1)\\ &  +(n+1)h_{n,k}(3)\}e^3] ]
    \end{split}
\end{equation}

The derivation of $\Delta x$ follows a similar procedure. Substituting Eq. \ref{cdinere} in Eq. \ref{ef} and considering only $\overline{\mathcal{A}}_n$ terms, 
\begin{equation}
\label{x1iner}
    \begin{split}
        &\Delta x  =  D_c\int_0^{2\pi}\sum_{n=0}^{\infty}\sum_{k=0}^{\lfloor \frac{n}{2}\rfloor} \overline{\mathcal{A}}_n (-1)^k\binom{n}{2k}[\cos{E}+(1+\cos^2{E})e \\& +\frac{1}{2}\{(2+2k-n) \cos{E} + (n+1)\cos^3{E}\} e^2  + \frac{1}{2}\{(2k-n) \\& +(2k+1)\cos^2{E}+(n+1)\cos^4{E}\} e^3]\sin^{2k}{E}\cos^{n-2k}{E} \\& \times \exp{(\beta x\cos{E})}dE
    \end{split}
\end{equation}

Using Eqs. \ref{hm} and \ref{gm} to integrate Eq. \ref{x1iner}, 
\begin{equation}
\label{x1iner2}
    \begin{split}
        &\Delta x  =  D_c\left[\sum_{{}^{l=0}_{n \in 2l}}^{\infty}\sum_{k=0}^{\lfloor \frac{n}{2}\rfloor} \overline{\mathcal{A}}_n (-1)^k\binom{n}{2k}[g_{n,k}(1)+(h_{n,k}(0) \right. \\& \left. +h_{n,k}(2))e +\frac{1}{2}\{(2+2k-n)g_{n,k}(1) + (n+1)g_{n,k}(3)\} e^2 \right.\\& \left. + \frac{1}{2}\{(2k-n)h_{n,k}(0)  +(2k+1)h_{n,k}(2)+(n+1)h_{n,k}(4)\}e^3] \right.\\& \left. + \sum_{{}^{\hspace{2mm}l=0}_{n \in 2l+1}}^{\infty}\sum_{k=0}^{\lfloor \frac{n}{2}\rfloor} \overline{\mathcal{A}}_n (-1)^k\binom{n}{2k} [h_{n,k}(1) +(g_{n,k}(0)+g_{n,k}(2))e \right. \\& \left.+ \frac{1}{2}\{(2+2k-n)h_{n,k}(1)+ (n+1)h_{n,k}(3)\} e^2 + \frac{1}{2}\{(2k-n) \right.\\& \left. \times g_{n,k}(0)+(2k+1)g_{n,k}(2)  +(n+1)g_{n,k}(4)\}e^3]\right]
    \end{split}
\end{equation}

The change in argument of perigee can be derived similarly as,
\begin{equation}
\label{x1iner3}
    \begin{split}
        &\Delta \omega  =  D_w\int_0^{2\pi}\sum_{n=0}^{\infty}\sum_{k=0}^{\lfloor \frac{n-1}{2}\rfloor} \overline{\mathcal{B}}_n (-1)^k\binom{n}{2k+1}[1+\cos{E}e \\ & +\frac{1}{2}\{(n+1)\cos^2{E}  - (n-2k-1)\} e^2 +  \frac{1}{2}\cos{E}\{(n+1)\cos^2{E} \\ & -(n-2k-1)\}e^3]  \sin^{2k+2}{E}\cos^{n-2k-1}{E} \exp{(\beta x\cos{E})}dE
    \end{split}
\end{equation}
Integrating the equation, the final form is given by
\begin{equation}
\label{omginer}
    \begin{split}
        &\Delta \omega  =  D_w\left[\sum_{{}^{l=0}_{n \in 2l}}^{\infty}\sum_{k=0}^{\lfloor \frac{n-1}{2}\rfloor} \overline{\mathcal{B}}_n (-1)^k\binom{n}{2k+1}[g_{n,k+1}(1)+h_{n,k+1}(2)e \right.\\& \left. +\frac{1}{2}\{(n+1)g_{n,k+1}(3) - (n-2k-1)g_{n,k+1}(1)\} e^2 + \frac{1}{2}\{(n+1)\right.\\& \left.  \times h_{n,k+1}(3)-(n-2k-1)g_{n,k+1}(1)\}e^3]+ \sum_{{}^{\hspace{2mm}l=0}_{n \in 2l+1}}^{\infty}\sum_{k=0}^{\lfloor \frac{n-1}{2}\rfloor} \overline{\mathcal{B}}_n (-1)^k \right. \\ & \left.\times \binom{n}{2k+1}[h_{n,k+1}(1)+g_{n,k+1}(2)e+\frac{1}{2}\{(n+1)h_{n,k+1}(3)\right. \\& \left.- (n-2k-1)h_{n,k+1}(1)\} e^2 + \frac{1}{2}\{(n+1)g_{n,k+1}(3) - (n-2k-1)\right.\\& \left.  \times h_{n,k+1}(1)\}e^3]\right]
    \end{split}
\end{equation}
For $n = 0$, the equations for semi-major axis and focal-length reduce to the original KH formulation given by Eqs. \ref{delakh} and \ref{delxkh} while the argument of perigee change reduces to zero. The average drag-coefficient that best approximates the higher order Fourier theory given by Eqs. \ref{a1iner2} and \ref{x1iner2} can be calculated using Eq. \ref{cdavg} as follows
\begin{equation}
\label{cdrho2}
\begin{split}
    C_{D0} & = \frac{\int_0^{2\pi}\rho C_D(E)dE}{\int_0^{2\pi}\rho dE}\\
    & = \dfrac{1}{I_0}\left[\sum_{{}^{l=0}_{n \in 2l}}^{\infty}\sum_{k=0}^{\lfloor \frac{n}{2}\rfloor} \overline{\mathcal{A}}_n (-1)^k\binom{n}{2k}[h_{n,k}(0) + \{(k-\dfrac{n}{2})h_{n,k}(0) \right. \\& \left. + \dfrac{n}{2}h_{n,k}(2)\}e^2]+ \sum_{{}^{l=0}_{n \in 2l+1}}^{\infty}\sum_{k=0}^{\lfloor \frac{n}{2}\rfloor} \overline{\mathcal{A}}_n (-1)^k\binom{n}{2k}[g_{n,k}(0)\right. \\& \left. + \{(k-\dfrac{n}{2})g_{n,k}(0) + \dfrac{n}{2}g_{n,k}(2)\}e^2] \right]
\end{split}
\end{equation}

\emph{\textbf{High eccentricity regime}}, $0.2\leq e<1$
\par 
Similar to the nadir pointing case, the body angle in the transformed variable is given by
\begin{equation}
    \sin{\phi} = \sqrt{\frac{1-(1-\lambda^2/z)^2}{1-e^2(1-\lambda^2/z)^2}}
\end{equation}
\begin{equation}
    \cos{\phi} = \sqrt{\frac{1-e^2}{1-e^2(1-\lambda^2/z)^2}}(1-\lambda^2/z)
\end{equation}
The drag-coefficient in the transformed variable is given by
\begin{equation}
\label{cdinerlam}
    \begin{split}
        & C_D(\lambda) =  \sum_{n=0}^{\infty} \left[\overline{\mathcal{A}}_n\left \{ \sum_{k=0}^{\lfloor \frac{n}{2}\rfloor}(-1)^k\binom{n}{2k}\left ( \sqrt{\frac{1-(1-\lambda^2/z)^2}{1-e^2(1-\lambda^2/z)^2}} \right)^{2k}\right.\right.\\&  \left.\left.\times \left( \sqrt{\frac{1-e^2}{1-e^2(1-\lambda^2/z)^2}}(1-\lambda^2/z) \right )^{n-2k}  \right\}  + \overline{\mathcal{B}}_n\left \{ \sum_{k=0}^{\lfloor \frac{n-1}{2}\rfloor}(-1)^k\right. \right. \\& \left.\left. \times \binom{n}{2k+1}  \left ( \sqrt{\frac{1-(1-\lambda^2/z)^2}{1-e^2(1-\lambda^2/z)^2}} \right)^{2k+1}\right. \right. \\ & \left. \left. \times \left( \sqrt{\frac{1-e^2}{1-e^2(1-\lambda^2/z)^2}}(1-\lambda^2/z) \right )^{n-2k-1}  \right\}  \right].
    \end{split}
\end{equation}

Substitute Eq. \ref{cdinerlam} in Eq. \ref{aeh1} and carrying out a power series expansion in $\lambda^2/z$,
\begin{equation}
\label{aehiner2}
    \begin{split}
        \Delta a =& 2\exp{(z)}\sqrt{2/z}D_c\frac{(1+e)^{3/2}}{(1-e)^{1/2}}\int_{0}^{\sqrt{2z}} \sum_{n=0}^{\infty}\sum_{k=0}^{\lfloor n/2\rfloor} \overline{\mathcal{A}}_n(-1)^k \\& \times \binom{n}{2k}\left(\frac{2}{1-e^2}\right)^k \left [(\lambda^2/z)^k + L_1(\lambda^2/z)^{k+1}+ L_2(\lambda^2/z)^{k+2} \right. \\& \left. + \mathcal{O}((\lambda^2/z)^{k+3}) \right ]\exp{(-\lambda^2)}d\lambda,
    \end{split}
\end{equation}

where $L_1$ and $L_2$ are functions of the summation indices $n$ and $k$, and the eccentricity $e$ and are given by
\begin{equation}
    L_1 = \frac{1}{4(1-e^2)}[(6k-4n+1)-8e+(3-6k)e^2],
\end{equation}
\begin{equation}
\begin{split}
    L_2 = & \frac{1}{32(1-e^2)^2}[\{4(2n-3k)^2+40k-24n+3\}+16(4n\\ &-6k-1)e
      - (72k^2+32k-24n-48kn-50)e^2 + 16(6k \\ & +1)e^3   +  ((2k-1)(18k+5)e^4 ] .
\end{split}
\end{equation}
Using Eq. \ref{gammfunc}, 
\begin{equation}
\begin{split}
    \Delta a = & D_c'\sum_{n=0}^{\infty}\sum_{k=0}^{\lfloor n/2\rfloor} \overline{\mathcal{A}}_n(-1)^k\binom{n}{2k}\left(\frac{2}{z(1-e^2)}\right)^k\left [\Gamma\left(\frac{2k+1}{2}\right) \right. \\ & \left. + \frac{L_1}{z}\Gamma\left(\frac{2k+3}{2}\right) + \frac{L_2}{z^2}\Gamma\left(\frac{2k+5}{2}\right)\right ].
\end{split}
\end{equation}

To derive $\Delta x$, substitute Eq. \ref{cdinerlam} in Eq. \ref{xeh1} and expand as a power series in $\lambda^2/z$ to obtain,
\begin{equation}
\label{xehiner2}
    \begin{split}
        \Delta x =& 2\exp{(z)}\sqrt{2/z}D_c\frac{(1+e)^{3/2}}{(1-e)^{1/2}}\int_{0}^{\sqrt{2z}} \sum_{n=0}^{\infty}\sum_{k=0}^{\lfloor n/2\rfloor} \overline{\mathcal{A}}_n(-1)^k\binom{n}{2k}  \\ & 
        \times \left(\frac{2}{1-e^2}\right)^k \left [(\lambda^2/z)^k + N_1(\lambda^2/z)^{k+1}+ N_2(\lambda^2/z)^{k+2} \right. \\ & \left. + \mathcal{O}((\lambda^2/z)^{k+3}) \right ]\exp{(-\lambda^2)}d\lambda,
    \end{split}
\end{equation}

where $N_1$ and $N_2$ given by
\begin{equation}
    N_1 = -\frac{1}{4(1-e^2)}[(4n-6k+3)+(6k+1)e^2],
\end{equation}
\begin{equation}
\begin{split}
    N_2 = & \frac{1}{32(1-e^2)^2}[\{4(2n-3k)^2+8(n-k)-5\}+32e-2(36k^2 \\ & +16k-28n  -24kn+7)e^2 + 32e^3+  (36k^2+40k+3)e^4 ]. 
\end{split}
\end{equation}
Using Eq. \ref{gammfunc}, 
\begin{equation}
\begin{split}
    \Delta x = & D_c'\sum_{n=0}^{\infty}\sum_{k=0}^{\lfloor n/2\rfloor}\overline{\mathcal{A}}_n(-1)^k\binom{n}{2k}\left(\frac{2}{z(1-e^2)}\right)^k\left [\Gamma\left(\frac{2k+1}{2}\right) \right. \\ & \left.+ \frac{N_1}{z}\Gamma\left(\frac{2k+3}{2}\right) + \frac{N_2}{z^2}\Gamma\left(\frac{2k+5}{2}\right)\right ].
\end{split}
\end{equation}

The change in argument of perigee can be similarly derived by retaining the $\overline{\mathcal{B}}_n$ terms, 

\begin{equation}
\begin{split}
    \Delta \omega = & 4\exp{(z)}D_w\sqrt{\dfrac{1+e}{1-e}}\int_{0}^{\sqrt{2z}}\sum_{n=0}^{\infty}\sum_{k=0}^{\lfloor \frac{n-1}{2}\rfloor}\overline{\mathcal{B}}_n(-1)^k \binom{n}{2k+1}\\ & \times \left(\dfrac{2}{1-e^2}\right)^{\frac{2k+1}{2}}\left[(\lambda^2/z)^\frac{2k+3}{2} + W_I(\lambda^2/z)^\frac{2k+5}{2}\right]d\lambda
    \end{split}
\end{equation}
where 
\begin{equation}
    W_I = -\frac{1}{4(1-e^2)}[(4n-6k-3)+4e+(6k+3)e^2]
\end{equation}

The integrated change is given by
\begin{equation}
\begin{split}
    \Delta \omega = & 2\exp{(z)}D_w\sqrt{\dfrac{1+e}{1-e}}\sum_{n=0}^{\infty}\sum_{k=0}^{\lfloor \frac{n-1}{2}\rfloor}\overline{\mathcal{B}}_n(-1)^k \binom{n}{2k+1} \\ & \times \left(\dfrac{2}{z(1-e^2)}\right)^{\frac{2k+1}{2}}\left[\frac{1}{z}\Gamma(k+2) + \frac{W_I}{z^2}\Gamma(k+3)\right]
\end{split}
\end{equation}

The density-averaged drag-coefficient can be derived as follows
\begin{equation}
\begin{split}
    C_{D0} & = \dfrac{\exp{z}}{\pi\sqrt{2z}I_0}\left[ \sum_{n=0}^{\infty}\sum_{k=0}^{\lfloor \frac{n}{2}\rfloor}\overline{\mathcal{A}}_n(-1)^k\binom{n}{2k}\left(\frac{2}{z(1-e^2)}\right)^{k} \right. \\ & \left. \times \left\{ \Gamma\left(\frac{2k+1}{2}\right) \right.\left. + \frac{C_{I1}}{z}\Gamma\left(\frac{2k+3}{2}\right)+ \frac{C_{I2}}{z}\Gamma\left(\frac{2k+5}{2}\right)\right\}\right]
\end{split}    
\end{equation}
where
\begin{equation}
    C_{I1} = -\frac{1}{4(1-e^2)}[(2n-2k-1)+(2n+2k+1)e^2],
\end{equation}
\begin{equation}
\begin{split}
    C_{I2} = & \frac{1}{32(1-e^2)^2}[(4k^2+16k+3)(1-e^2)^2 +4n(3e^4+2ke^4 \\ & +8e^2-2k-3)+(1+e^2)^2n^2] .
\end{split}
\end{equation}

\subsection{Body-Orbit double Fourier (BODF) model}
\label{bodf}
In developing the theory for the BFF model, the Fourier coefficients were assumed to be constant in the orbit. But since the drag-coefficient is a function of ambient parameters, the body-fixed Fourier coefficients are periodic functions of the eccentric anomaly. This allows the body-fixed Fourier coefficients to be expressed as Fourier series expansions around the eccentric anomaly. 
\begin{equation}
\label{bodfa}
    \overline{\mathcal{A}}_n(E) = \sum_{m=0}^{\infty} (\overline{\mathbb{A}}_{mn}\cos{mE} + \overline{\mathbb{B}}_{mn}\sin{mE}),
\end{equation}

\begin{equation}
\label{bodfb}
    \overline{\mathcal{B}}_n(E) = \sum_{m=0}^{\infty} (\overline{\mathbb{C}}_{mn}\cos{mE} + \overline{\mathbb{D}}_{mn}\sin{mE}).
\end{equation}

Therefore, the total drag-coefficient can be expressed as a body-orbit double Fourier (BODF) model \citep{jgcd_drag},
\begin{equation}
\begin{split}
    C_d =  \sum_{m=0}^{\infty}\sum_{n=0}^{\infty} & (\overline{\mathbb{A}}_{mn}\cos{mE}\cos{n\phi} + \overline{\mathbb{B}}_{mn}\sin{mE}\cos{n\phi} \\
    & + \overline{\mathbb{C}}_{mn}\cos{mE}\sin{n\phi} + \overline{\mathbb{D}}_{mn}\sin{mE}\sin{n\phi}).
\end{split}
\end{equation}
Since the sinusoidal orbit terms are zero under the assumptions of the theory, the drag coefficient can be simplified to
\begin{equation}
\begin{split}
    C_d =  \sum_{m=0}^{\infty}\sum_{n=0}^{\infty} ( \overline{\mathbb{A}}_{mn}\cos{mE}\cos{n\phi} + \overline{\mathbb{C}}_{mn}\cos{mE}\sin{n\phi}). 
\end{split}
\end{equation}
Instead of re-deriving the analytical change for the nadir-pointing and inertially stabilized cases, an approximation is made for the drag-coefficient. The orbit-fixed terms are averaged over the orbit weighted by the density in order to obtain more accurate body-fixed Fourier coefficients. Therefore, the body-fixed Fourier coefficients can be written as
\begin{equation}
\label{Amnavg}
\begin{split}
    \overline{\mathcal{A}}_{n0} & = \frac{\int_0^{2\pi}\rho \overline{\mathcal{A}}_n(E)dE}{\int_0^{2\pi}\rho dE}\\
    & = \frac{\int_0^{2\pi}\rho \sum_{m=0}^{\infty}\overline{\mathbb{A}}_{mn}\cos{mE}dE}{\int_0^{2\pi}\rho dE}\\
    & = \frac{\sum_{m=0}^{\infty}\overline{\mathbb{A}}_{mn} I_m}{I_0}.
\end{split}
\end{equation}

\begin{equation}
\label{Bmnavg}
\begin{split}
    \overline{\mathcal{B}}_{n0} & = \frac{\int_0^{2\pi}\rho \overline{\mathcal{B}}_n(E)dE}{\int_0^{2\pi}\rho dE}\\
    & = \frac{\int_0^{2\pi}\rho \sum_{m=0}^{\infty}\overline{\mathbb{C}}_{mn}\cos{mE}dE}{\int_0^{2\pi}\rho dE}\\
    & = \frac{\sum_{m=0}^{\infty}\overline{\mathbb{C}}_{mn} I_m}{I_0}.
\end{split}
\end{equation}
The body-fixed Fourier coefficients calculated using Eqs. \ref{Amnavg} and \ref{Bmnavg} can be used in the theory developed in Section \ref{bff} for a more accurate computation of the change in the orbital elements.

\section{Circular orbits}
\label{circ}
Under the assumptions of this work, the drag-coefficient variation due to ambient parameters is zero at a constant altitude. Therefore, only the zeroth order coefficient remains in the OFF model. The drag-coefficient can still vary due to attitude and therefore, the higher order BFF coefficients are still non-zero.
For the nadir pointing profile, $\phi = 0$ and the $C_D$ remains constant. But for the inertially stabilized profile, $\phi = E$. Therefore, the change in semi-major axis can be written from Eq. \ref{a_lpe} as 
\begin{equation}
\label{a_circ}
\begin{split}
    \Delta a & = -a^2\delta'\int_0^{2\pi}\frac{(1+e\cos{E})^{3/2}}{(1-e\cos{E})^{1/2}} \sum_{n=0}^{\infty} (\overline{\mathcal{A}}_{n}\cos{nE} + \overline{\mathcal{B}}_{n}\sin{nE}) \\ & \times \rho dE\\
    & =  -2\pi a^2\delta'\rho\overline{\mathcal{A}}_{0}.
\end{split}    
\end{equation}
since density is constant for a circular orbit. Therefore, the higher-order Fourier coefficients do not contribute to the change in orbital elements for an inertially stabilized profile in a circular orbit.

\section{Validation results}
\label{valid}
The theory developed in Sections \ref{off} and \ref{bff} is validated through comparisons with numerical integration of simulated satellite trajectories. The satellite orbits are simulated under the assumptions of the King-Hele theory. Only the two-body and atmospheric drag forces are considered in the dynamics.  A spherically symmetric and exponentially decaying atmosphere is assumed with a constant scale height. For the OFF model, a spherical satellite is considered such that there are no variations in the attitude. For the BFF model, non-spherical satellites are considered with the only variations considered in the drag-coefficient being due to attitude, unless stated otherwise. The drag-coefficients are modeled using the diffuse reflection incomplete accommodation (DRIA) model that linearly combines drag-coefficients based on clean surfaces and satellite surfaces completely covered by atomic oxygen \citep{walker}. Note that the model is not valid for altitudes greater than 500 km. Since the variation of drag-coefficient is not well understood for higher altitudes, the DRIA model is used for all altitudes. With future developments in drag-coefficient modeling for higher altitudes, a different model can be used in the current framework with no changes to the developed theory. All ambient parameters are modeled using NRLMSISE-00 \cite{pic} as the atmospheric model. The qualitative results are independent of the specific attributes of the atmosphere and the satellite surface; therefore, the details have been left out. The errors between the analytically and numerically computed changes in semi-major axis and focal length are compared for the Fourier theory and the original King-Hele theory with three constant drag-coefficients -  the zeroth-order Fourier coefficient, the drag-coefficient evaluated at perigee and the derived density-averaged drag-coefficient, summarized in Table \ref{kh_cases}. The orbital elements and satellite parameters in Table \ref{offmod} remain constant for all the cases. 

\begin{table}
\caption{Nomenclature for the graphical results}
\label{kh_cases}
\begin{tabular}{@{}ll@{}}
\toprule
\multicolumn{1}{c}{\textbf{Case}} & \multicolumn{1}{c}{\textbf{Description}}                                                                        \\ \midrule
Fourier $C_D$                        & Full Fourier theory developed here                                                                              \\
KH: Averaged $C_D$                   & \begin{tabular}[c]{@{}l@{}}Original King-Hele formulation \\ with derived density-averaged $C_D$\end{tabular}      \\
KH: Perigee $C_D$                    & \begin{tabular}[c]{@{}l@{}}Original King-Hele formulation \\ with $C_D$ evaluated at perigee\end{tabular}          \\
KH: Order 0 $C_D$                    & \begin{tabular}[c]{@{}l@{}}Original King-Hele formulation \\ with zeroth order Fourier coefficient\end{tabular} \\ \bottomrule
\end{tabular}
\end{table}

\begin{table}
\caption{Simulation parameters common for all the cases}
\label{offmod}
\begin{tabular}{@{}lll@{}}
\toprule
\multicolumn{2}{c}{Parameter}                                                                                                                                   & \multicolumn{1}{c}{Value}          \\ \midrule
\multirow{3}{*}{\begin{tabular}[c]{@{}l@{}}Orbital\\elements\end{tabular}} 
                                                                                 & $i_0$                                                                   & $65^\circ$              \\
                                                                                 & $\Omega_0$ & $60^\circ$             \\
                                                                                 & $\omega_0$                                                          & $40^\circ$             \\
\multirow{2}{*}{\begin{tabular}[c]{@{}l@{}}Satellite \\ parameters\end{tabular}} & $m$                                                                        & 500 kg                             \\
                                                                                 & $S$                                                        & 10 m$^2$            \\ \cmidrule(l){1-3} 
\end{tabular}
\end{table}

\subsection{Test cases for OFF model}
\begin{figure*}[H]
\centering
\begin{minipage}{0.5\textwidth}
  \centering
  \includegraphics[width=1\linewidth]{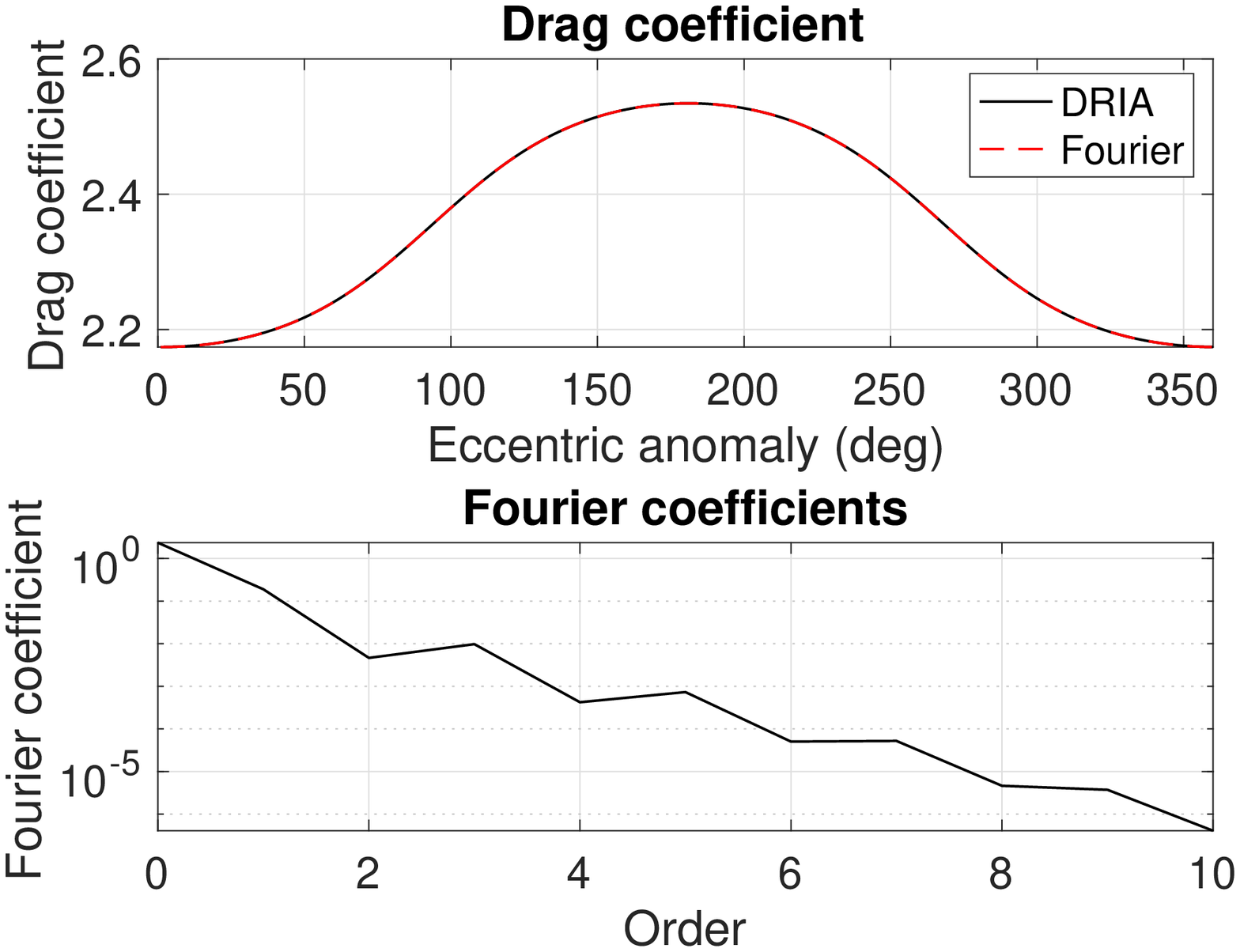}
  (a)
\end{minipage}%
\begin{minipage}{0.5\textwidth}
  \centering
  \includegraphics[width=1\linewidth]{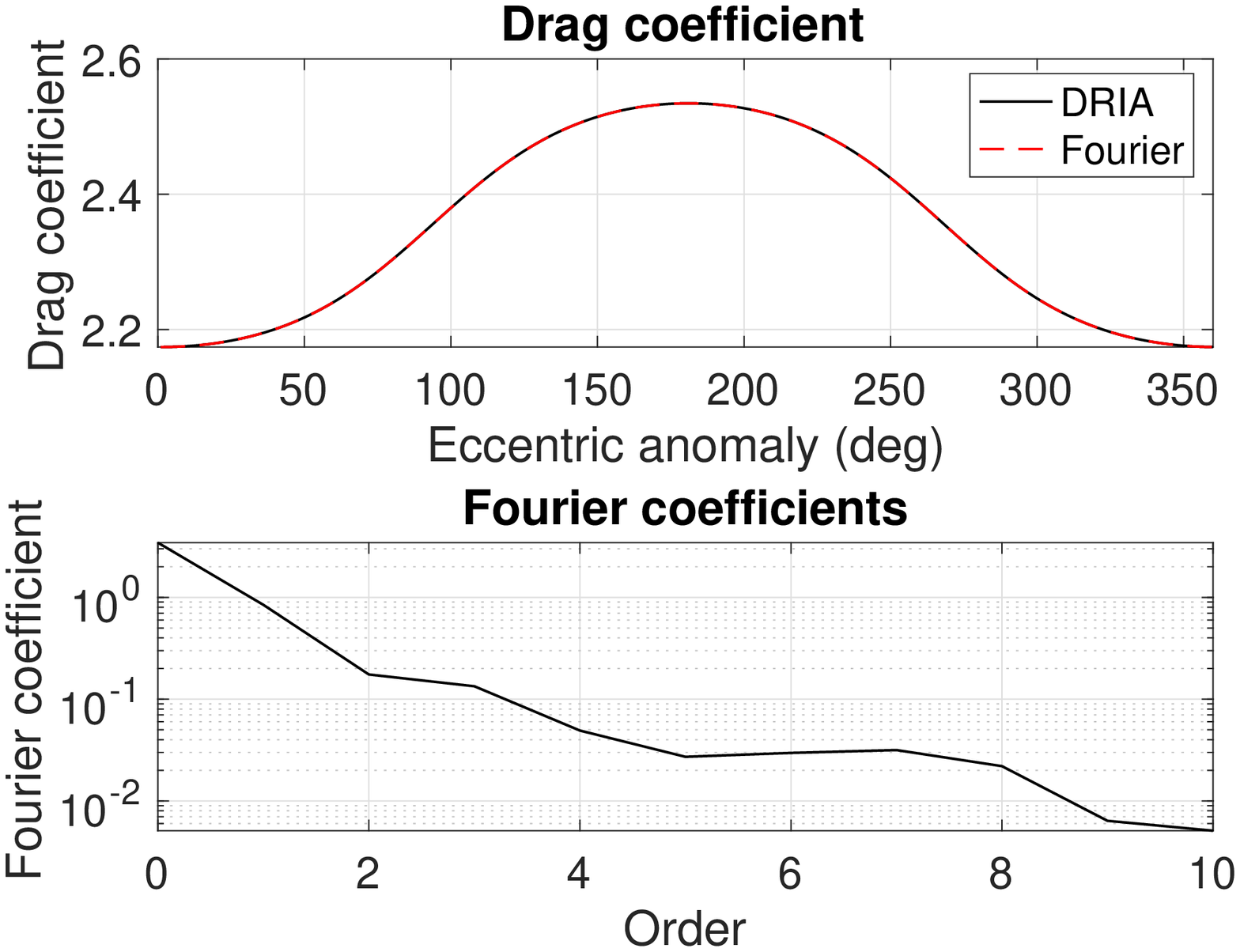}
  (b)
\end{minipage}
\caption{Diffuse reflection incomplete accommodation (DRIA) modeled drag-coefficient and Fourier coefficients for the OFF model in the (a) low eccentricity regime;  (b) high eccentricity regime.}
\label{cd_off}
\end{figure*}

A spherical satellite with perigee at 300 km and apogee at 500 km and 7000 km for low and high eccentricity regimes respectively is considered for the OFF theory. The density parameters at the perigee for both cases are $\rho_{p0} = 1.9417e-11$ kg/m$^3$ and $H = 49.23$ km corresponding to a mean solar activity level ($F_{10.7} = 150$ s.f.u). The drag coefficient for the low and high eccentricity cases along with the Fourier coefficients are plotted in Fig. \ref{cd_off}. The Fourier series approximates the drag-coefficient at lower eccentricities more accurately than at higher eccentricities and the Fourier coefficients decrease more rapidly for higher orders in the former. In the high eccentricity case, the drag-coefficient from $E = 40^\circ$ to $E = 320^\circ$ does not affect the orbit since the altitude within that range is greater than 1000 km. 

\begin{figure*}[H]
\centering
\begin{minipage}{0.5\textwidth}
  \centering
  \includegraphics[width=1\linewidth]{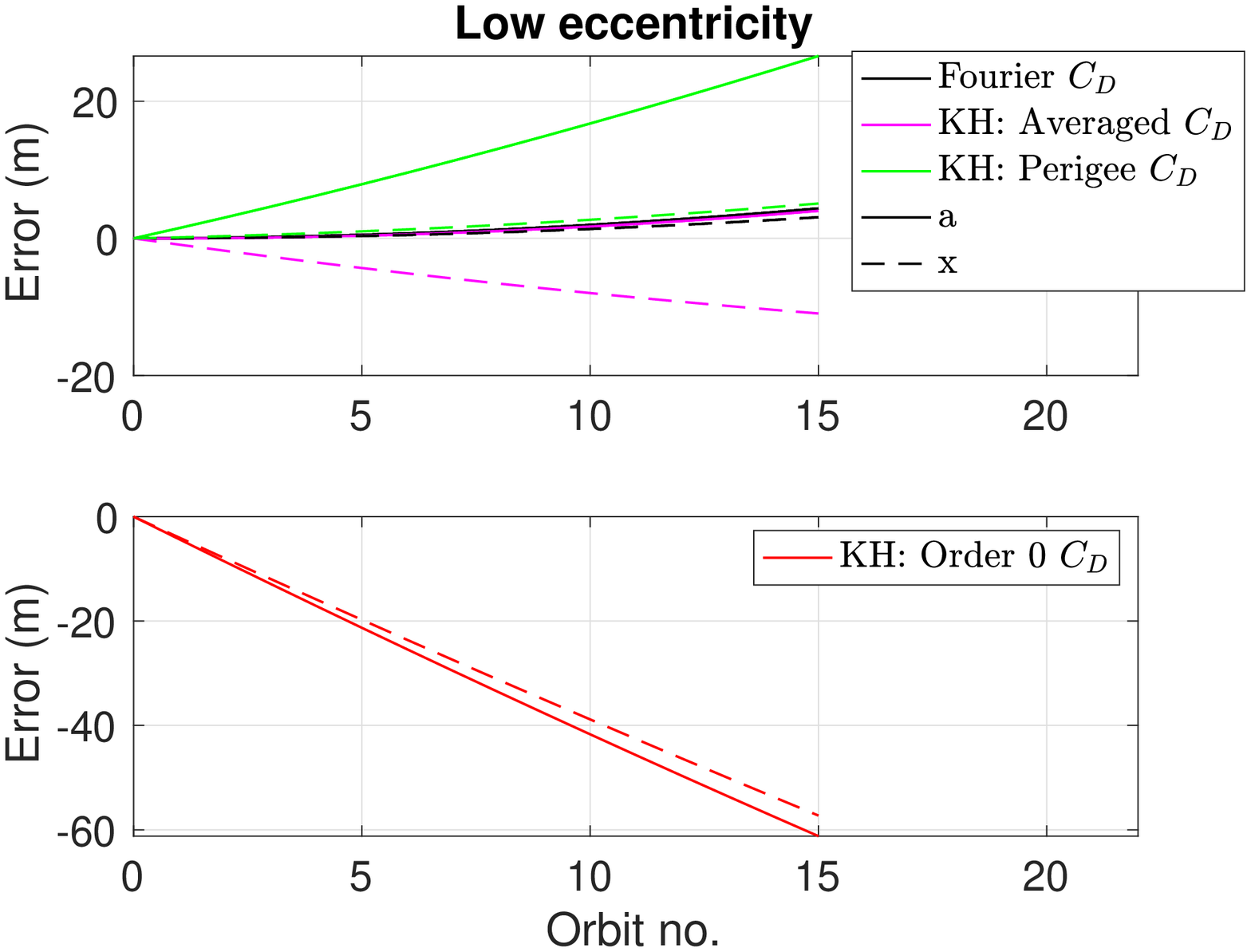}
  (a)
\end{minipage}%
\begin{minipage}{0.5\textwidth}
  \centering
  \includegraphics[width=1\linewidth]{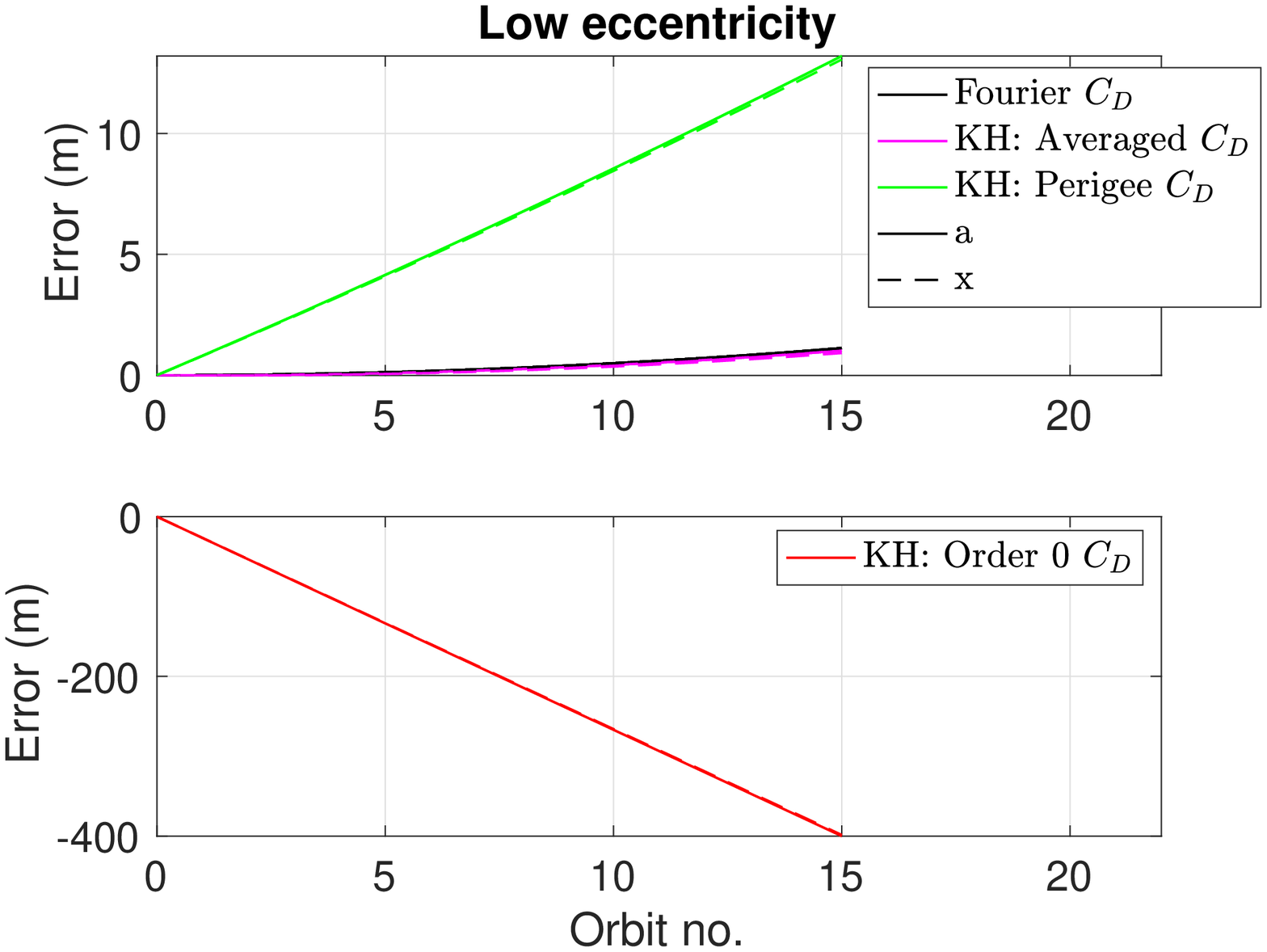}
  (b)
\end{minipage}
\caption{Error between analytical and numerical changes in semi-major axis and focal length for the the OFF model and the original King-Hele (KH) theory with three constant drag-coefficients (density-averaged, perigee and order 0 Fourier) in (a) low eccentricity regime and (b) high eccentricity regime}
\label{fig_off}
\end{figure*}

Time-profiles of the errors in the analytically computed change in semi-major axis and focal-length compared to the numerical results are plotted in Fig. \ref{fig_off}  for low and high eccentricities. The figures depict errors for the Fourier theory as well as the original King-Hele theory with three constant drag-coefficients. The error for the order 0 drag-coefficient is plotted separately as it is much larger than the other errors. It can be seen that the results of the full Fourier theory and the derived average drag-coefficient are similar though the full Fourier theory gives a more accurate focal-length change in the low eccentricity regime. They both perform an order of magnitude better than the perigee $C_D$.  

\begin{figure*}[H]
\centering
\begin{minipage}{0.5\textwidth}
  \centering
  \includegraphics[width=1\linewidth]{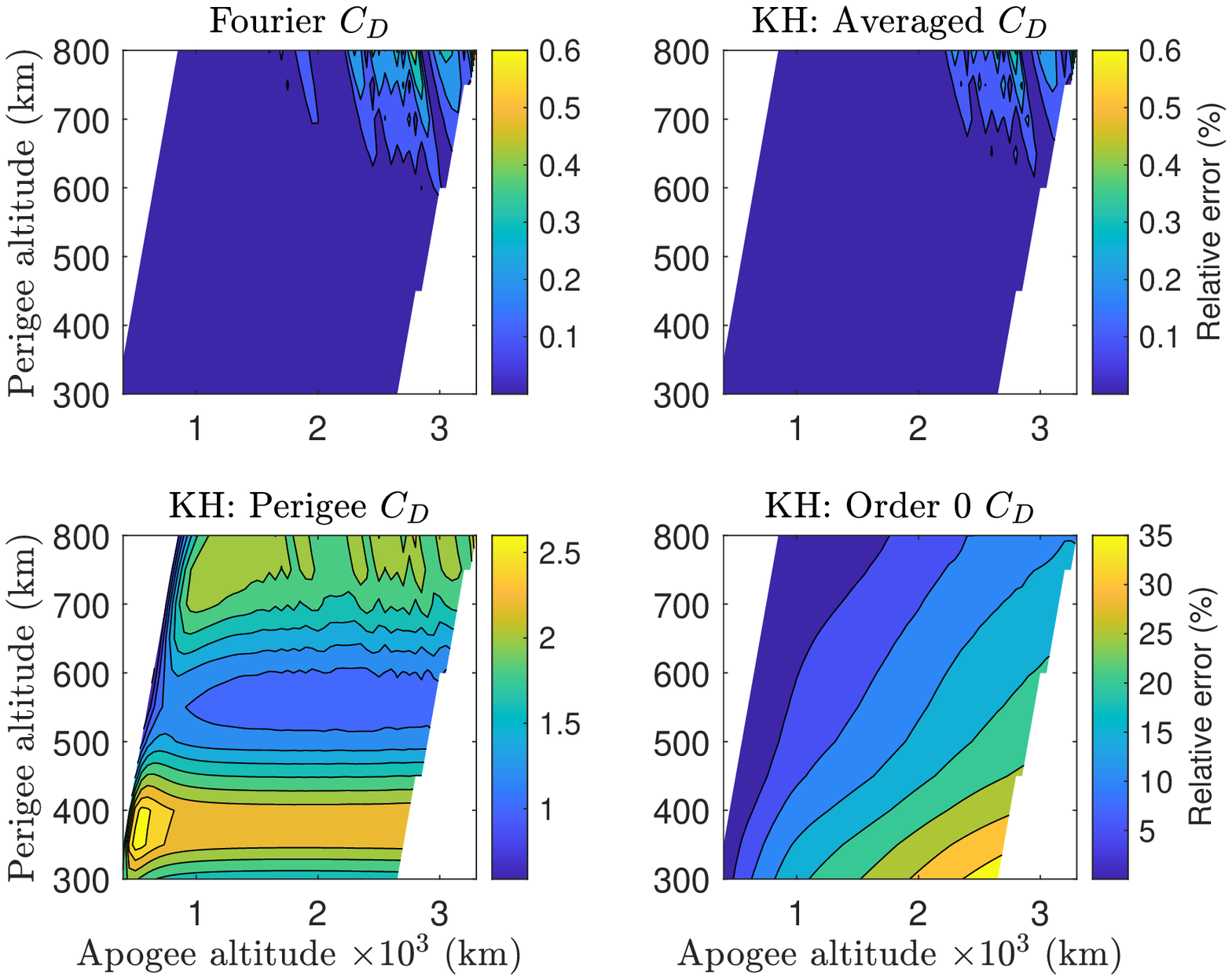}
  (a)
\end{minipage}%
\begin{minipage}{0.5\textwidth}
  \centering
  \includegraphics[width=1\linewidth]{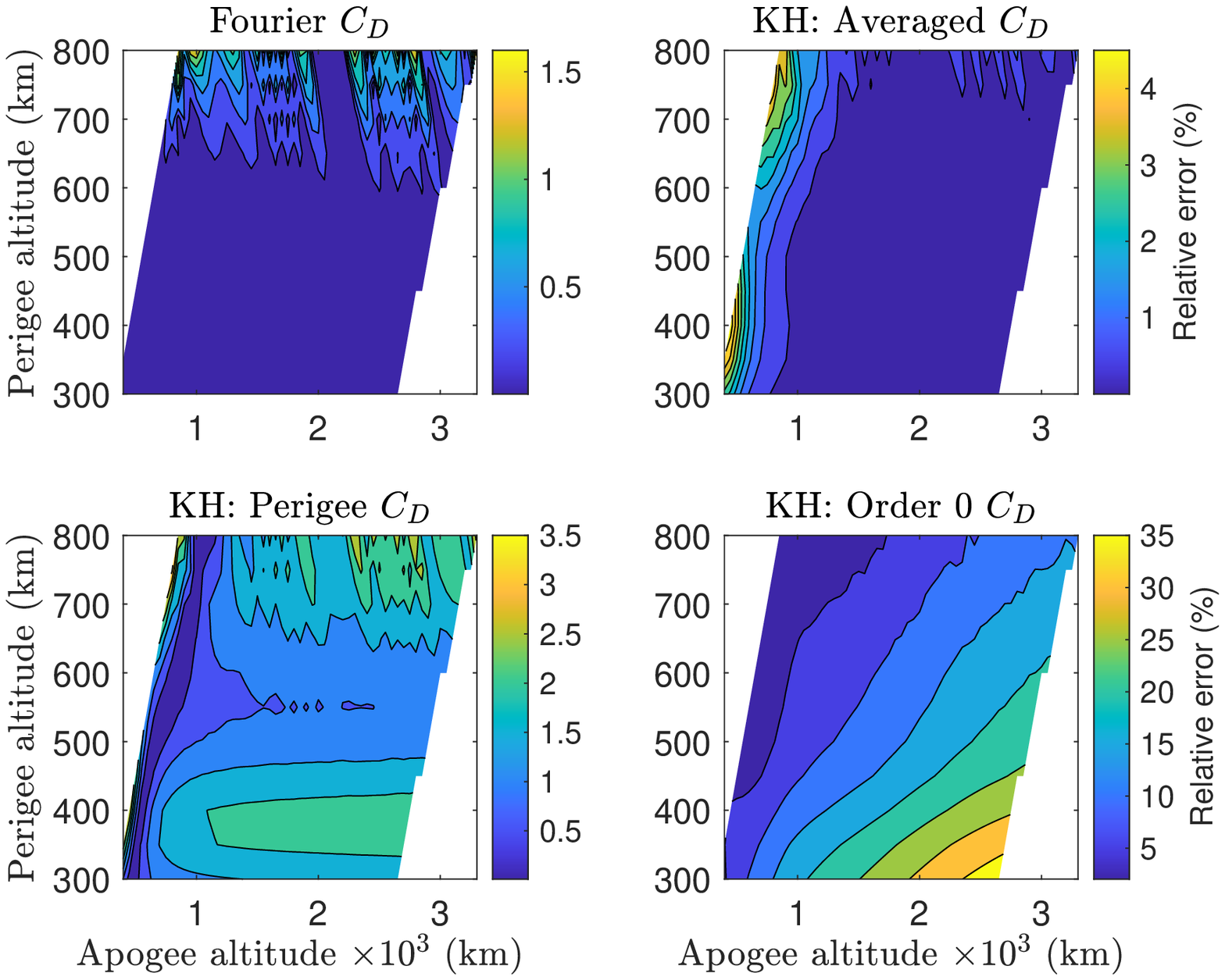}
  (b)
\end{minipage}
\caption{Relative error in analytically computed change in (a) semi-major axis and (b) focal length compared to numerical results for OFF model and original King-Hele (KH) theory with three constant drag-coefficients (density-averaged, perigee and order 0 Fourier) in low eccentricity regime}
\label{off_grid_lowEcc}
\end{figure*}

\begin{figure*}[H]
\centering
\begin{minipage}{0.5\textwidth}
  \centering
  \includegraphics[width=1\linewidth]{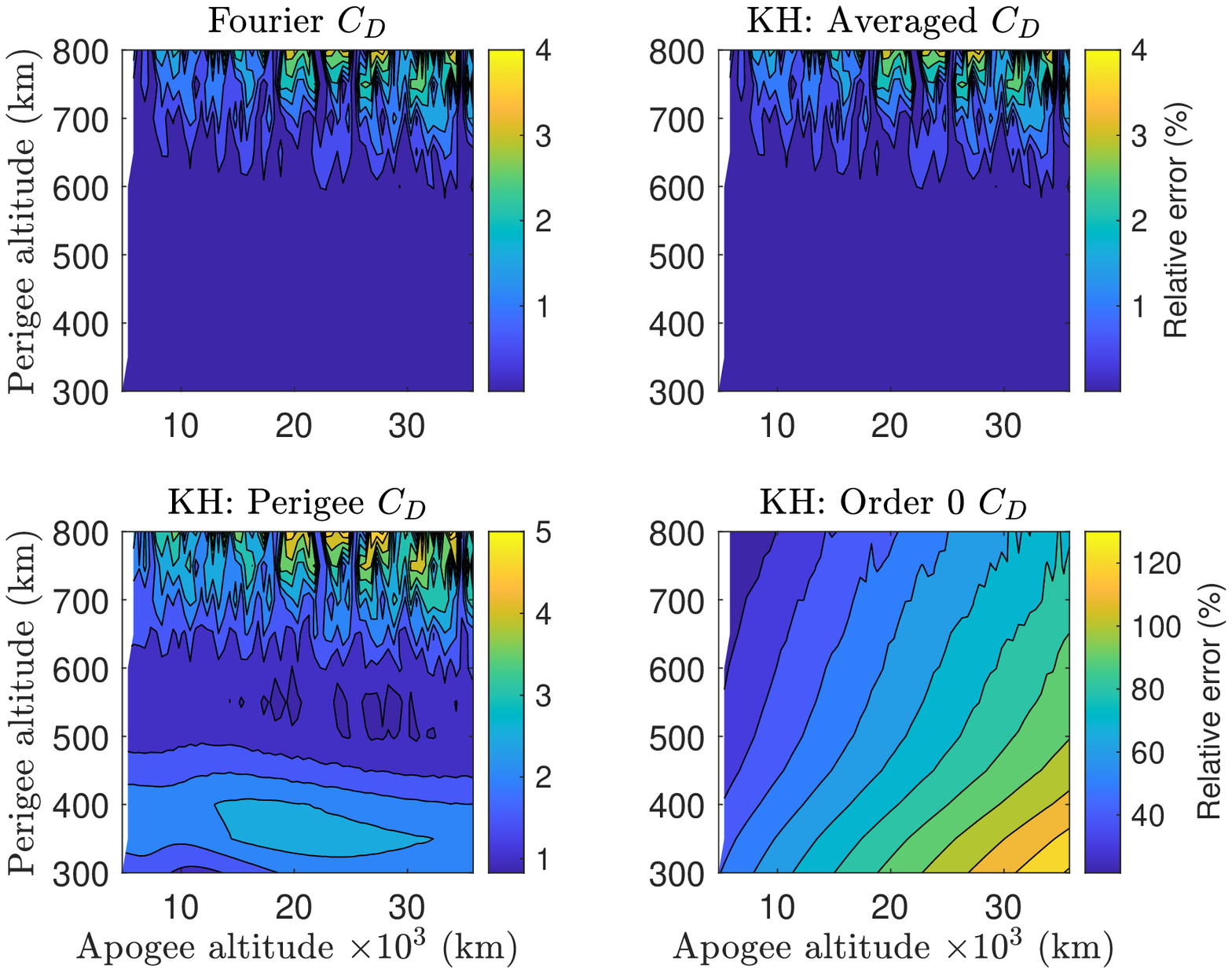}
  (a)
\end{minipage}%
\begin{minipage}{0.5\textwidth}
  \centering
  \includegraphics[width=1\linewidth]{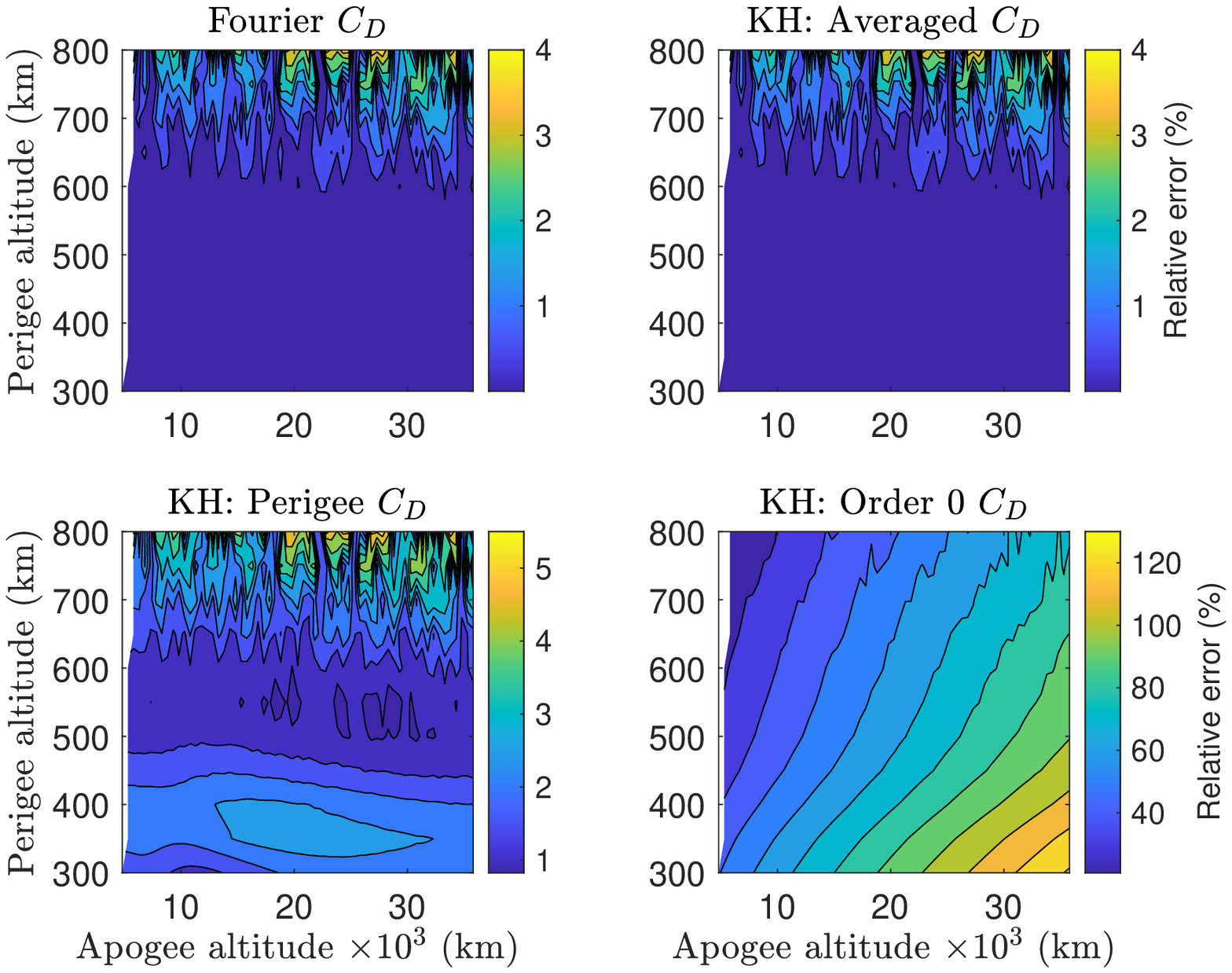}
  (b)
\end{minipage}
\caption{Relative error in analytically computed change in (a) semi-major axis and (b) focal length compared to numerical results for OFF model and original King-Hele (KH) theory with three constant drag-coefficients (density-averaged, perigee and order 0 Fourier) in high eccentricity regime}
\label{off_grid_highEcc}
\end{figure*}

The relative errors in the analytical semi-major axis and focal-length over a single orbital period are computed for a grid of perigee and apogee heights with a constraint of $0.01<e<0.15$ in the low eccentricity regime and $ 0.25<e<0.75 $ in the high eccentricity regime to avoid truncation errors. The errors are plotted in Fig. \ref{off_grid_lowEcc} and \ref{off_grid_highEcc} for low and high eccentricity regimes respectively. The performances of the full Fourier theory and the original King-Hele formulation with the derived average $C_D$ are similar to each other except for focal-length for low eccentricity regime. The relative errors are largest for high perigees and apogees since the change in the orbital elements over an orbital period is very small at such high altitudes. The relative errors for perigee $C_D$ and order 0 $C_D$ are worse throughout the grid.

\subsection{Test cases for the BFF model}
\begin{figure*}[H]
\centering
\begin{minipage}{0.5\textwidth}
  \centering
  \includegraphics[width=1\linewidth]{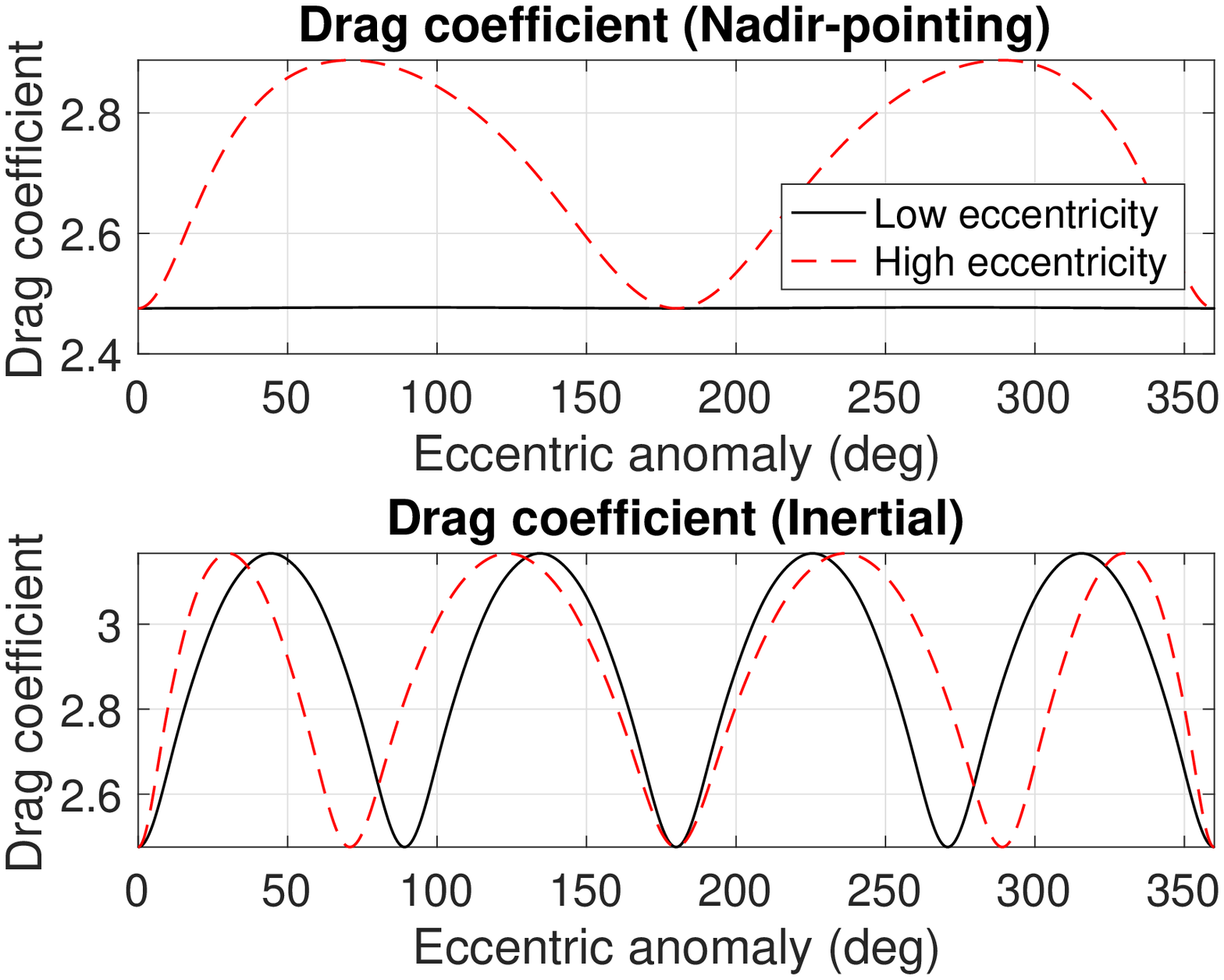}
  (a)
\end{minipage}%
\begin{minipage}{0.5\textwidth}
  \centering
  \includegraphics[width=1\linewidth]{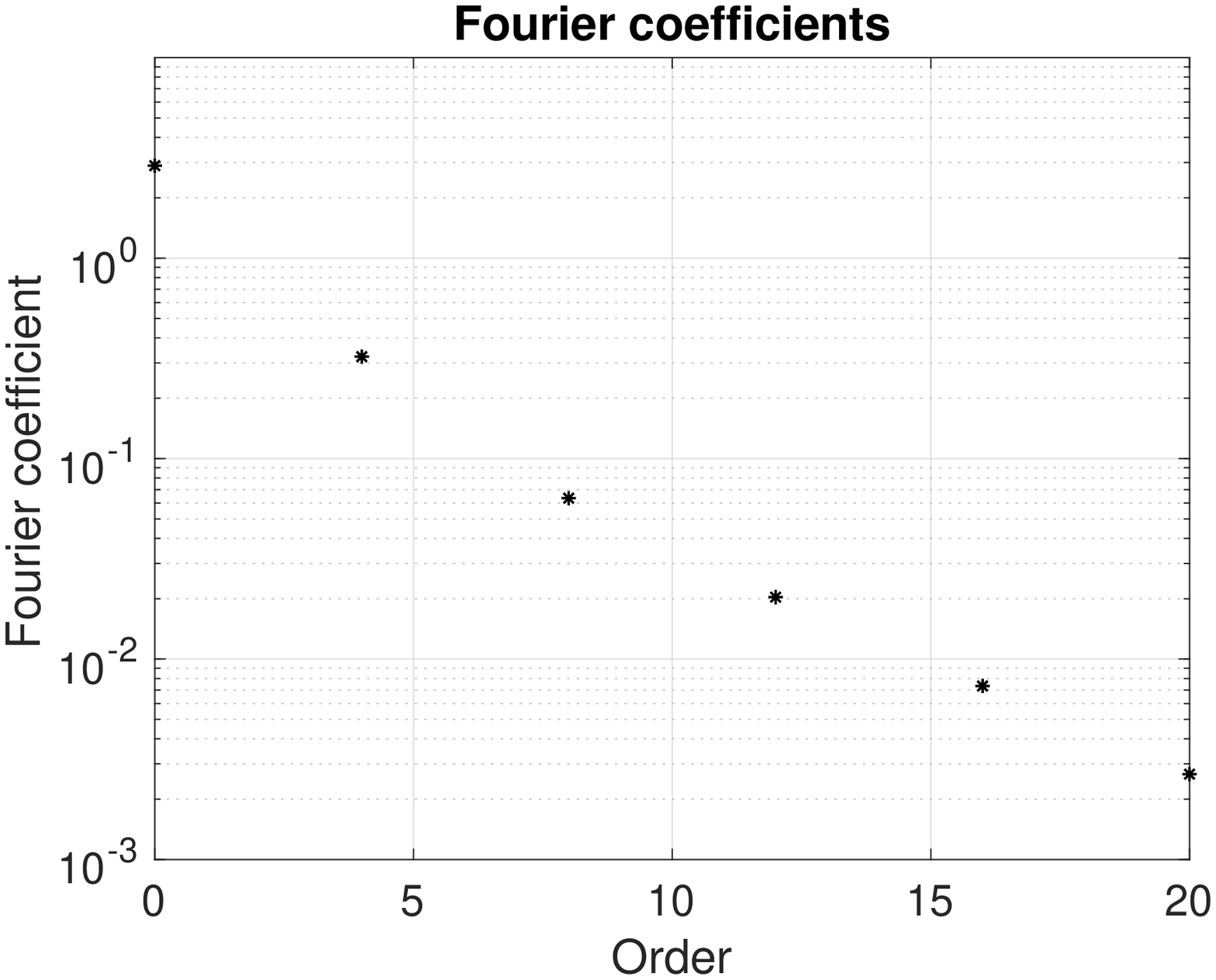}
  (b)
\end{minipage}
\caption{Diffuse reflection incomplete accommodation (DRIA) modeled drag-coefficients for the BFF model in the nadir pointing and inertially stabilized cases; (b) Fourier coefficients for the BFF model.}
\label{cd_bff}
\end{figure*}

To validate the BFF model, a symmetric cubical satellite with equal properties for all the six surfaces is considered with the perigee and apogee altitudes same as the previous case. The drag coefficients for the nadir-pointing and inertially stabilized cases are plotted in Fig. \ref{cd_bff}. The variation in the drag-coefficient for the nadir-pointing case is very small for the low eccentricity regime since the flight-path angle is very small. On the other hand, the drag-coefficients for the inertial case are similar for both eccentricity regimes since $\phi$ undergoes a complete rotation. For a symmetric cubical satellite, only the cosine Fourier coefficients with orders that are multiples of four are non-zero. All the sine Fourier coefficients are zero due to symmetry. 

\begin{figure*}[H]
\centering
\begin{minipage}{0.5\textwidth}
  \centering
  \includegraphics[width=1\linewidth]{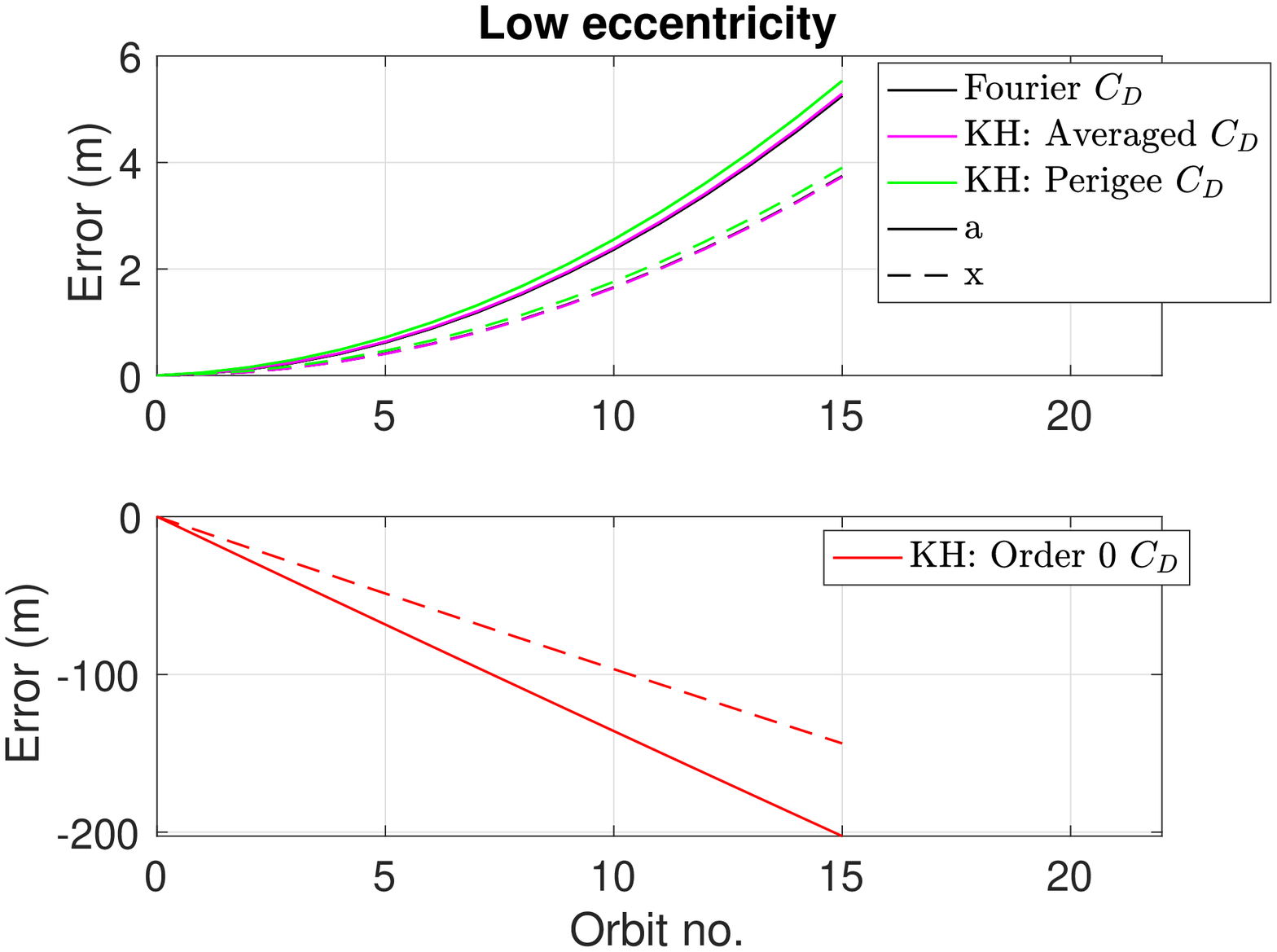}
  (a)
\end{minipage}%
\begin{minipage}{0.5\textwidth}
  \centering
  \includegraphics[width=1\linewidth]{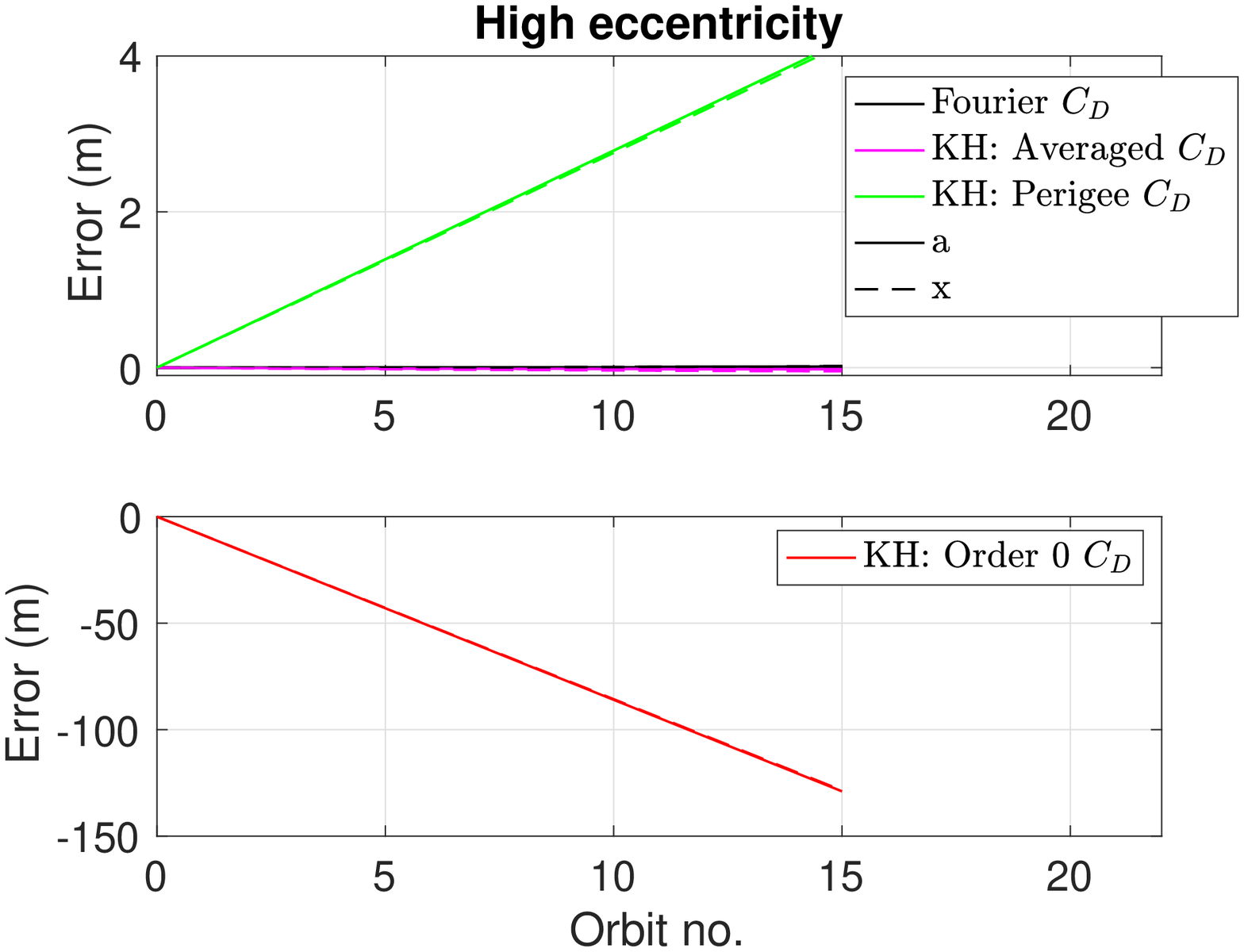}
  (b)
\end{minipage}
\caption{Error between analytical and numerical changes in semi-major axis and focal length for the the BFF model and the original King-Hele (KH) theory with three constant drag-coefficients (density-averaged, perigee and order 0 Fourier) in (a) low eccentricity regime and (b) high eccentricity regime for a nadir-pointing satellite}
\label{fig_bff_nadir}
\end{figure*}

\begin{figure*}[H]
\centering
\begin{minipage}{0.5\textwidth}
  \centering
  \includegraphics[width=1\linewidth]{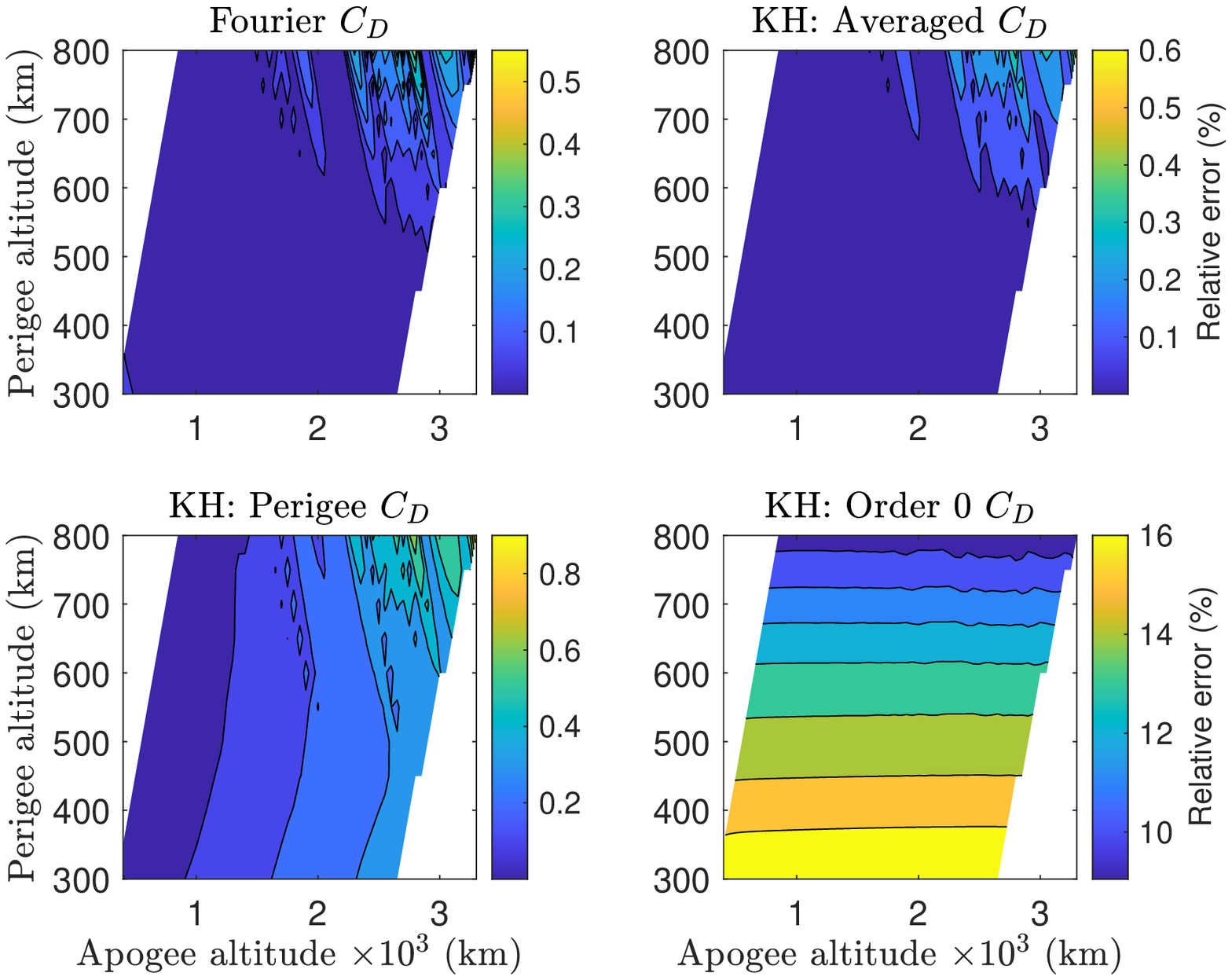}
  (a)
\end{minipage}%
\begin{minipage}{0.5\textwidth}
  \centering
  \includegraphics[width=1\linewidth]{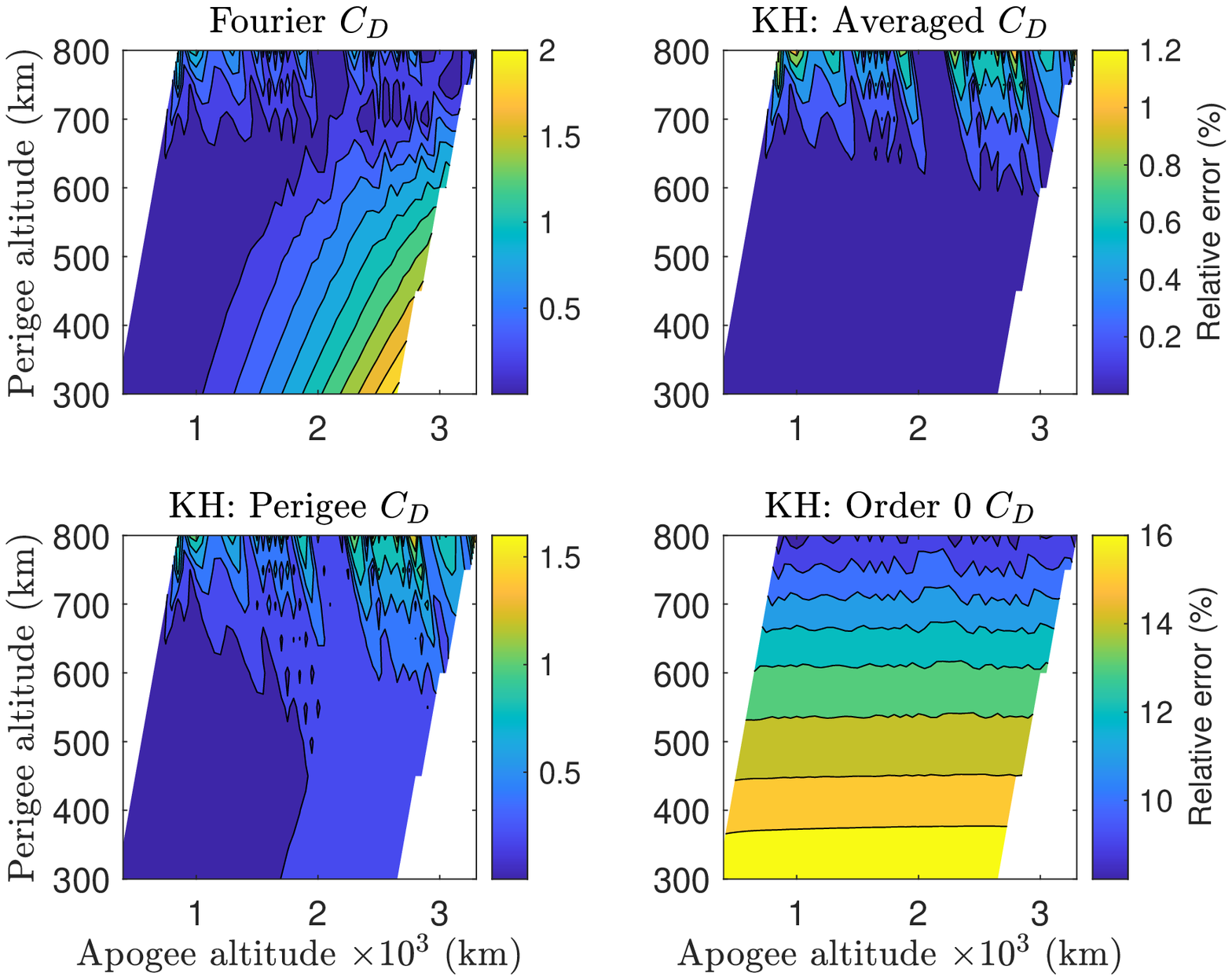}
  (b)
\end{minipage}
\caption{Relative error in analytically computed change in (a) semi-major axis and (b) focal length compared to numerical results for BFF model and original King-Hele (KH) theory with three constant drag-coefficients (density-averaged, perigee and order 0 Fourier) in low eccentricity regime for a nadir pointing profile}
\label{bff_grid_lowEcc}
\end{figure*}

\begin{figure*}[H]
\centering
\begin{minipage}{0.5\textwidth}
  \centering
  \includegraphics[width=1\linewidth]{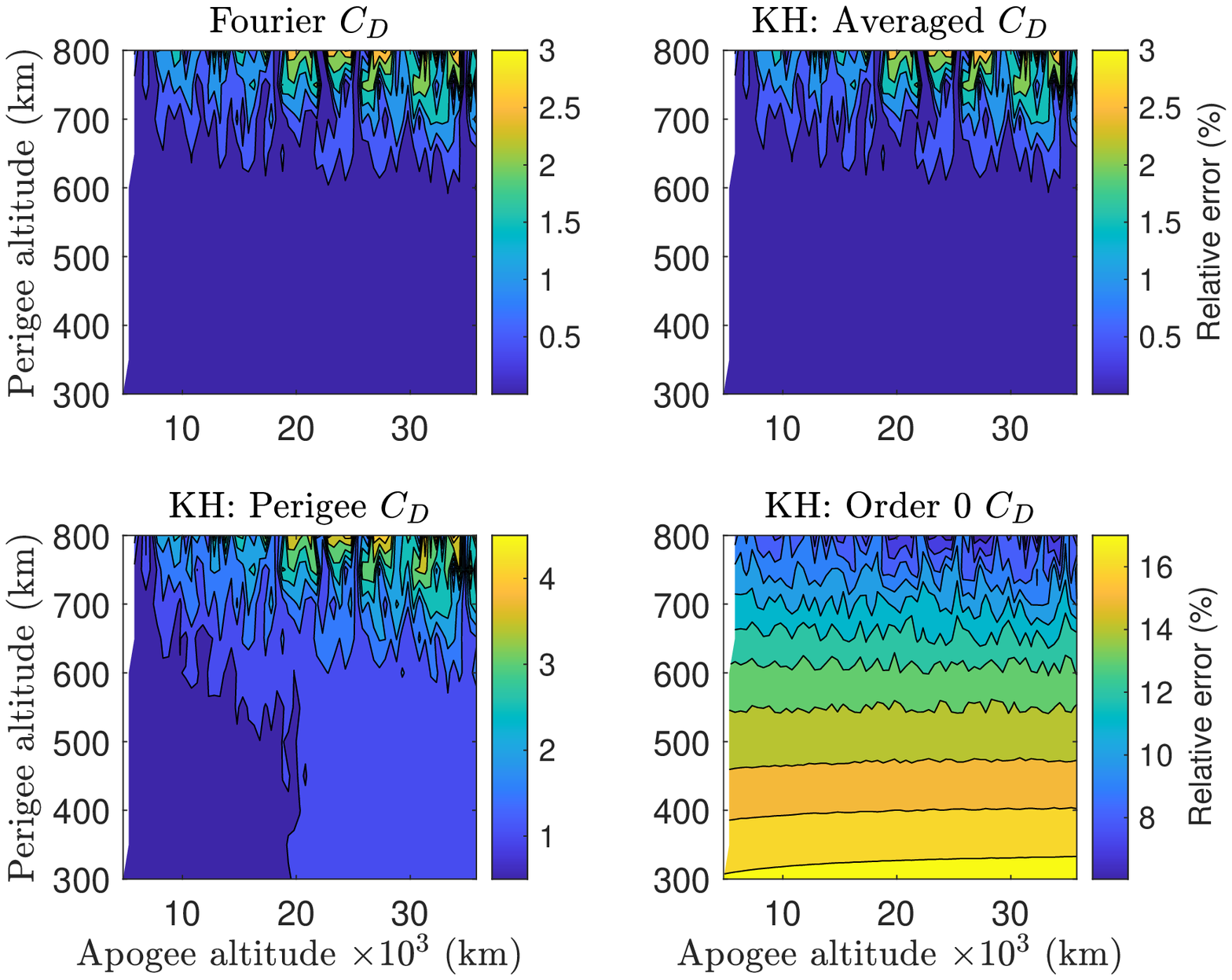}
  (a)
\end{minipage}%
\begin{minipage}{0.5\textwidth}
  \centering
  \includegraphics[width=1\linewidth]{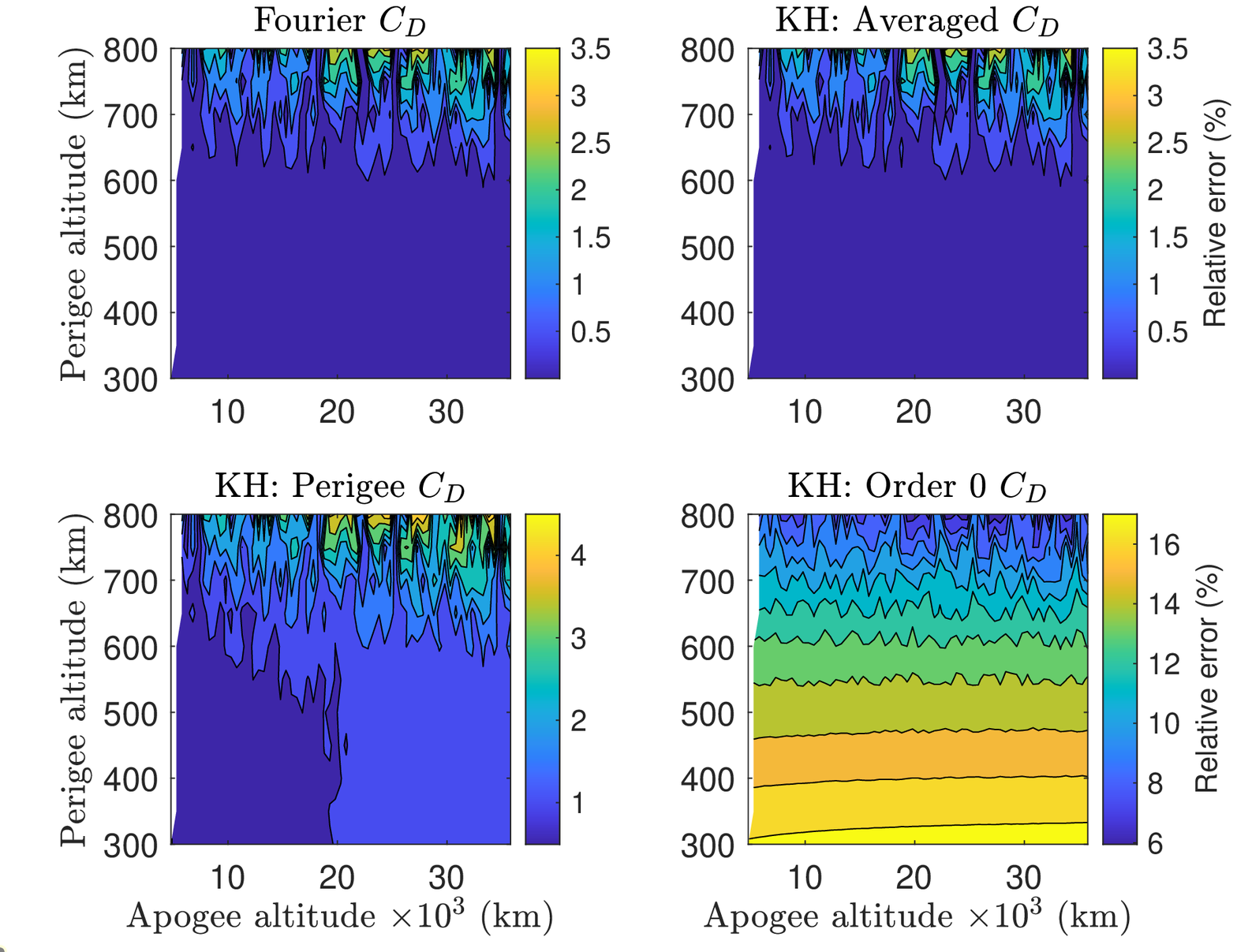}
  (b)
\end{minipage}
\caption{Relative error in analytically computed change in (a) semi-major axis and (b) focal length compared to numerical results for BFF model and original King-Hele (KH) theory with three constant drag-coefficients (density-averaged, perigee and order 0 Fourier) in high eccentricity regime for a nadir-pointing profile}
\label{bff_grid_highEcc}
\end{figure*}

The errors between the Fourier theory and the numerical results are compared with the original formulation with the three constant drag-coefficients in Fig. \ref{fig_bff_nadir} for low and high eccentricities. The averaged $C_D$ and the full Fourier theory have similar errors in the both the eccentricity regimes.  This is also demonstrated by relative errors over a grid of perigee and apogee altitudes in Figs. \ref{bff_grid_lowEcc} and \ref{bff_grid_highEcc}. It should be noted that the variation of drag-coefficients is very small for a nadir-pointing profile as seen in Fig. \ref{cd_bff}. In the high eccentricity regime, most of the variation is in higher altitudes, which has a negligible contribution to the orbit. Therefore, the averaged and full King-Hele theory are expected to perform similarly. 

\begin{figure*}[H]
\centering
\begin{minipage}{0.5\textwidth}
  \centering
  \includegraphics[width=1\linewidth]{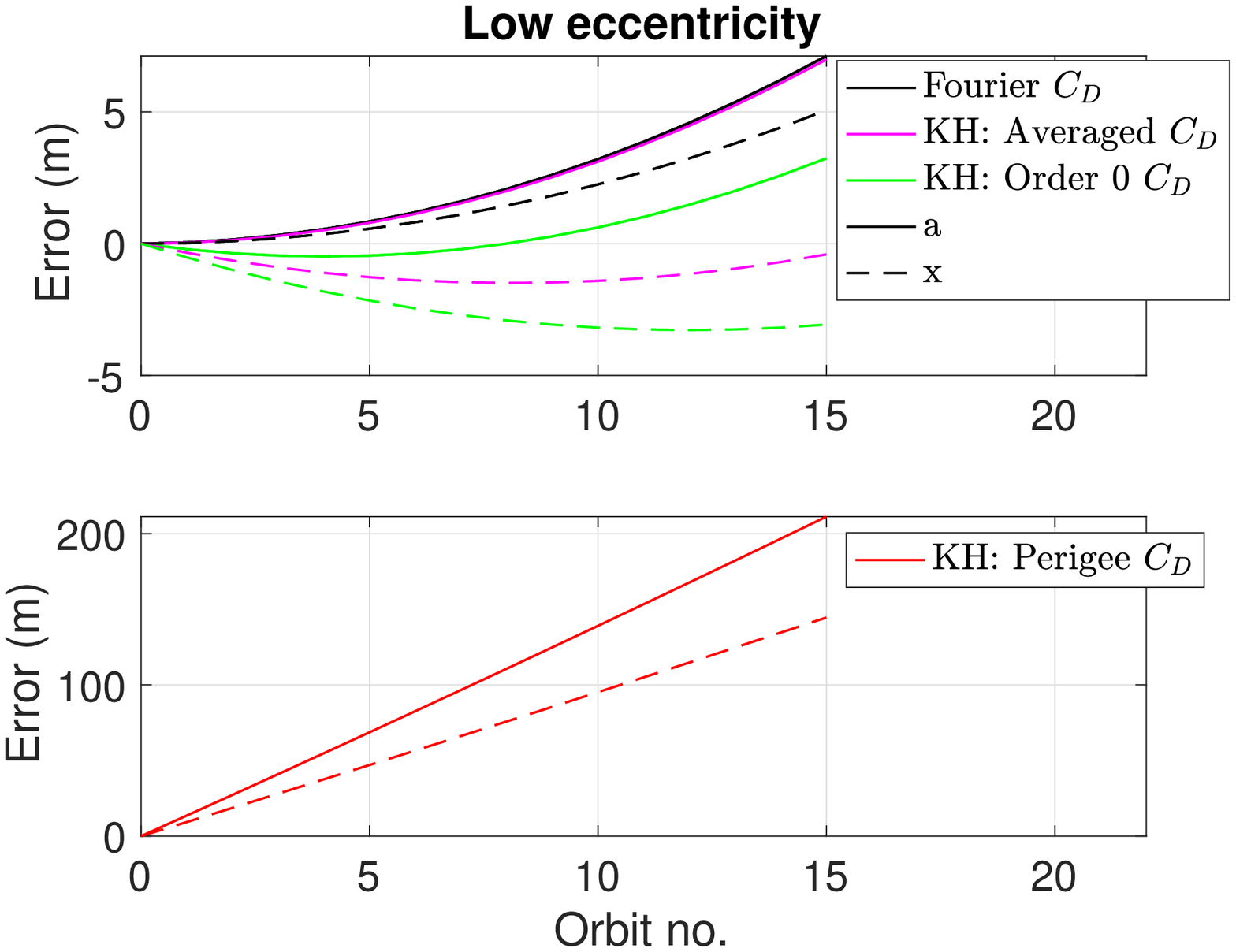}
  (a)
\end{minipage}%
\begin{minipage}{0.5\textwidth}
  \centering
  \includegraphics[width=1\linewidth]{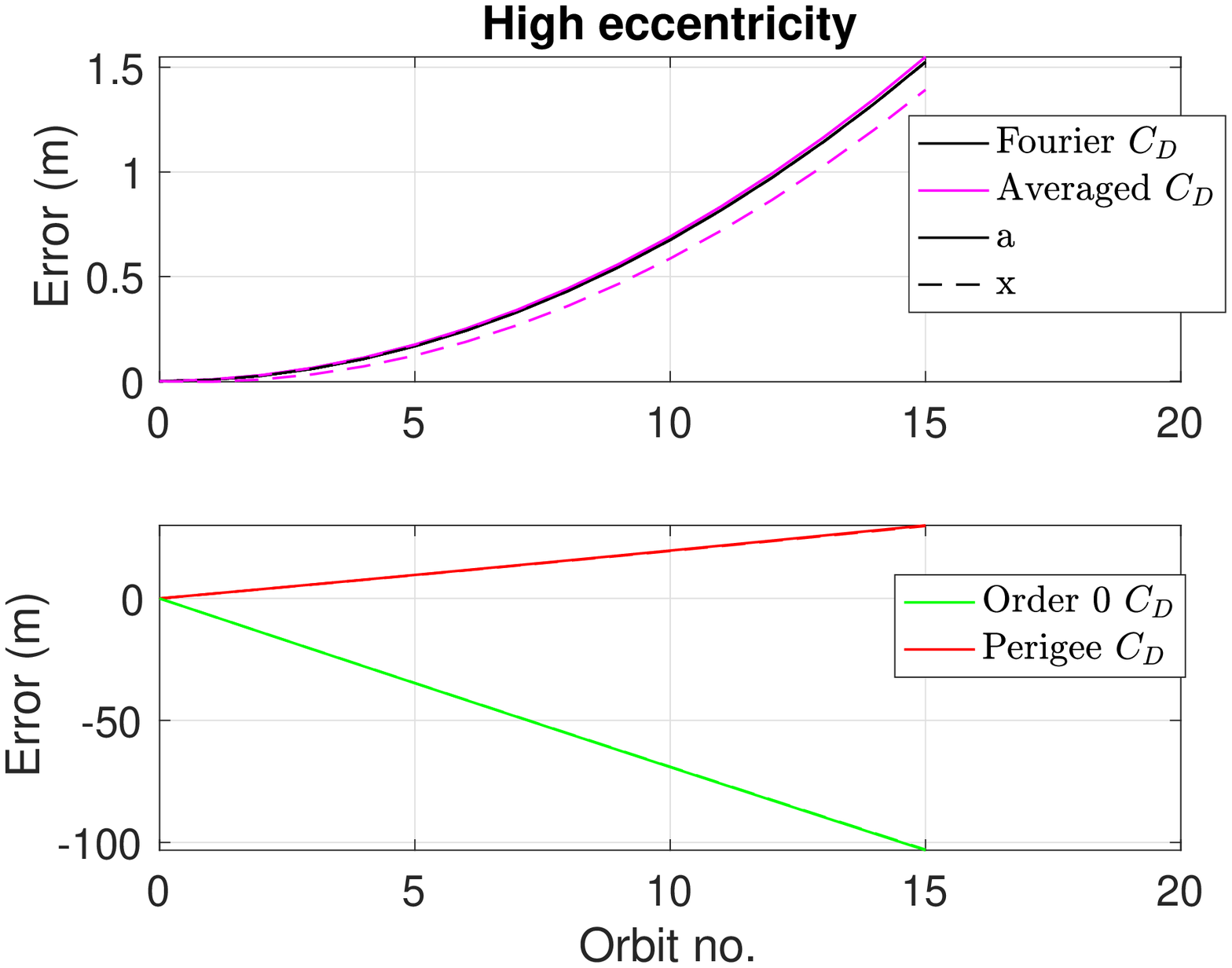}
  (b)
\end{minipage}
\caption{Error between analytical and numerical changes in semi-major axis and focal length for the the BFF model and the original King-Hele (KH) theory with three constant drag-coefficients (density-averaged, perigee and order 0 Fourier) in (a) low eccentricity regime and (b) high eccentricity regime for an inertially stabilized satellite}
\label{fig_bff_iner}
\end{figure*}

\begin{figure*}[H]
\centering
\begin{minipage}{0.5\textwidth}
  \centering
  \includegraphics[width=1\linewidth]{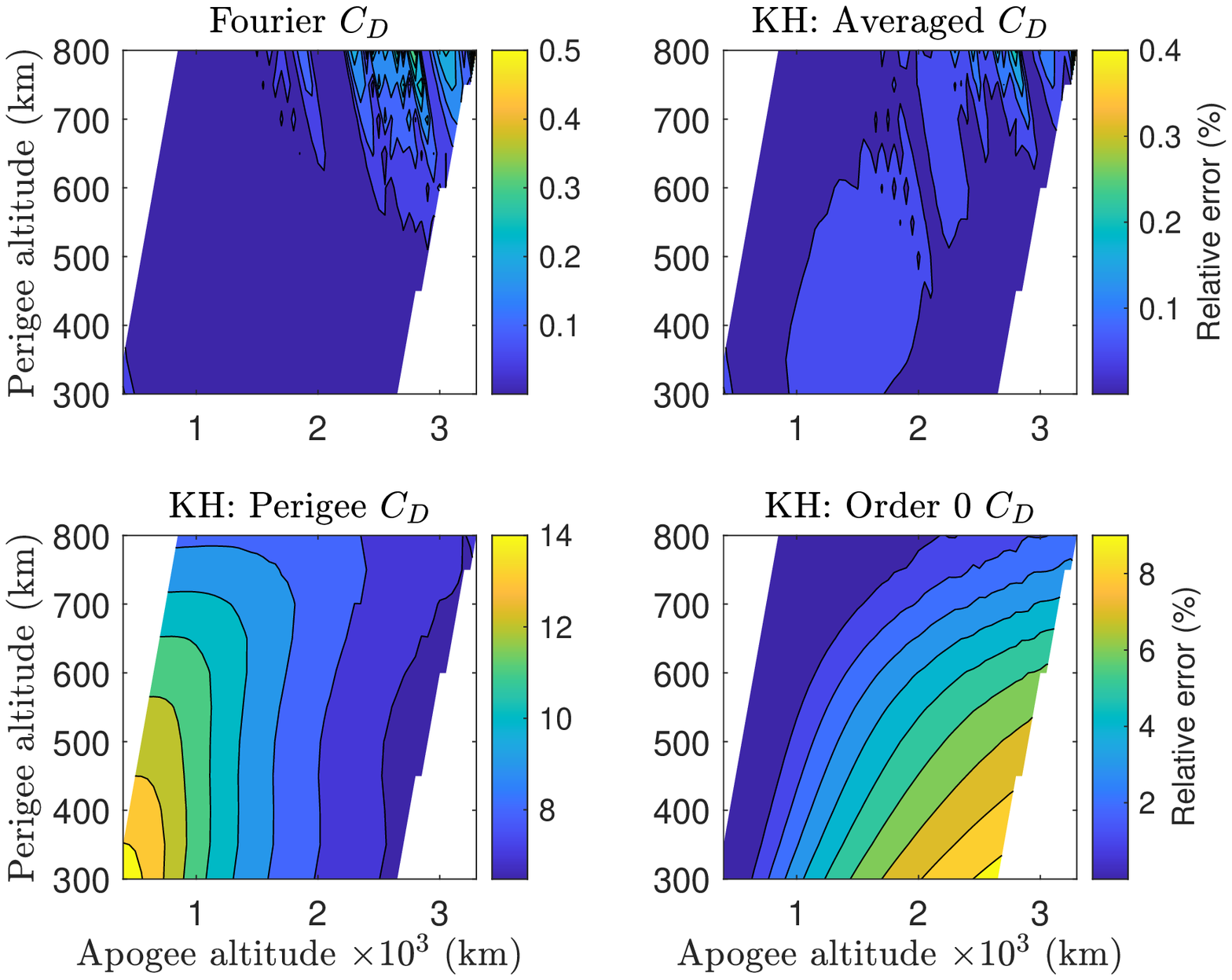}
  (a)
\end{minipage}%
\begin{minipage}{0.5\textwidth}
  \centering
  \includegraphics[width=1\linewidth]{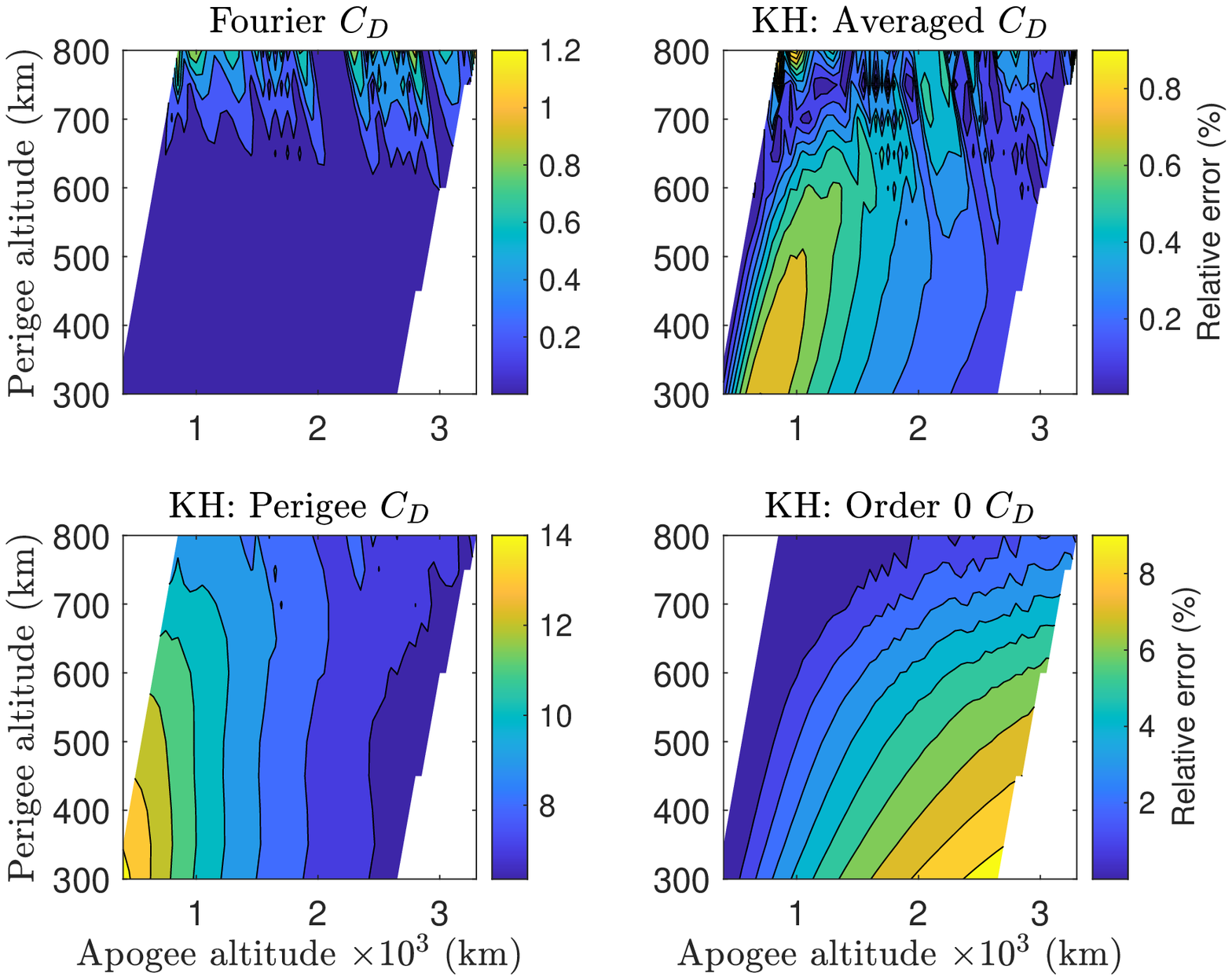}
  (b)
\end{minipage}
\caption{Relative error in analytically computed change in (a) semi-major axis and (b) focal length compared to numerical results for BFF model and original King-Hele (KH) theory with three constant drag-coefficients (density-averaged, perigee and order 0 Fourier) in low eccentricity regime for an inertially stabilized profile}
\label{bff_grid_lowEcc_iner}
\end{figure*}

\begin{figure*}[H]
\centering
\begin{minipage}{0.5\textwidth}
  \centering
  \includegraphics[width=1\linewidth]{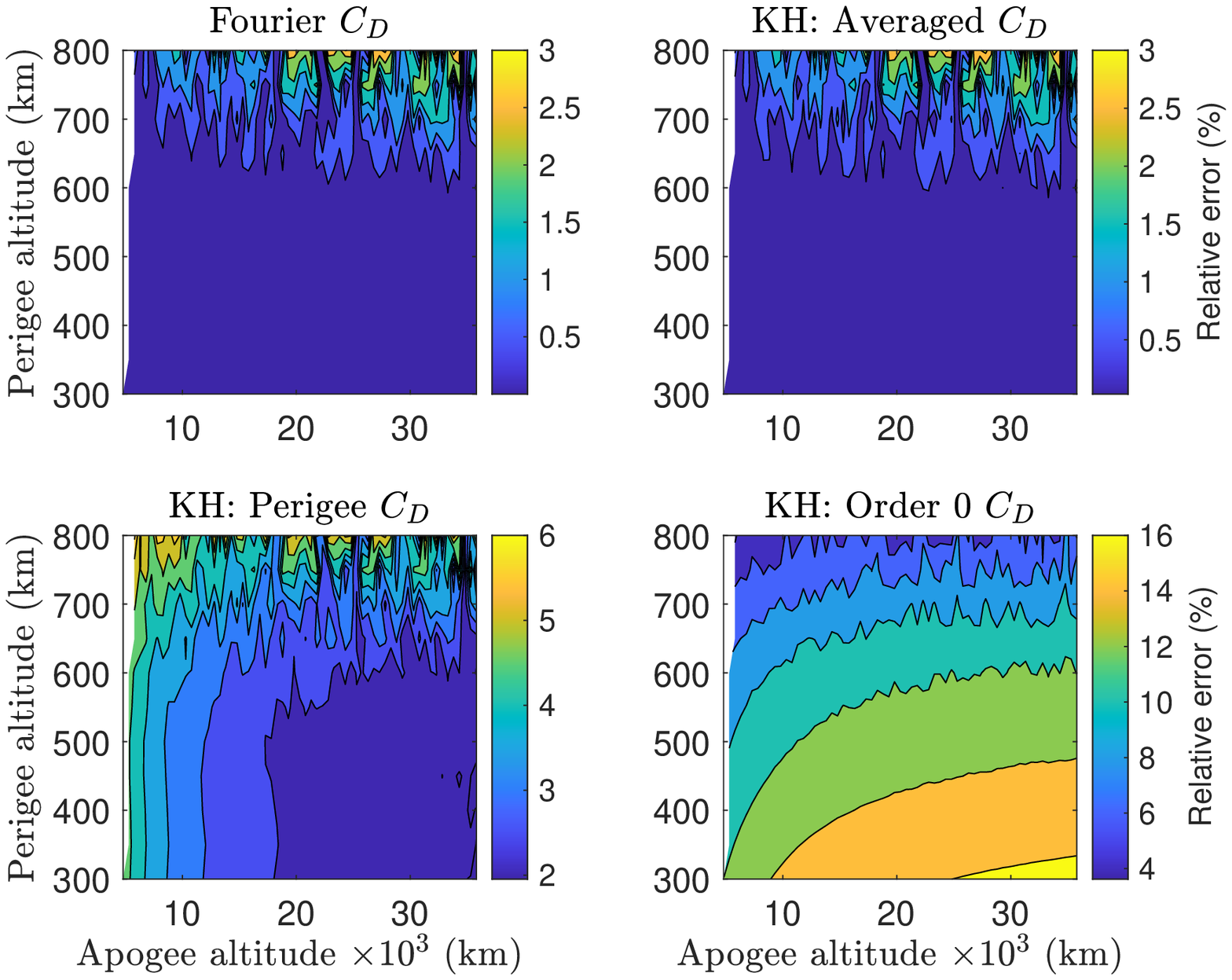}
  (a)
\end{minipage}%
\begin{minipage}{0.5\textwidth}
  \centering
  \includegraphics[width=1\linewidth]{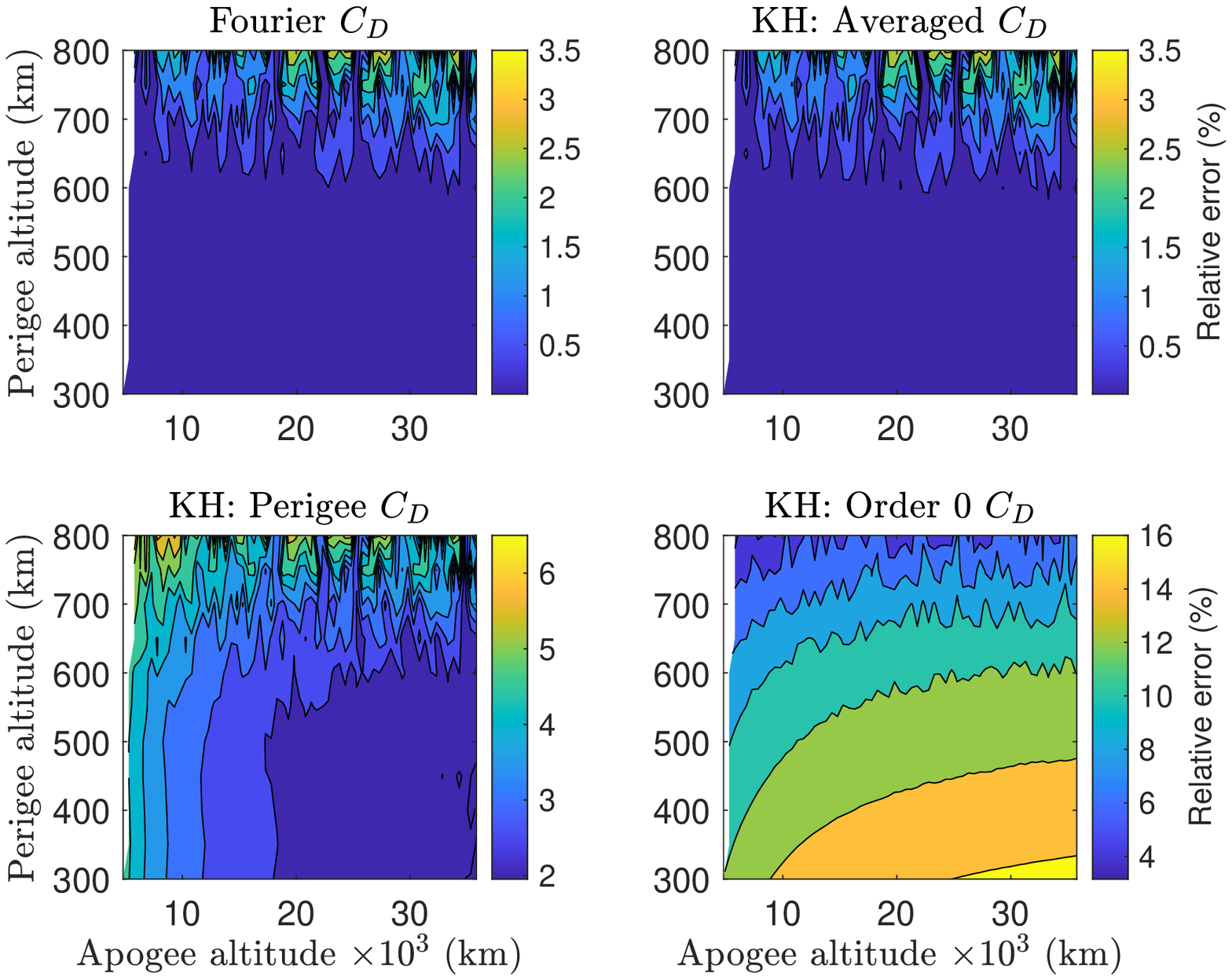}
  (b)
\end{minipage}
\caption{Relative error in analytically computed change in (a) semi-major axis and (b) focal length compared to numerical results for BFF model in and original King-Hele (KH) theory with three constant drag-coefficients (density-averaged, perigee and order 0 Fourier) in high eccentricity regime for an inertially stabilized profile}
\label{bff_grid_highEcc_iner}
\end{figure*}

For the inertially stabilized case, the averaged $C_D$ performs better than the full Fourier theory for the particular perigee and apogee heights considered as shown in Fig. \ref{fig_bff_iner}. Over a grid of perigee and apogee altitudes, the full Fourier theory has a larger variation of relative errors in Figs. \ref{bff_grid_lowEcc_iner} and \ref{bff_grid_highEcc_iner}. But overall, it performs better than the averaged drag-coefficient; the errors for the full theory are smaller for 65.7 \% cases of the grid for semi-major axis and 77 \% cases for focal-length in the low eccentricity regime. 

\begin{figure*}[H]
\centering
\begin{minipage}{0.5\textwidth}
  \centering
  \includegraphics[width=1\linewidth]{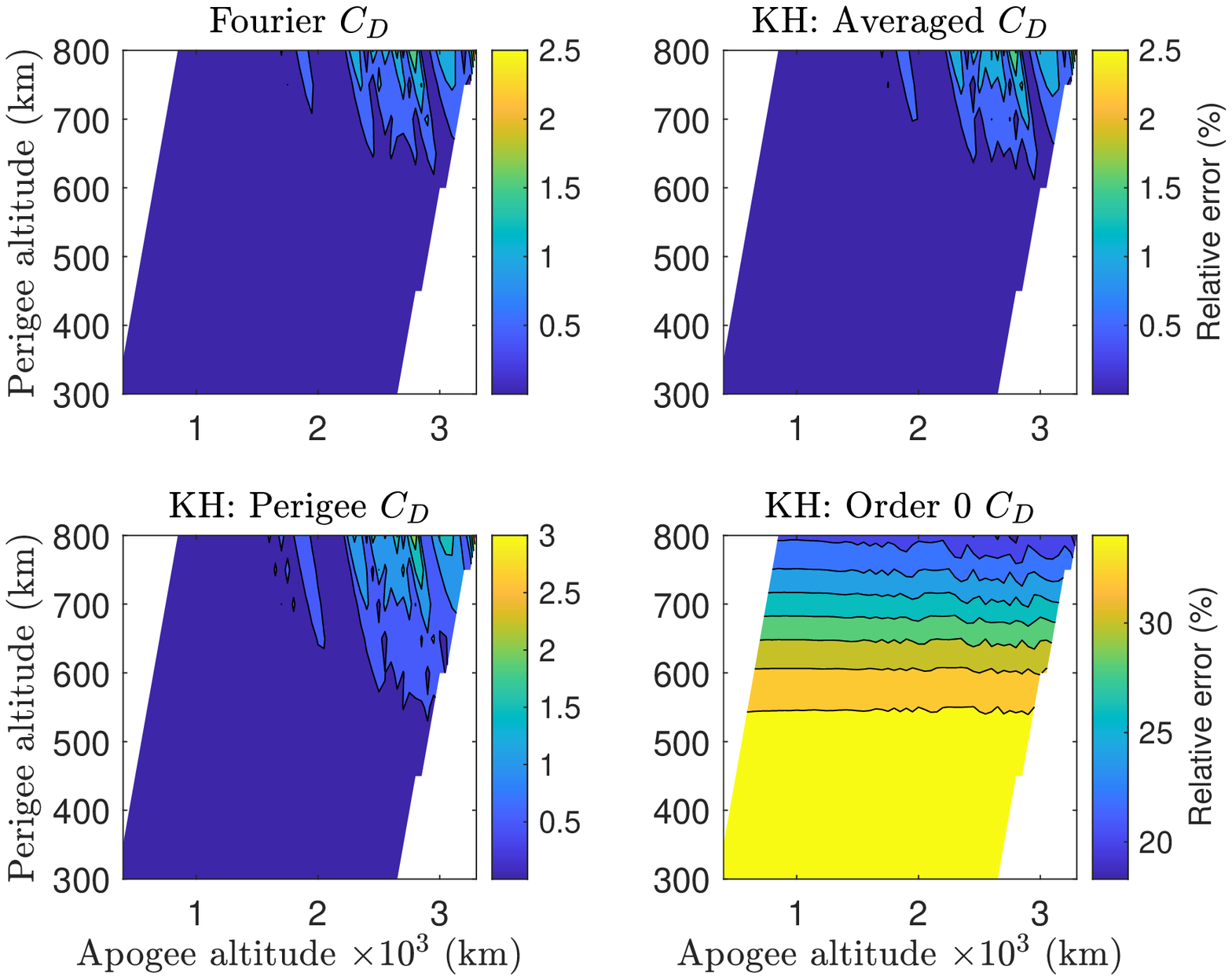}
  (a)
\end{minipage}%
\begin{minipage}{0.5\textwidth}
  \centering
  \includegraphics[width=1\linewidth]{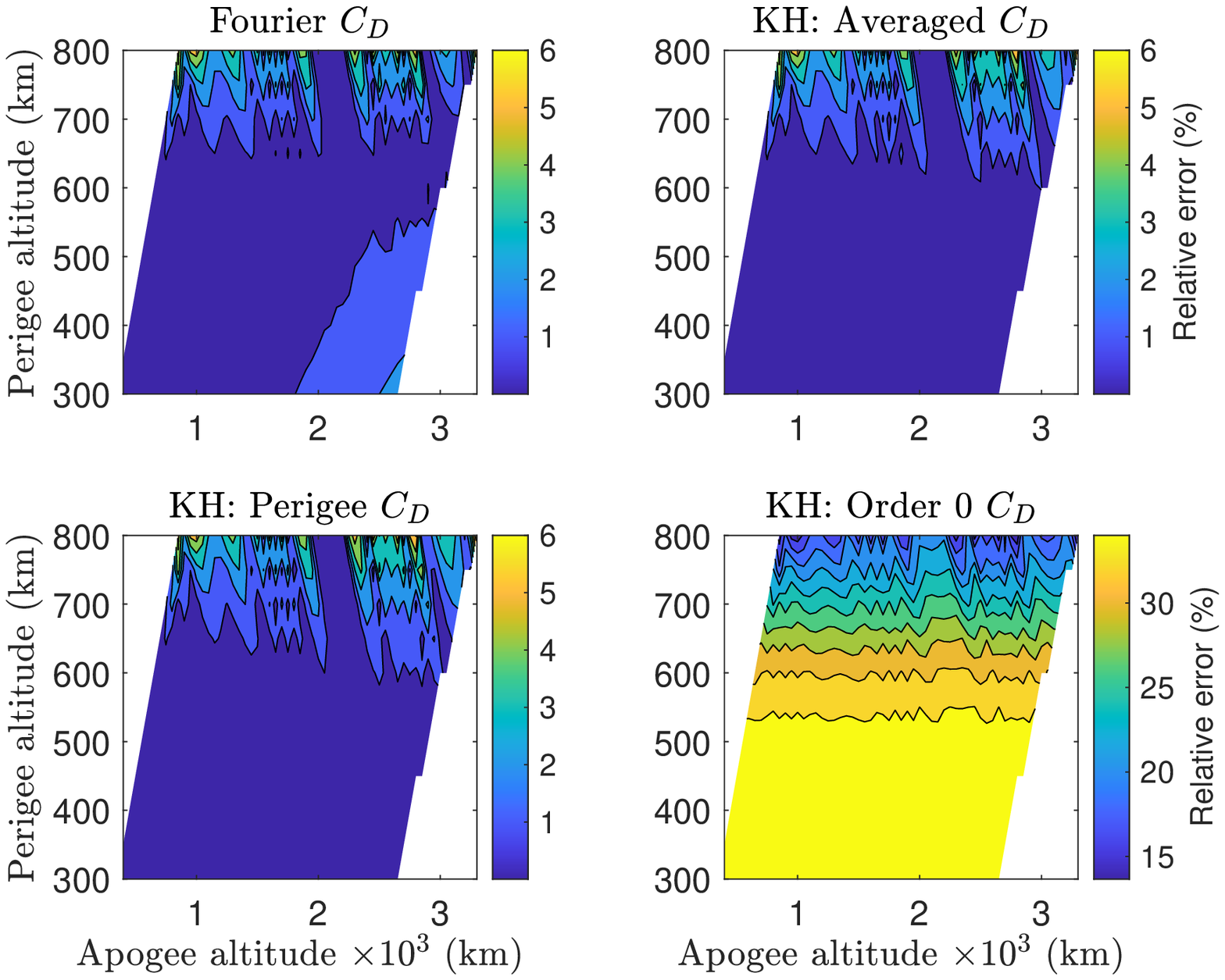}
  (b)
\end{minipage}
\begin{minipage}{0.25\textwidth}
  \centering
  \includegraphics[width=1\linewidth]{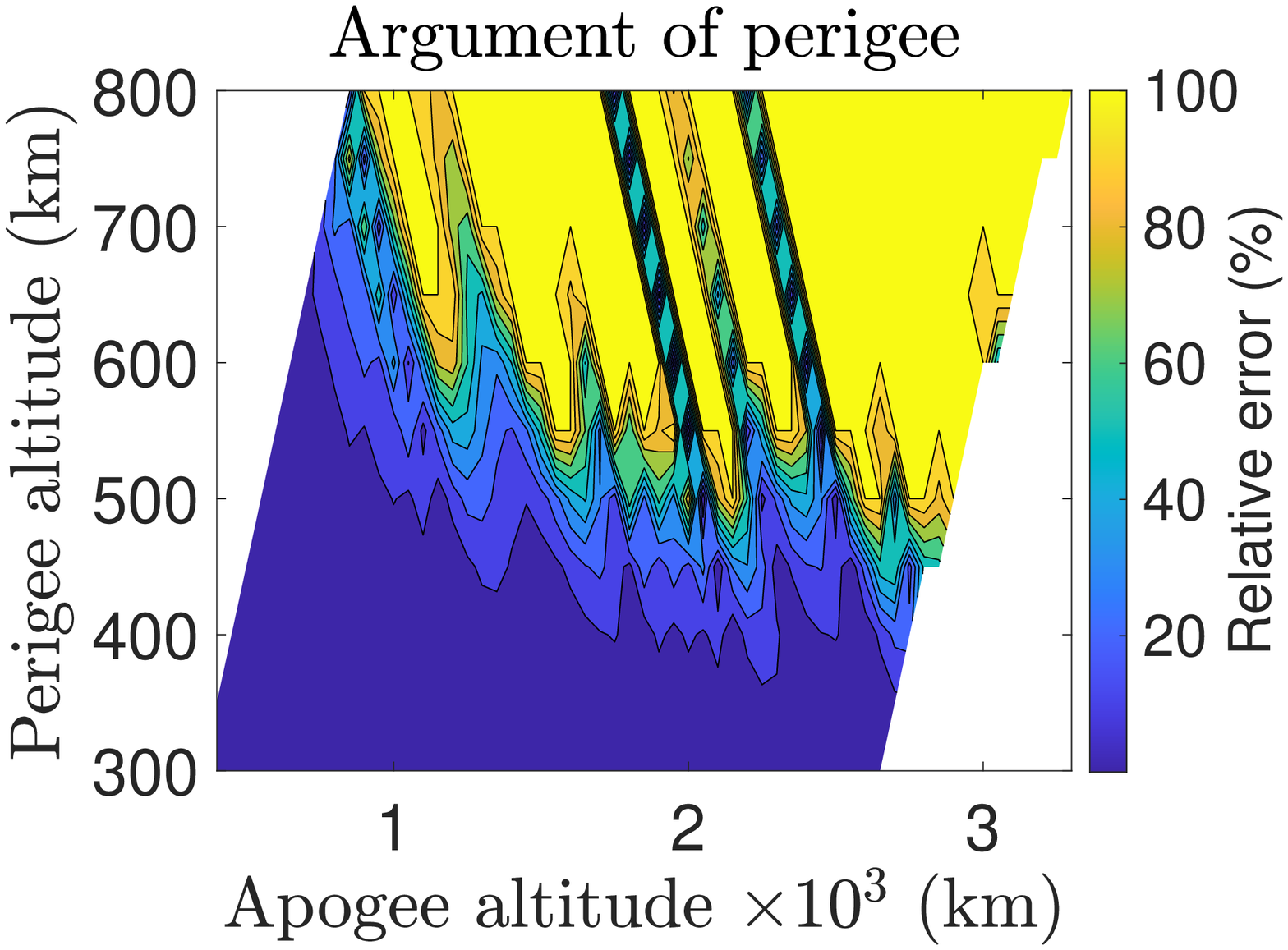}
  (c)
\end{minipage}
\caption{Relative error in analytically computed change in (a) semi-major axis, (b) focal length and (c) argument of perigee compared to numerical results for BFF model in and original King-Hele (KH) theory with three constant drag-coefficients (density-averaged, perigee and order 0 Fourier) in low eccentricity regime for a nadir-pointing asymmetrical satellite}
\label{bff_grid_lowEcc_nadir_asymm}
\end{figure*}

In order to test the theory for the argument of perigee change, an asymmetrical satellite is considered with non-zero $\overline{\mathcal{B}}_n$. The satellite is considered to be of half-trapezoidal shape with one face inclined at $45^\circ$. All the six surfaces are considered to have different material properties such that the satellite is asymmetric in the body frame. The semi-major axis, focal-length and argument of perigee errors for low eccentricity regime are plotted in Fig. \ref{bff_grid_lowEcc_nadir_asymm}. The full Fourier theory performs better than the averaged drag-coefficient for 95 \% of the cases for semi-major axis but for 41 \% of the cases for focal-length. For a constant drag-coefficient, the argument of perigee change is zero. Therefore, 100 \% relative errors are obtained with the original King-Hele formulation. With the full Fourier theory, the errors are less than 100 \% for around 61 \% of the cases. The results are not shown for high-eccentricity regime because the change in argument of perigee over an orbit is negligible.

\begin{figure*}[H]
\centering
\begin{minipage}{0.5\textwidth}
  \centering
  \includegraphics[width=1\linewidth]{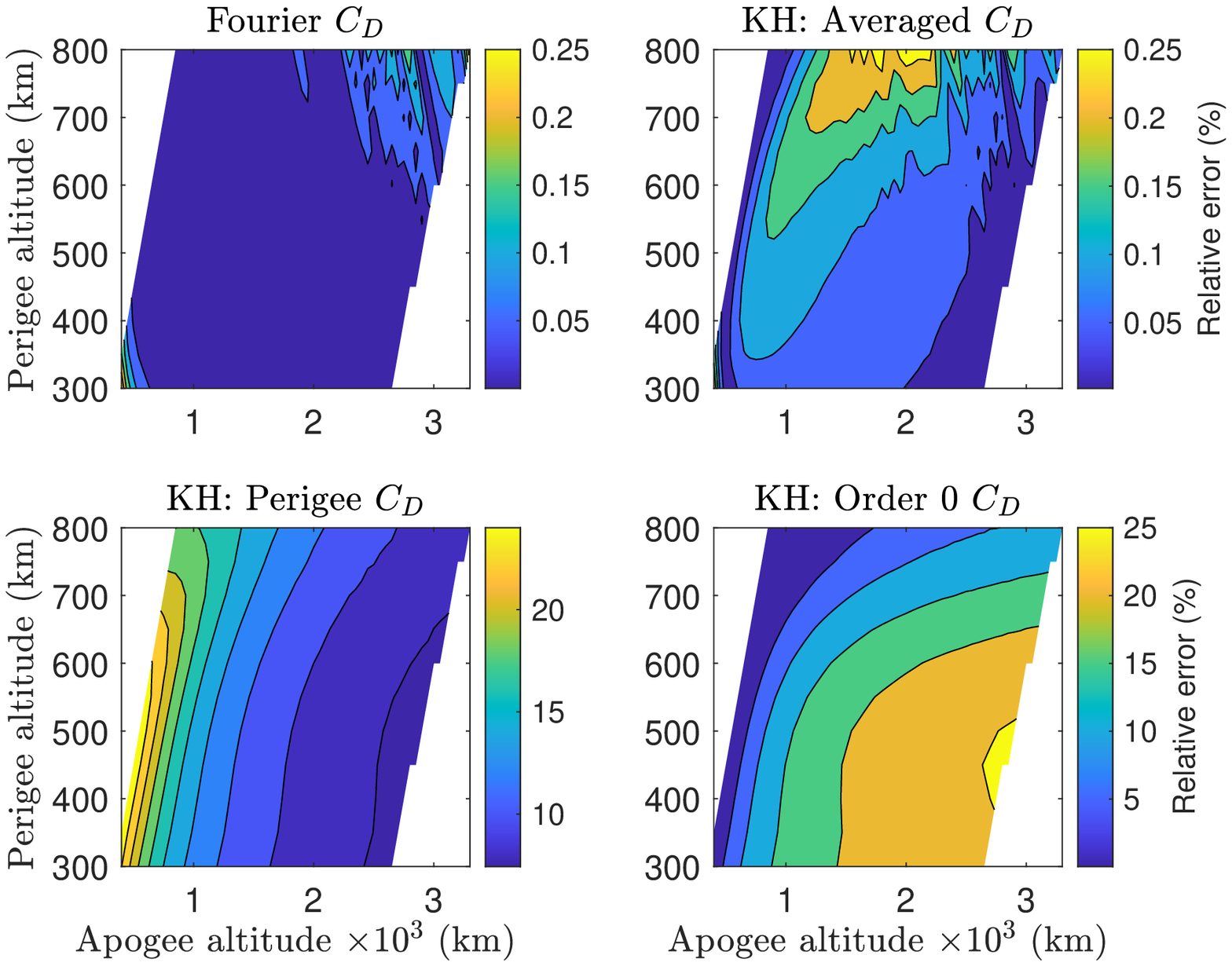}
  (a)
\end{minipage}%
\begin{minipage}{0.5\textwidth}
  \centering
  \includegraphics[width=1\linewidth]{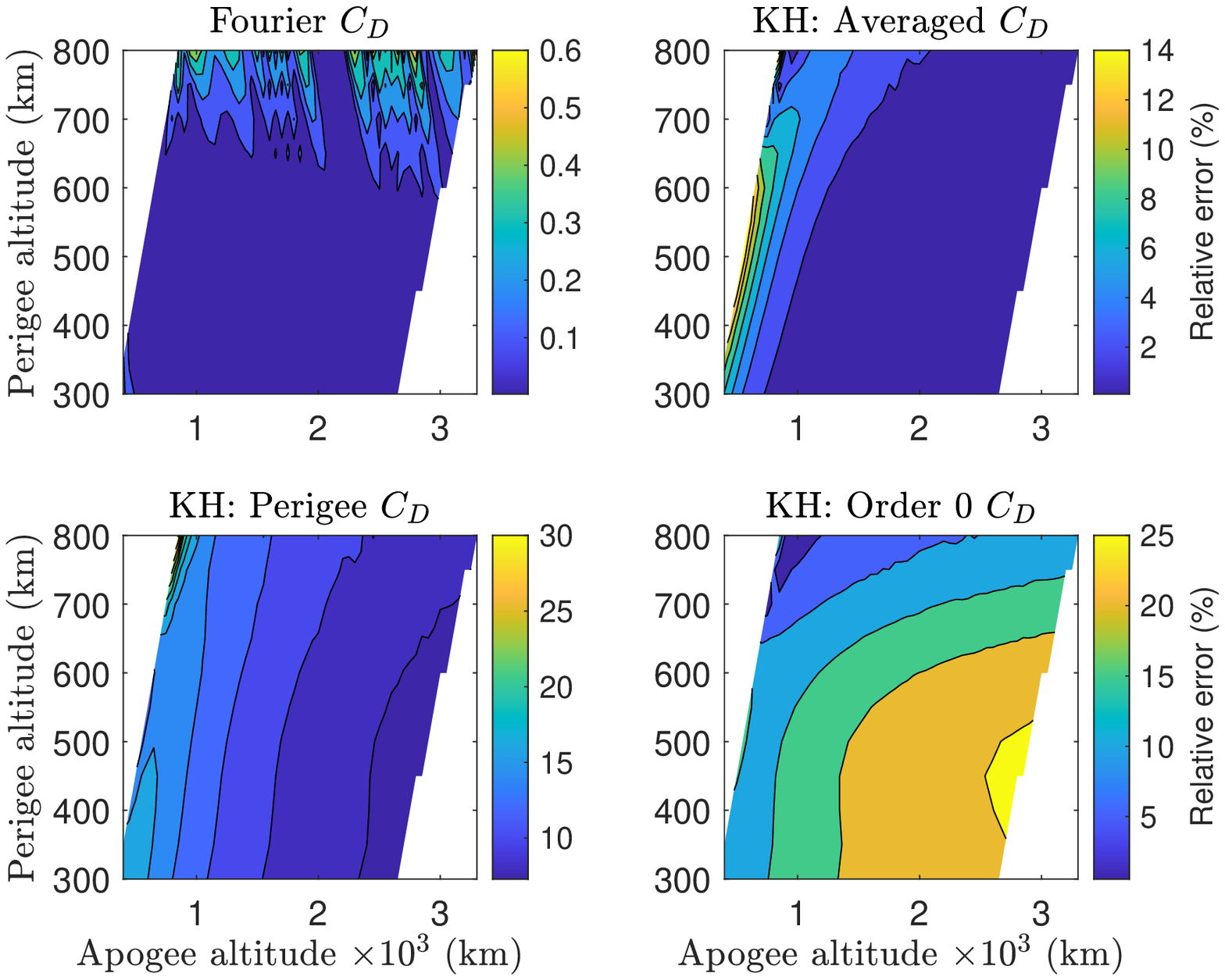}
  (b)
\end{minipage}
\begin{minipage}{0.25\textwidth}
  \centering
  \includegraphics[width=1\linewidth]{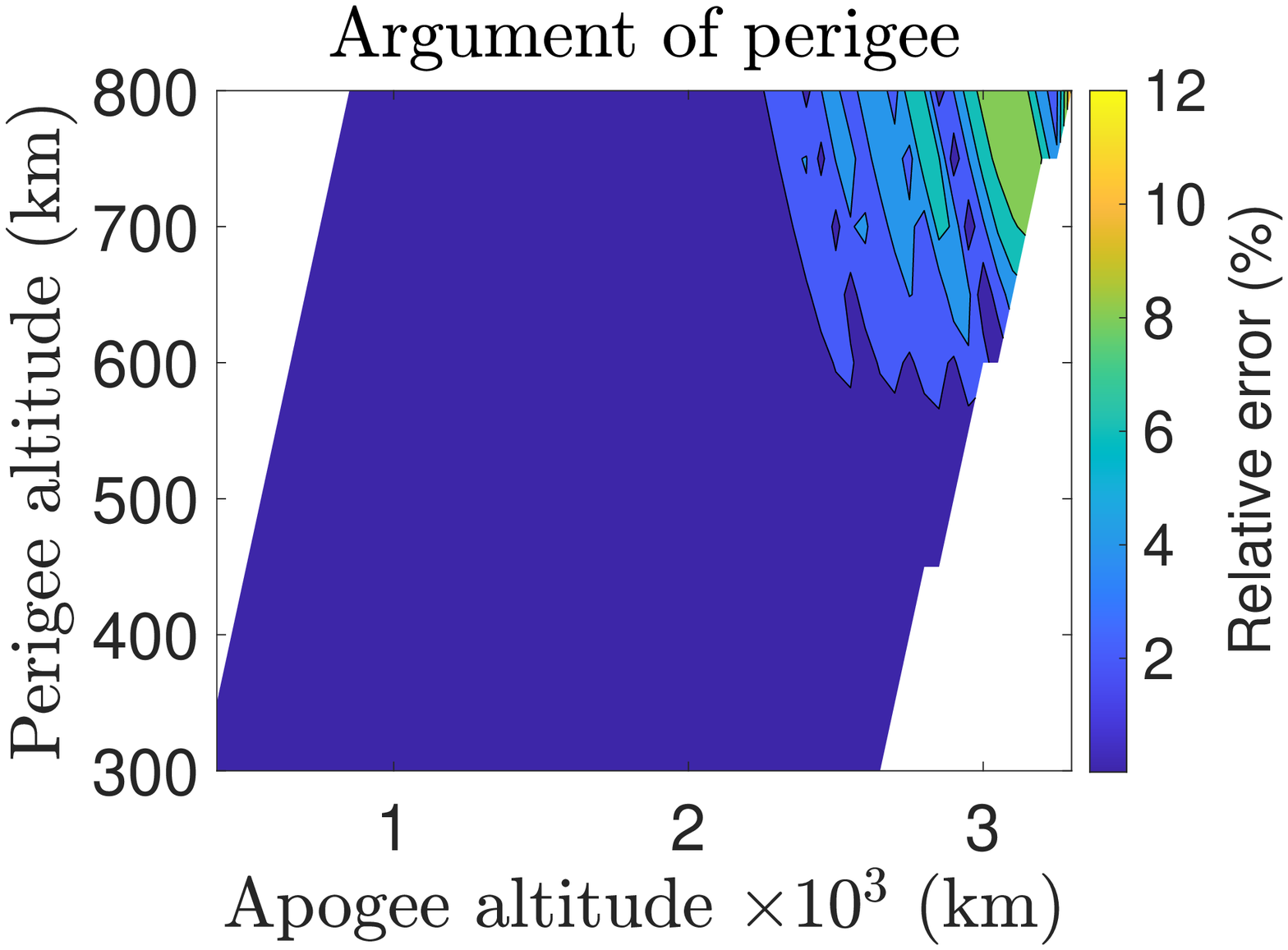}
  (c)
\end{minipage}
\caption{Relative error in analytically computed change in (a) semi-major axis, (b) focal length and (c) argument of perigee compared to numerical results for BFF model in and original King-Hele (KH) theory with three constant drag-coefficients (density-averaged, perigee and order 0 Fourier) in low eccentricity regime for an inertially stabilized asymmetrical satellite}
\label{bff_grid_lowEcc_iner_asymm}
\end{figure*}

The inertially stabilized case is more interesting since the argument of perigee change is larger in this case due to larger variations in the drag-coefficient. The errors in the orbital elements are plotted in Fig. \ref{bff_grid_lowEcc_iner_asymm}. It is evident that the full Fourier theory performs better than the averaged drag-coefficient for all the orbital elements. Similar to the nadir-pointing case, the argument of perigee variation is negligible in the high-eccentricity regime and therefore, the results have not been shown here.

\subsection{Test cases for the BODF model}
\begin{figure*}[H]
\centering
\begin{minipage}{0.5\textwidth}
  \centering
  \includegraphics[width=1\linewidth]{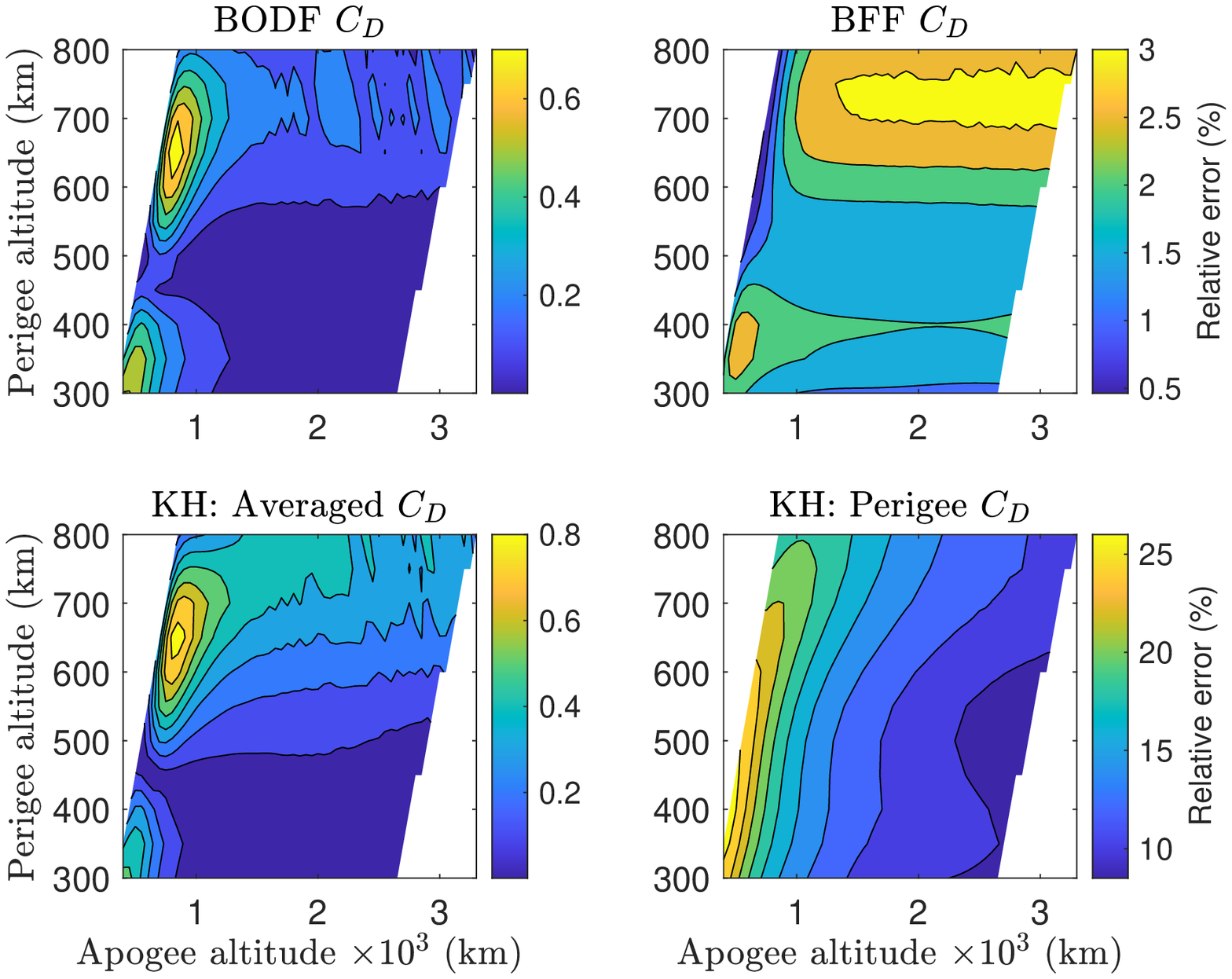}
  (a)
\end{minipage}%
\begin{minipage}{0.5\textwidth}
  \centering
  \includegraphics[width=1\linewidth]{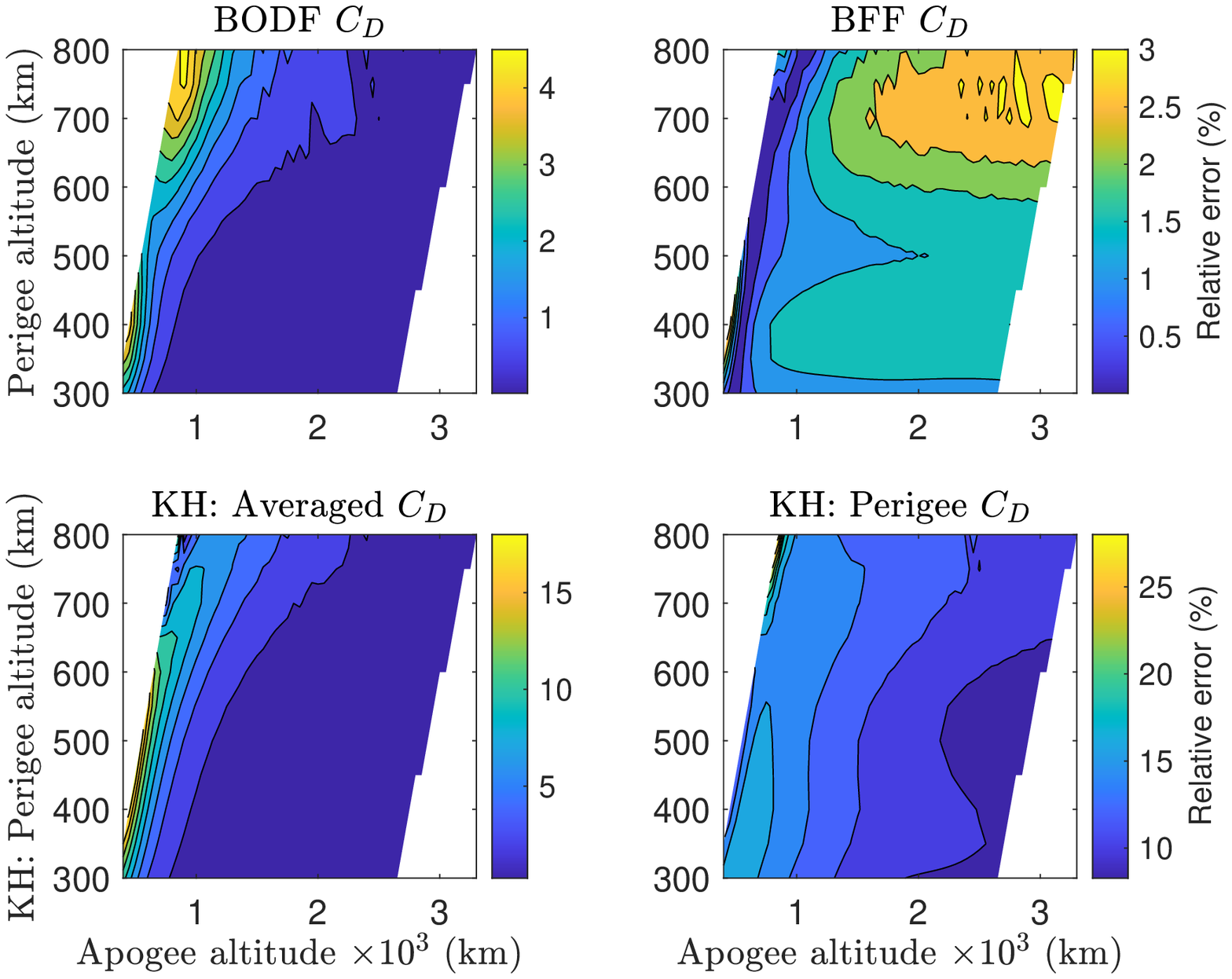}
  (b)
\end{minipage}
\begin{minipage}{0.25\textwidth}
  \centering
  \includegraphics[width=1\linewidth]{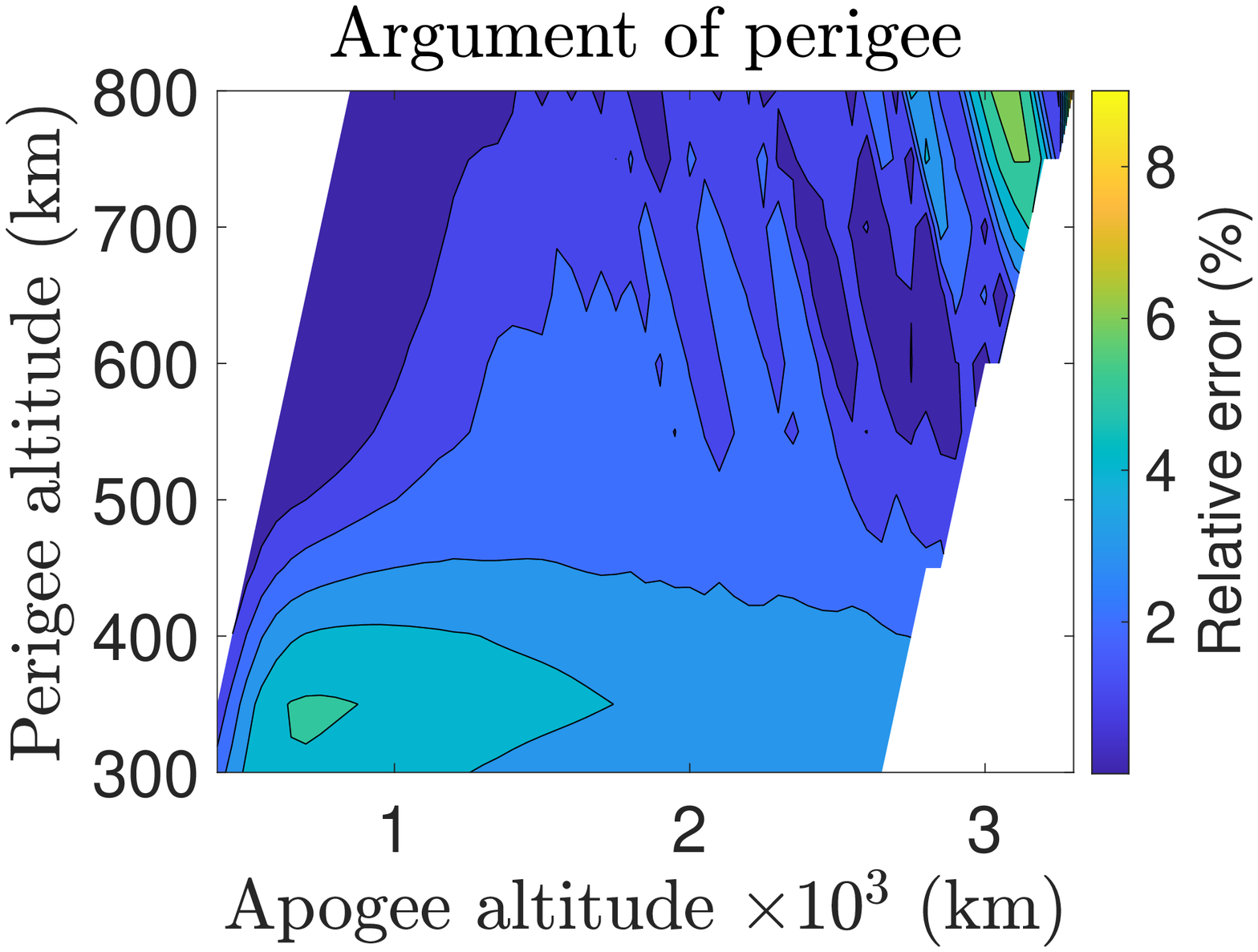}
  (c)
\end{minipage}
\caption{Relative error in analytically computed change in (a) semi-major axis, (b) focal length and (c) argument of perigee compared to numerical results for BODF model, BFF model with coefficients evaluated at perigee and and original King-Hele (KH) theory with two constant drag-coefficients (density-averaged and perigee) in low eccentricity regime for an inertially stabilized profile}
\label{bodf_grid_lowEcc_iner}
\end{figure*}

\begin{figure*}[H]
\centering
\begin{minipage}{0.5\textwidth}
  \centering
  \includegraphics[width=1\linewidth]{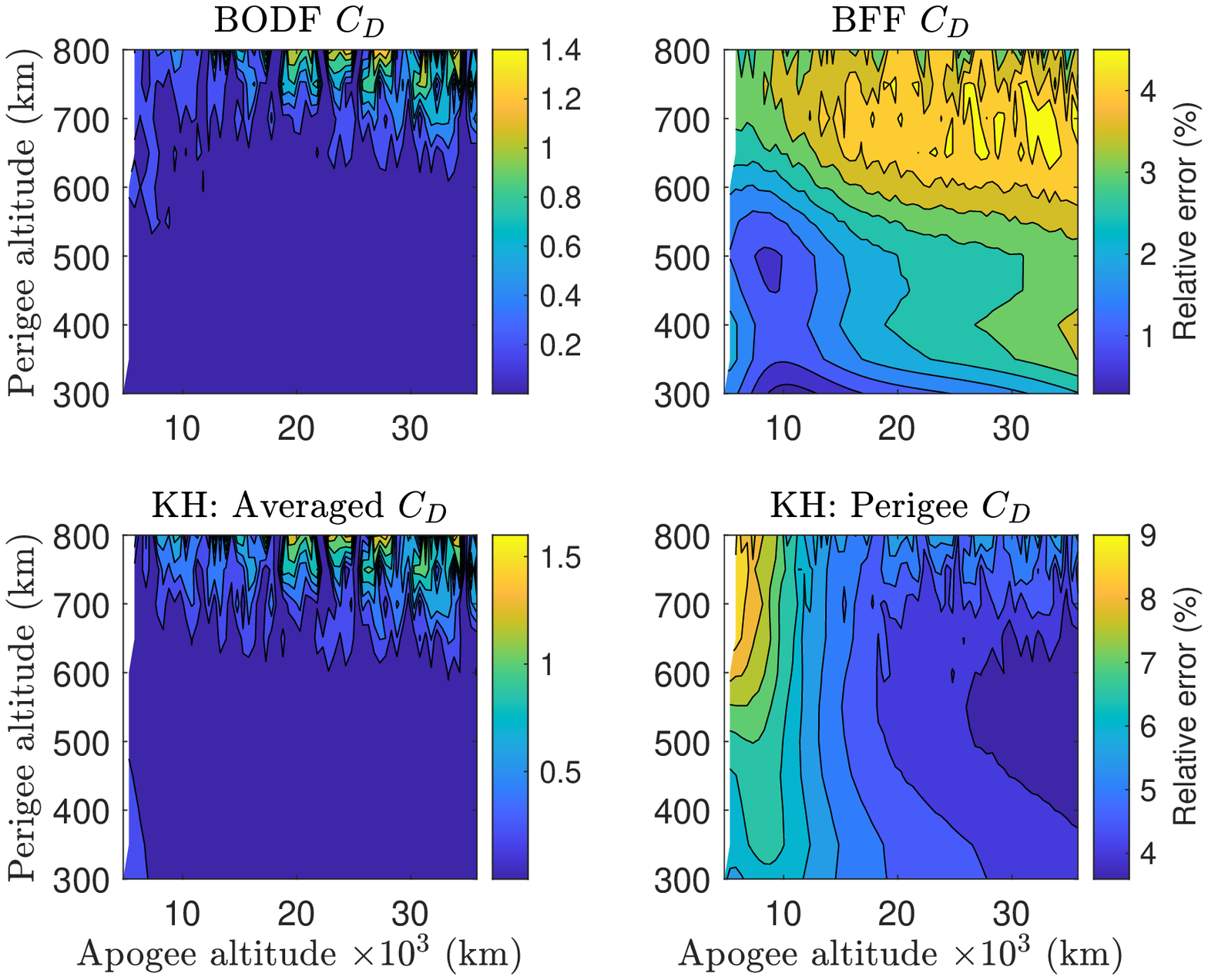}
  (a)
\end{minipage}%
\begin{minipage}{0.5\textwidth}
  \centering
  \includegraphics[width=1\linewidth]{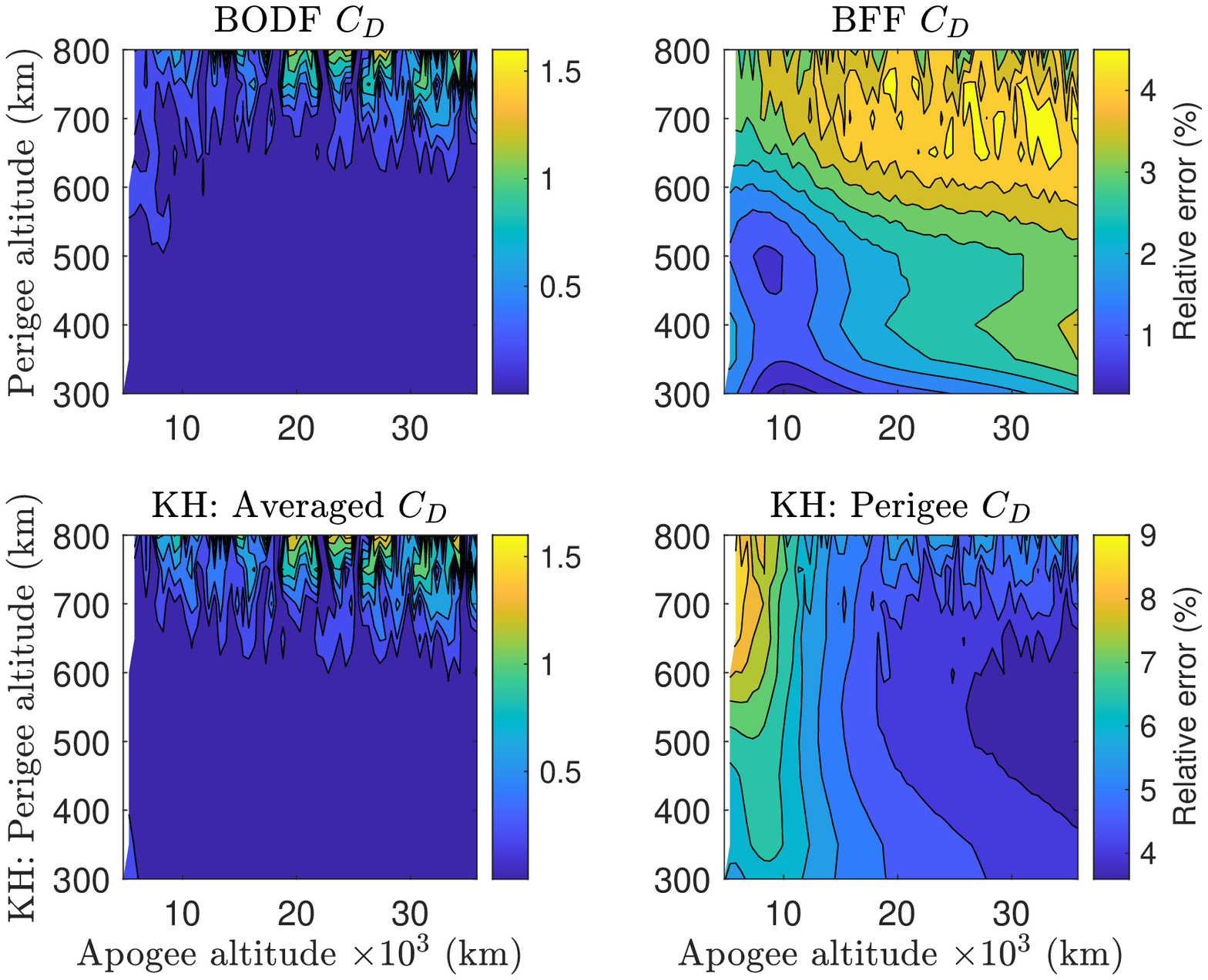}
  (b)
\end{minipage}
\begin{minipage}{0.25\textwidth}
  \centering
  \includegraphics[width=1\linewidth]{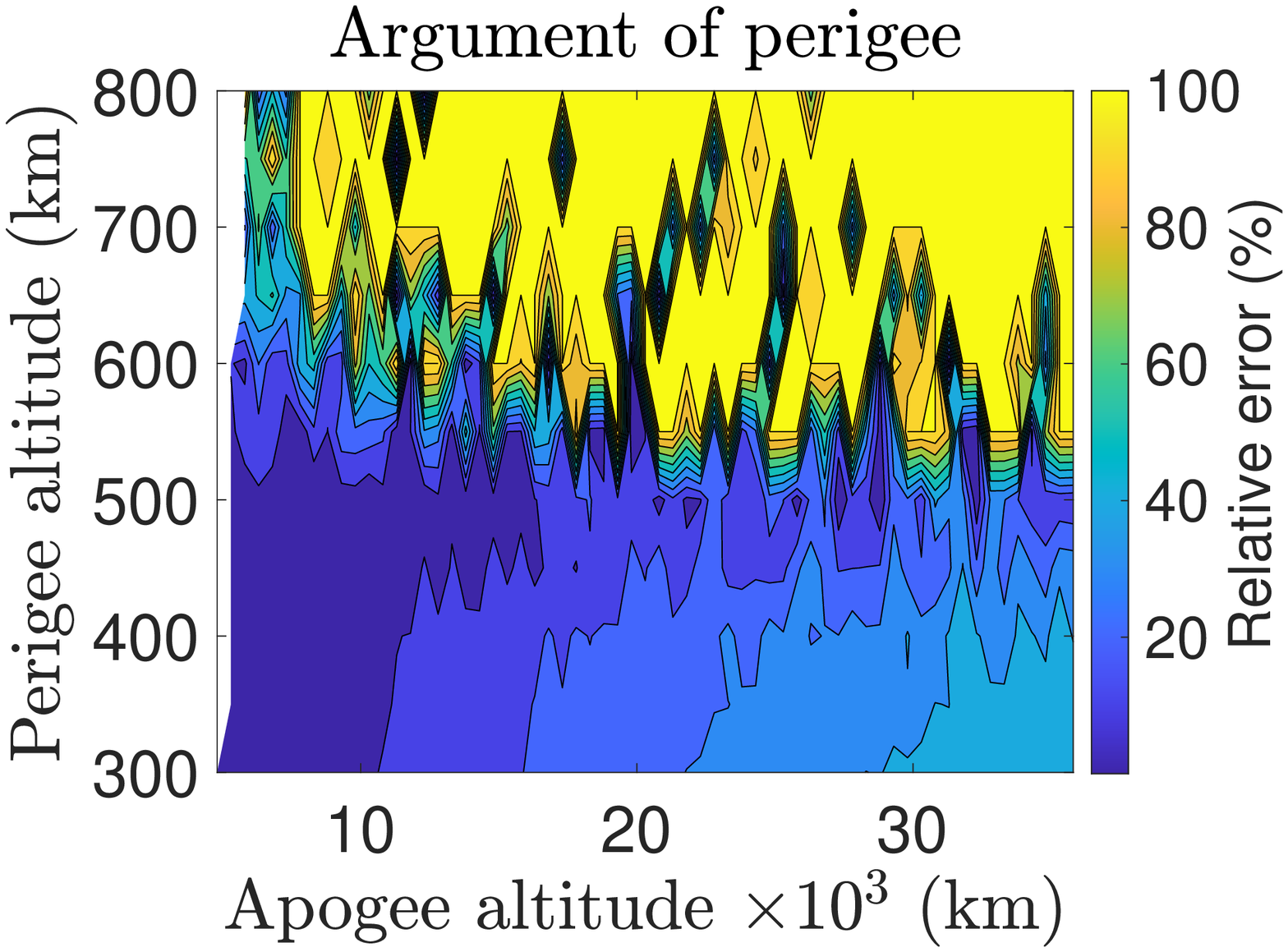}
  (c)
\end{minipage}
\caption{Relative error in analytically computed change in (a) semi-major axis, (b) focal length and and (c) argument of perigee compared to numerical results for BODF model, BFF model with coefficients evaluated at perigee and and original King-Hele (KH) theory with two constant drag-coefficients (density-averaged and perigee) in high eccentricity regime for an inertially stabilized profile}
\label{bodf_grid_highEcc_iner}
\end{figure*}

In order to validate the BODF model, the asymmetrical satellite introduced for the BFF model is considered.  The results of BODF are compared with BFF for which the Fourier coefficients are evaluated at perigee. Figs. \ref{bodf_grid_lowEcc_iner} and \ref{bodf_grid_highEcc_iner} plot the relative errors for the BODF model compared to BFF model and constant drag-coefficients for low and high eccentricity regimes. The argument of perigee errors are calculated only for the BODF model. It can be seen that in both cases, BODF has the highest accuracy in maximum areas of the grid, followed by the averaged drag-coefficient except for focal length in low eccentricity regime. Simply averaging the BFF coefficients weighted by density over the orbit can improve the prediction performance over a constant set of BFF coefficients evaluated at perigee. 

The simulation results for BFF, OFF and BODF models clearly demonstrate that capturing the periodic variation of the drag-coefficient in orbit can lead to improvements in predicting the evolution of the orbital elements. For many of the cases, the density-averaged drag-coefficient derived from the full Fourier theory is as accurate as the full Fourier theory. Therefore, if using the original King-Hele theory, the density-averaged drag-coefficients derived in Eqs. \ref{cdrho}, \ref{cdrho1} and \ref{cdrho2} should be used depending on the case. It should be noted that for periodic attitude profiles, such as the cases considered here, the drag-coefficient variation due to both ambient parameters and attitude can be captured using OFF model. But the theory developed here considers the BFF model separately since the BFF coefficients are physically different from OFF coefficients and are fixed to the body-frame. Therefore, they don't have to be evaluated for different orbital parameters unlike the OFF coefficients.

\section{Conclusion}
\label{conc}
This paper addresses the problem with a constant drag-coefficient in the King-Hele theory and derives a modified theory with a time-varying drag-coefficient. Under the assumptions of the original King-Hele formulation for a symmetric exponentially decaying atmosphere with a constant scale height, the drag-coefficient dependence on ambient parameters is periodic and can be expressed as a Fourier series in the orbit-fixed frame. Similarly, the variation of the drag-coefficient with orientation of the velocity vector in the body frame can be captured using a Fourier series expansion in the body-frame. Using these two models, the King-Hele theory is extended to include the variation of drag-coefficient in the averaging equations. An approximate framework is provided to capture the dependence of the drag-coefficient on both body and orbit dependent factors. In the original King-Hele theory and subsequent modifications, the constant drag-coefficient that should be used is not explicitly stated since the variation of drag-coefficient is not considered. This paper provides an analytical formula for the constant drag-coefficient that approximates the full Fourier theory most accurately. The developed theory predicts secular changes in the argument of perigee for an asymmetrical satellite with periodic attitude variations whereas the original King-Hele formulation states the change to be zero under the assumptions of the theory. The simulation results for the body-fixed Fourier (BFF), orbit-fixed Fourier (OFF) and body-orbit double Fourier (BODF) models demonstrate that the predictions of orbital element evolution can be improved by allowing the drag-coefficient to vary in the averaging integrals. The improvements can be orders of magnitude depending on the constant drag-coefficient being used. This development can lead to improvements in estimation of orbital lifetimes and derivation of densities from orbit decay data. The theory developed for OFF model can be used for satellites with no variations in attitude or whose attitude profiles are unknown. The BFF model with the Fourier coefficients averaged in the orbit can be used for a general case with a known attitude profile. Additionally, the averaged equations for OFF and BFF models can be used in semi-analytical theories of satellite orbit propagation by considering the slow variation of the Fourier coefficients due to their dependence on the semi-major axis and eccentricity that will be addressed in future work. 

The theory developed in this paper is independent of the underlying physical model for the drag-coefficient being used. Therefore, future developments in drag-coefficient modeling can be used to improve estimates of the Fourier coefficients, especially at higher altitudes, that can be then used in this theory to increase fidelity of orbital element predictions. Moreover, the theory can be easily supplemented with other modifications developed in literature such as extensions to accommodate generic atmospheric density models and adapting the theory to non-singular elements, to provide a complete analytical theory for a satellite in an atmosphere.

\section*{Data Availability}
The synthetic data that support the findings of this study are available from the corresponding author, Vishal Ray, upon reasonable request.
 



\bibliographystyle{mnras}
\bibliography{example} 
S 


\bsp	
\label{lastpage}
\end{document}